\documentclass[preprint,12pt]{elsarticle}

\usepackage{lineno}
\usepackage[T1]{fontenc}
\usepackage[unicode]{hyperref}
\usepackage{amssymb}
\usepackage{amsmath}
\usepackage{bm}
\newcommand{\degree}{$^\circ$}
\usepackage{float}
\usepackage{subcaption}
\usepackage{booktabs}
\usepackage{multirow}
\usepackage{xcolor}
\usepackage[margin=1in]{geometry}
\usepackage{makecell}
\usepackage[final]{changes}

\journal{Journal}

\begin{document}

\begin{frontmatter}

\title{Machine learning based surrogate models for microchannel heat sink optimization}

\author[inst1,inst2]{Ante Sikirica}
\ead{ante.sikirica@uniri.hr}

\author[inst1,inst2]{Luka Grbčić}
\ead{luka.grbcic@riteh.hr}

\author[inst1,inst2]{Lado Kranjčević\texorpdfstring{\corref{cor1}}{}}
\ead{lado.kranjcevic@riteh.hr}

\cortext[cor1]{Corresponding author}

\affiliation[inst1]{organization={Center for Advanced Computing and Modelling, University of Rijeka},
            addressline={Radmile Matejčić 2}, 
            city={Rijeka},
            postcode={51000 Rijeka}, 
            country={Croatia}}

\affiliation[inst2]{organization={Faculty of Engineering, University of Rijeka},
            addressline={Vukovarska 58}, 
            city={Rijeka},
            postcode={51000 Rijeka}, 
            country={Croatia}}

\begin{abstract}
Microchannel heat sinks are an efficient cooling method for semiconductor packages. However, to properly cool increasingly complex and thermally dense circuits, microchannel designs should be improved and expanded on. In this paper, microchannel designs with secondary channels and with ribs are investigated using computational fluid dynamics and are coupled with a multi-objective optimization algorithm to determine and propose optimal solutions based on observed thermal resistance and pumping power. A workflow that combines Latin hypercube sampling, machine learning-based surrogate modeling and multi-objective optimization is proposed. Random forests, gradient boosting algorithms and neural networks were considered during the search for the best surrogate. We demonstrated that tuned neural networks can make accurate predictions and be used to create an acceptable surrogate model. Optimized solutions show a negligible difference in overall performance when compared to the conventional optimization approach. Additionally, solutions are calculated in one-fifth of the original time. Generated designs attain temperatures that are lower by more than 10\% under the same pressure limits as a convectional microchannel design. When limited by temperature, pressure drops are reduced by more than 25\%. Finally, the influence of each design variable on the thermal resistance and pumping power was investigated by employing the SHapley Additive exPlanations technique. Overall, we have demonstrated that the proposed framework has merit and can be used as a viable methodology in microchannel heat sink design optimization.
\end{abstract}

\begin{keyword}
Microchannel heat sink \sep Machine learning \sep Multi-objective optimization \sep Computational fluid dynamics
\end{keyword}

\end{frontmatter}

\section{Introduction}
\label{sec:introduction}

Microfabrication technology improvements over the last several decades have enabled the production of dense semiconductor packages. However, the resulting increase in thermal density poses a problem as it leads to higher device temperatures, which can cause failure. As technology advances, heat dissipation will become a major performance-limiting factor, thus the development of adequate cooling is of paramount importance.

Tuckerman and Pease \cite{tuckerman1981} introduced microchannel heat sinks (MCHS) as a method for ensuring effective cooling of integrated circuits. These early MCHS designs featured rectangular channels embedded in a thermally conductive substrate. MCHS with rectangular, trapezoidal, and triangular cross-sections were studied by Gunnasegaran et al. \cite{gunnasegaran2010} and Wang et al. \cite{wang2016}. Rectangular cross-sections were deemed optimal, particularly when the aspect ratio of the cross-section is $\approx$ 10. Moreover, it was demonstrated that by reducing the hydraulic diameter, their effectiveness can be increased even further. Recently, Vasilev et al. \cite{vasilev2019} investigated rectangular MCHS designs. The authors have concluded that by modifying the spatial orientation of channels while maintaining the same pumping power, the thermal resistance can be lowered. Xia et al. \cite{xia2015a} investigated various avenues for improving the conventional rectangular MCHS. Different cross-sections, header shapes, and inflow/outflow locations were evaluated. Due to flow disturbance, channels with reentrant cavities performed better in terms of thermal performance. Later, Xia et al. \cite{xia2015b} numerically and experimentally evaluated designs with corrugated channels.

Li et al. \cite{li2016} investigated MCHS with rectangular ribs and triangular cavities. Lower observed temperatures were attributed to flow disturbance and the interruption and redevelopment of the hydrodynamic and thermal boundary layers. A paper by Ghani et al. \cite{ghani2017a} explored the properties of MCHS with sinusoidal cavities and ribs. Bayrak et al. \cite{bayrak2019} presented a comparative numerical analysis that included conventional as well as designs with cavities and ribs. Bidirectional ribs were evaluated in a paper by Wang et al. \cite{wang2019}. The combination of vertical and spanwise ribs provided better thermal performance, albeit with a notable pressure drop. The use of slanted passages to induce secondary flow was investigated by Kuppusamy et al. \cite{kuppusamy2014}, Ghani et al. \cite{ghani2017b} and Memon et al. \cite{memon2020}. Introduced topological changes produced flow disturbance. All authors reported a reduction in thermal resistance and pressure drop. A study by Japar et al. \cite{japar2018} combined cavities, ribs and secondary channels. The proposed composite design, according to the research, has excellent thermal performance.

Recent design improvements include the use of wavy microchannels \cite{sakanova2015, lin2017}, double-layered \cite{hung2012, wu2014, leng2015} and porous MCHS \cite{chuan2015, li2019}. A combined double-layered semi-porous MCHS design was investigated using a multi-objective (MO) optimization approach in a paper by \cite{wang2020} which additionally included an overview of previous design improvement studies. A comparative overview of proposed MCHS designs has been conducted by Lu and Vafai \cite{lu2016} and Sidik et al. \cite{sidik2017}. Ramesh et al. \cite{ramesh2021} conducted an overview of experimental and numerical studies in the field. Recently, a review by Gao et al. \cite{gao2022} summarized all advancements and methodological approaches used for microchannel modeling.

Applied geometrical and topological modifications to MCHS are typically arbitrarily chosen to provide an improvement. Consequently, optimization is required to provide balanced designs for thermal performance and pressure drop. The simplest MCHS optimization workflows combine analytical expressions or surrogate models with single objective iterative processes or optimization algorithms \cite{ma2017, li2018}. A contemporary alternative is topology optimization \cite{lv2018, yan2019, zeng2019, ozguc2021}.

Currently, the dominant workflow employed to optimize MCHS designs incorporates response surface method (RSM) and Non-dominated Sorting Genetic Algorithm (NSGA-II). Based on an initial set of numerical (or analytical) results, a relationship between variables and objectives is established and used in a multi-objective analysis. Papers by Normah et al. \cite{normah2015}, Xia et al. \cite{xia2016}, Shi et al. \cite{shi2019} and Yang and Cao \cite{yang2020} employed analogous methodology with pumping power and thermal resistance designated as optimization objectives. Alperen and Sertac \cite{alperen2020} and Kose et al. \cite{kose2022} considered the use of pumping power and Nusselt number. In a comparative study, Hadad et al. \cite{hadad2019} noted that the Jaya algorithm outperforms Faced Centered Central Composite Design and NSGA-II-based approaches. Kwanda \cite{tartibu2020} recommended using the Improved Augmented $\epsilon$-Constraint Method as it was able to outperform The Strength Pareto Evolutionary Algorithm 2 and NSGA-II. A recent study by Polat and Cadirci \cite{polat2022} suggested the use of artificial neural networks (ANN) coupled with NSGA-II algorithm to generate the Pareto front of an MCHS with diamond-shaped pin fins.

Balachandar et al. \cite{balachandar2015} combined ANN and genetic algorithm to determine the optimal fin geometry. The network was trained using CFD results. A paper by Yang et al. \cite{yang2021} considered a conditional generative adversarial network pix2pix trained on a CFD-generated dataset to propose optimized geometries of micro-pin fin heat sinks. Both studies employed small datasets and lack extensive assessment of developed ANN surrogates. Studies by Mohammadpour et al. \cite{mohammadpour2021, mohammadpour2022} combined CFD and machine learning algorithms to generate optimal designs. In \cite{mohammadpour2021}, support vector regression and PSO were combined and led to a design that reduces temperature by 4 K. Subsequently, the k-nearest neighbors algorithm was used to improve performance in MCHS with double synthetic jets. Improvements of up to 15 K were reported. Multi-objective workflow combining CFD, ANN and NSGA-II algorithm was considered by Sui et al. \cite{sui2022} to optimize the absorber geometries in membrane-based microchannels. Optimized geometry reduced pressure by more than six times. Recently, Bayer et al. \cite{bayer2022} used artificial neural networks trained on CFD data to optimize wavy, double-layered MCHS with porous ribs. Authors reported an improvement of up to 1.2 K. It is important to note, however, that the majority of papers dealing with ML and MCHS lack the assessment depth needed to properly validate a model. This includes dataset partitioning, analysis of the influence of dataset size, and often detailed model tuning that addresses overfitting. Although there has been progress in this aspect, a concise set of guidelines should be defined and adhered to, to ensure the validity of proposed ML models.

As the complexity of electronic chips grows, ensuring adequate heat dissipation will be an increasingly challenging task. Thus, the purpose of this work is to present and propose a methodology for optimizing MCHS designs that is based on high-performance computing and machine learning (ML). The proposed approach should allow for the rapid generation of optimized MCHS designs for a given purpose. The use of machine learning instead of traditional RSM enables a computationally less expensive procedure that, if conducted per accepted guidelines, should ensure overall validity with increased accuracy. The inclusion of learning curves should ensure validity for any considered regression models. Furthermore, ML allows seamless analysis of feature importance thus defining key variables that affect the performance. Initially, a workflow is defined that combines three-dimensional computational fluid dynamics (CFD) simulations with the NSGA-II optimization algorithm. Latin hypercube sampling (LHS) is subsequently used to define a set of design points for which numerical simulations are conducted. Obtained data are thereafter used to train and fine-tune a machine learning model. Common machine learning algorithms such as Random Forests, Gradient Boosting algorithms and ANNs are evaluated. Based on observations, recommendations on appropriate ML choices are given. Finally, the trained machine learning model is coupled with the NSGA-II algorithm, and the resulting Pareto front is compared to the front obtained using the traditional methodology. Microchannel heat sinks with secondary channels and ribs are considered in this paper. Relevant design variables in the optimization process are determined using the novel SHapley Additive exPlanations (SHAP) \cite{lundberg2017} technique. \added{Thus obtained insights are noteworthy even if the ML-based approach is not employed and should hence facilitate and direct design optimization studies to consider and optimize the most relevant design parameters.} Optimal designs based on thermal resistance and pumping power are reported. The presented methodology should be considered as a guideline for future ML implementations when optimizing MCHS.

\section{Problem definition and methodology}
\label{sec:problem_definition_methodology}

\subsection{MCHS models}

\begin{figure}[H]
\centering
\begin{subfigure}[b]{0.495\textwidth}
    \centering
    \includegraphics[trim={3.5cm 1.25cm 3.5cm 1.25cm}, clip, width=\textwidth]{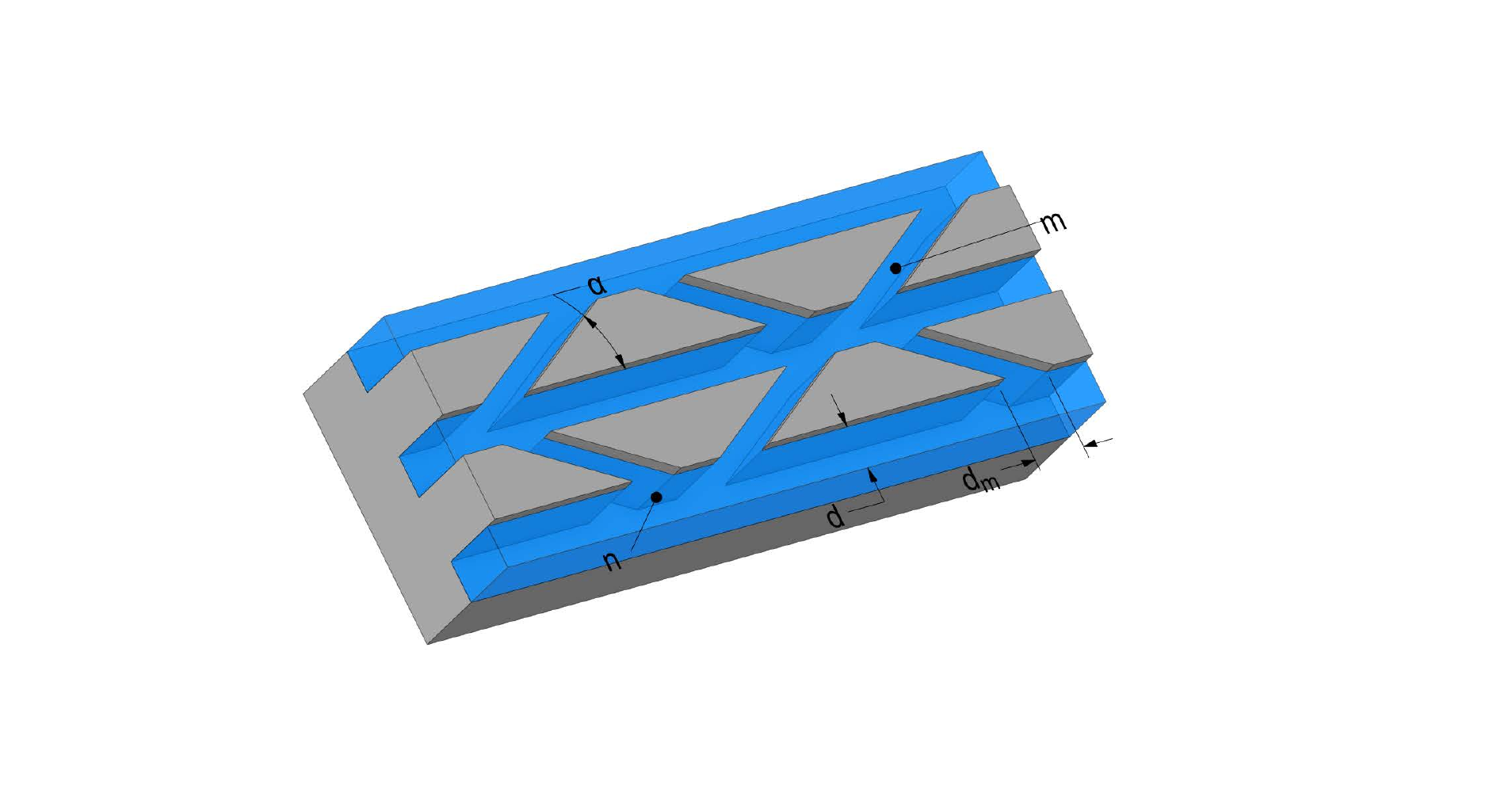}
    \caption{}
    \label{fig:channel_geometry:a}
\end{subfigure}
\begin{subfigure}[b]{0.495\textwidth}
    \centering
    \includegraphics[trim={3.5cm 1.25cm 3.5cm 1.25cm}, clip, width=\textwidth]{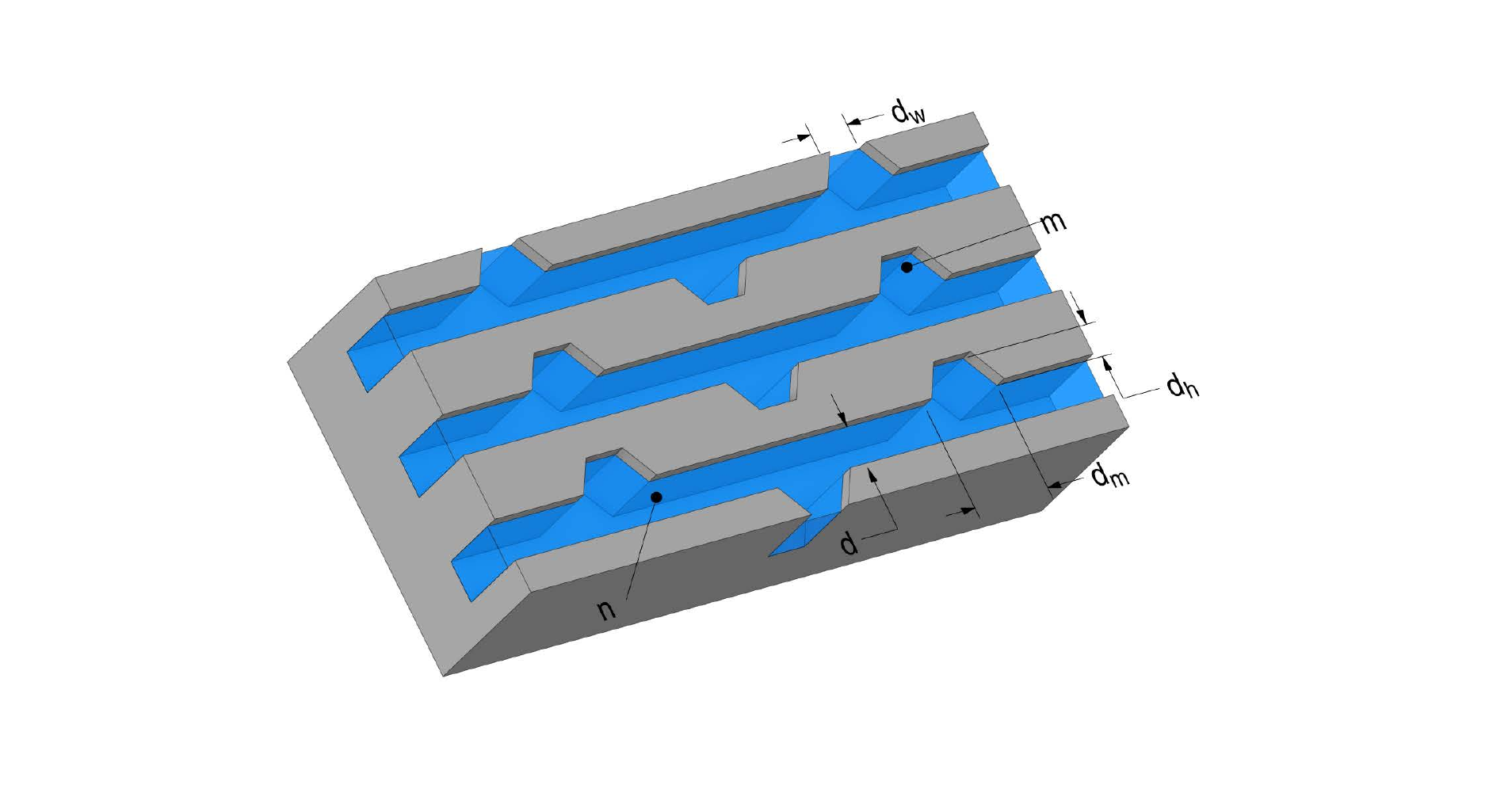}
    \caption{}
    \label{fig:channel_geometry:b}
\end{subfigure}
\caption{Considered MCHS designs; (a) MCHS with secondary channels (SC-MC), (b) MCHS with ribs(R-MC).}
\label{fig:channel_geometry}
\end{figure}

Two distinct MCHS models were considered in this paper. They are based on a conventional microchannel heat sink defined by Xia et al. \cite{xia2015a}. The adoption of a design was done to ensure comparability. The adopted design was modified to include slanted passages and ribs. Figure \ref{fig:channel_geometry} shows the geometric configurations of the proposed microchannel variants. The model with secondary channels will henceforth be referred to as SC-MC, while the model with ribs will be referred to as R-MC.

\begin{figure}[H]
\centering
\begin{subfigure}[b]{0.495\textwidth}
    \centering
    \includegraphics[trim={4.5cm 0.5cm 4.5cm 0.5cm}, clip, width=\textwidth]{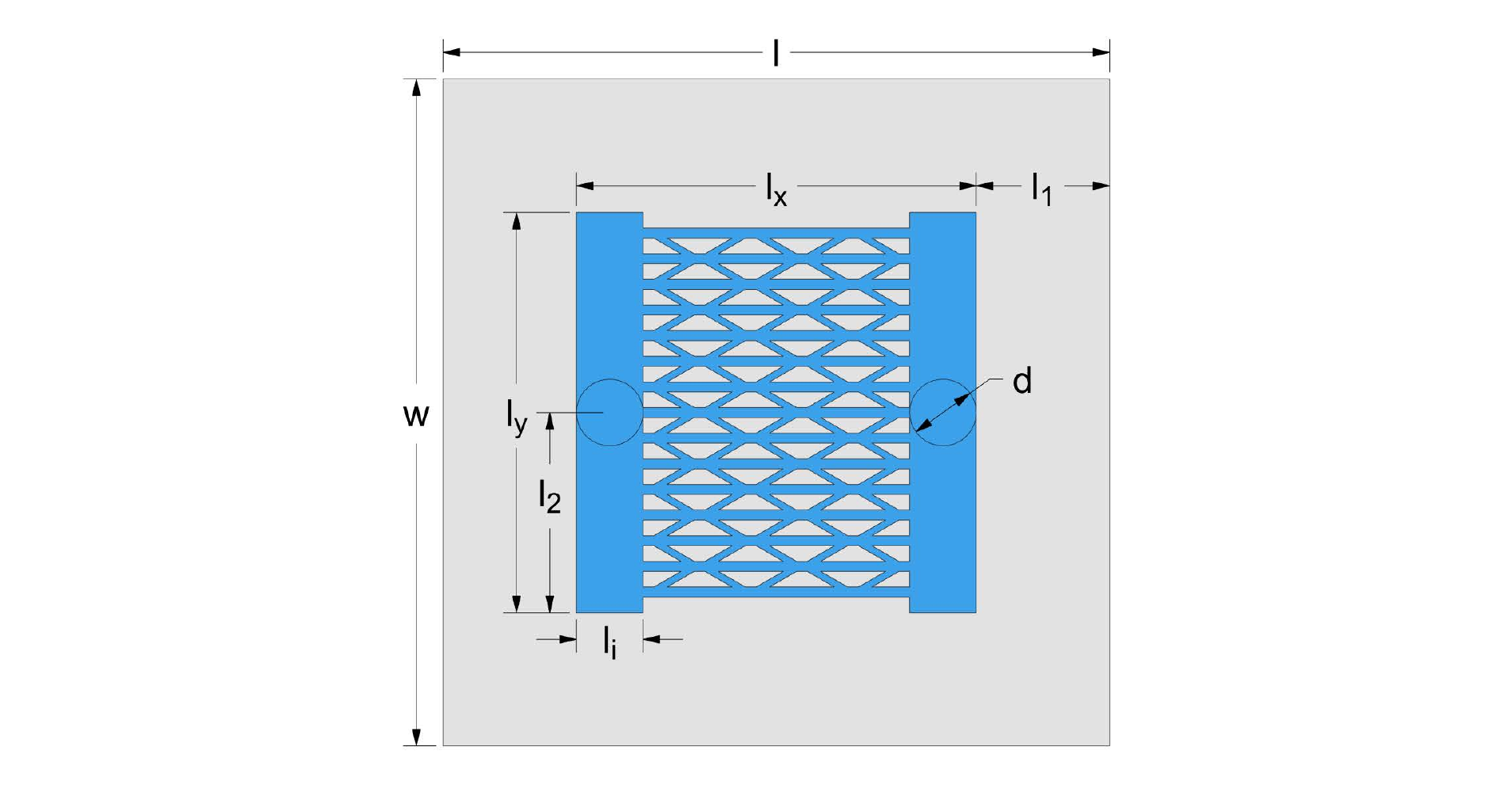}
    \caption{}
    \label{fig:sections:a}
\end{subfigure}
\begin{subfigure}[b]{0.495\textwidth}
    \centering
    \includegraphics[trim={4.5cm 0.5cm 4.5cm 0.5cm}, clip, width=\textwidth]{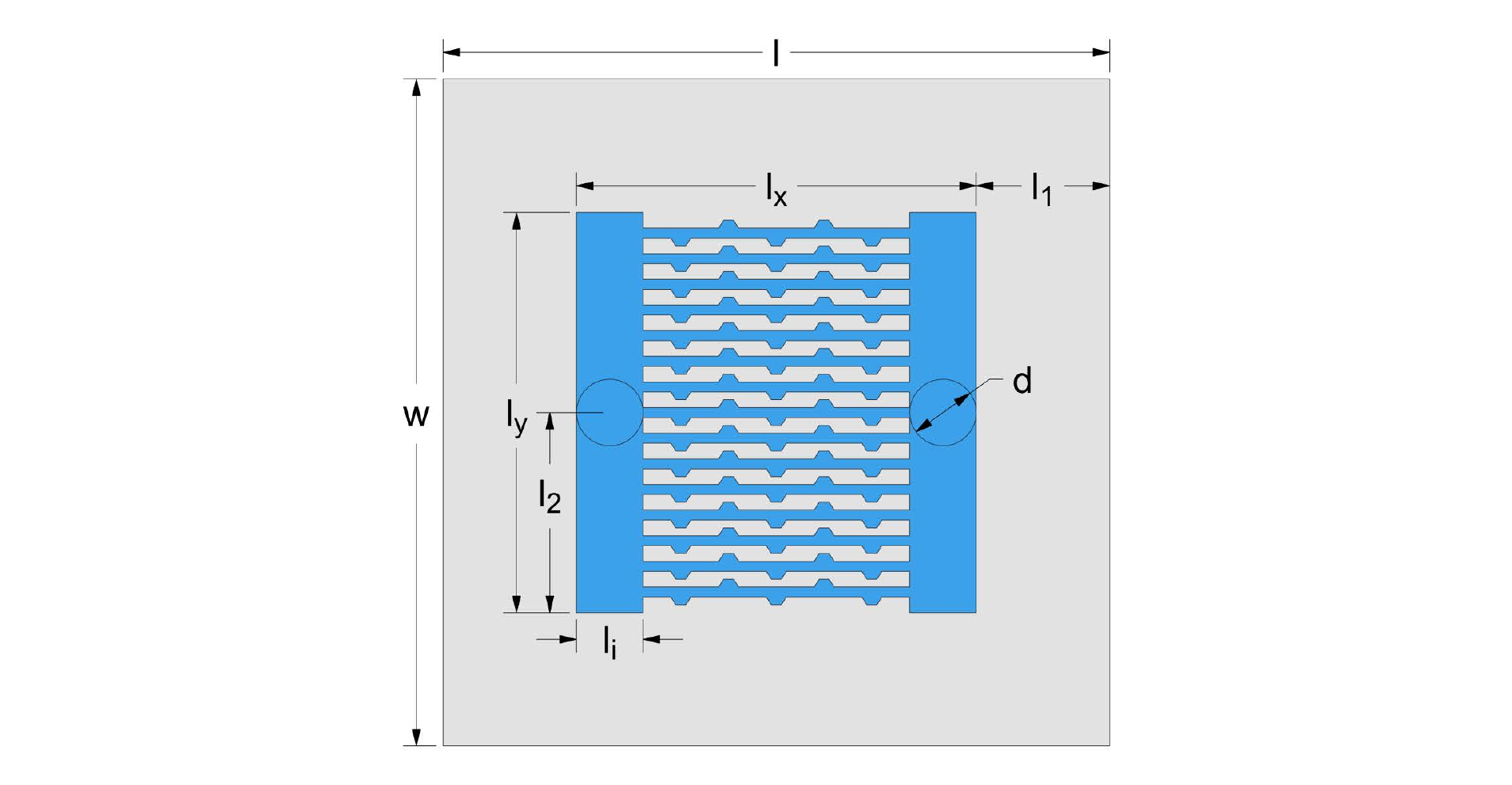}
    \caption{}
    \label{fig:sections:b}
\end{subfigure} \\
\begin{subfigure}[b]{0.495\textwidth}
    \centering
    \includegraphics[trim={4.5cm 3.8cm 4.5cm 4cm}, clip, width=\textwidth]{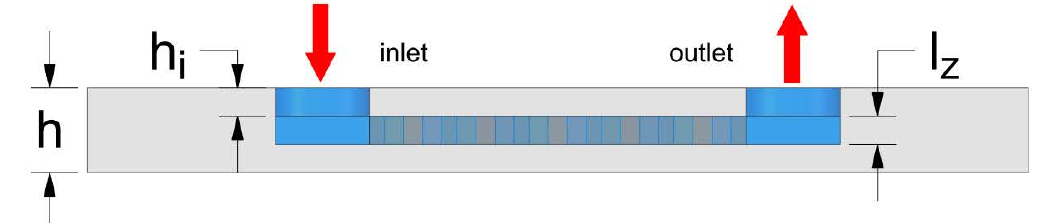}
    \caption{}
    \label{fig:sections:c}
\end{subfigure}
\caption{Sectional views at the $xy$ plane where $z=0$; (a) SC-MC, (b) R-MC. $xz$ cross section at $y=0$ (c). }
\label{fig:sections}
\end{figure}

The overall dimensions of the heat sinks are 10 mm $\times$ 10 mm $\times$ 0.9 mm. The coolant, water, enters the MCHS at the inlet, located centrally in the header region as shown in Figure \ref{fig:sections:c}. The location of the outlet is symmetrical to the inlet. The height of the channels is 0.3 mm. Depending on the channel width, the hydraulic diameter of the main channels is 0.15 mm $\leq d_h \leq$ 0.3 mm. Parameters depicted in Figure \ref{fig:channel_geometry} correspond to optimization variables and will hence be discussed later. All other relevant geometric characteristics are given in Table \ref{tab:geometric_characteristics}.

\begin{table}[H]
\centering
\footnotesize
\caption{Geometric characteristics of the MCHS.}
\label{tab:geometric_characteristics}
\begin{tabular}{@{}lccccccccccc@{}}
\toprule
 & $l$ & $w$ & $h$ & $l_x$ & $l_y$ & $l_z$ & $l_1$ & $l_2$ & $l_i$ & $h_i$ & $d_i$ \\ \midrule
Value (mm) & 10 & 10 & 0.9 & 6 & 6 & 0.3 & 2 & 3 & 1 & 0.3 & 1 \\ \bottomrule
\end{tabular}
\end{table}

Sectional views of the three-dimensional models of the MCHS are given in Figure \ref{fig:sections}. These figures include all noted geometric characteristics and are representative of the computational domain. 

\subsection{Numerical modelling and assumptions}
\label{subsec:numerical_modelling}

The primary assumption in this study is that the flow inside the MCHS is laminar and incompressible. A single Newtonian fluid was studied. Radiation heat transfer, gravity and other body forces were neglected. Viscous dissipation has been taken into account. The problem is three-dimensional and all the results were observed at a steady state. The heat sink is made of silicon. The working fluid is water. Both materials were assumed to have constant thermophysical properties i.e. values at 298 K for density, specific heat capacity, thermal conductivity and dynamic viscosity were used. This was an intentional simplification. Material properties are given in Table \ref{tab:material_characteristics}.

\begin{table}[H]
\centering
\footnotesize
\caption{Thermophysical properties of materials in this study.}
\label{tab:material_characteristics}
\begin{tabular}{@{}lcccc@{}}
\toprule
 & $\rho$ (kg m$^{-3}$) & $c_p$ (J kg$^{-1}$ K$^{-1}$) & $k$ (W m$^{-1}$ K$^{-1}$) & $\mu$ (kg m$^{-1}$ s$^{-1}$) \\ \midrule
Solid & 2329 & 712 & 148 & - \\
Fluid & 996.6 & 4178 & 0.609 & 9.027 $\cdot 10^{-4}$ \\ \bottomrule
\end{tabular}
\end{table}

The adopted inflow flow rate ranges from 100 ml$\cdot$min$^{-1}$ to 200 ml$\cdot$min$^{-1}$. For the purposes of the optimization procedure, however, $Q = 150$ ml$\cdot$min$^{-1}$ has been chosen as a reference and if not stated otherwise, is the flow rate in the reported results. The bottom of the computational domain (MCHS) has been subjected to a uniform heat flux $q_w = 2 \cdot 10^{6}$ W m$^{-2}$. All remaining heat sink walls were considered adiabatic.

Based on noted assumptions, governing equations, namely the continuity, momentum and energy equations can be defined:
\begin{linenomath}
\begin{gather}
    \nabla \cdot \bm{u} = 0 \\
    \rho_f ( \bm{u} \cdot \nabla) \bm{u} = - \nabla p + \mu_f \nabla^2 \bm{u} \\
    \rho_f c_{p,f} ( \bm{u} \cdot \nabla T_f ) = k_f \nabla^2 T_f \\
    k_s \nabla^2 T_s  = 0
\end{gather}
\end{linenomath}
where $\bm{u}$ is velocity, $p$ pressure, $T$ temperature, $k_s$ thermal conductivity of the solid, while $\rho_f$, $\mu_f$, $c_{p,f}$ and $k_f$ are density, specific heat capacity, thermal conductivity and dynamic viscosity of the fluid, respectively. 

CFD simulations of conjugated heat transfer have been conducted using Ansys Fluent 20 \cite{fluent}. A coupled algorithm was used to solve the momentum and continuity equations. Second order accurate scheme was employed for the viscous and second-order upwind scheme for the remaining terms. The maximum number of iterations was 5000. Convergence was assumed if the residuals for the energy fall below $10^{-9}$ and below $10^{-4}$ for the remaining variables. Additionally, an external convergence-control mechanism was used. The mechanism ensured that the standard derivation of the last 25\% of iterations, for converged cases, was below 3\%. A constant volume flow rate, $Q$, and water temperature $T_f = 298$ K were set at the inlet:
\begin{linenomath}
\begin{gather}
    u_{in} = \frac{Q}{A_{in}},~T_{f,in} = 298 K.
\end{gather}
\end{linenomath}
At the outlet, a constant gauge pressure $p = 0$ Pa was used:
\begin{linenomath}
\begin{gather}
    p_{out} = p_{atm}.
\end{gather}
\end{linenomath}
The bottom was assigned a constant heat flux $q_w$:
\begin{linenomath}
\begin{gather}
    k_s \frac{\partial T_s}{\partial z} = - q_w.
\end{gather}
\end{linenomath}
The remaining external walls were considered thermally insulated i.e. $q = 0$ W m$^{-2}$. The interface between the solid and the fluid was treated as a thermally coupled wall and a no-slip condition for the fluid was specified:
\begin{linenomath}
\begin{gather}
    -k_s \frac{\partial T_s}{\partial n} = -k_f \frac{\partial T_f}{\partial n}\\
    u_{f,w} = 0.
\end{gather}
\end{linenomath}

\subsection{Automated mesh generation}

\begin{figure}[H]
\centering
\begin{subfigure}[b]{0.495\textwidth}
    \centering
    \includegraphics[trim={4cm 4cm 6cm 6cm}, clip, width=\textwidth]{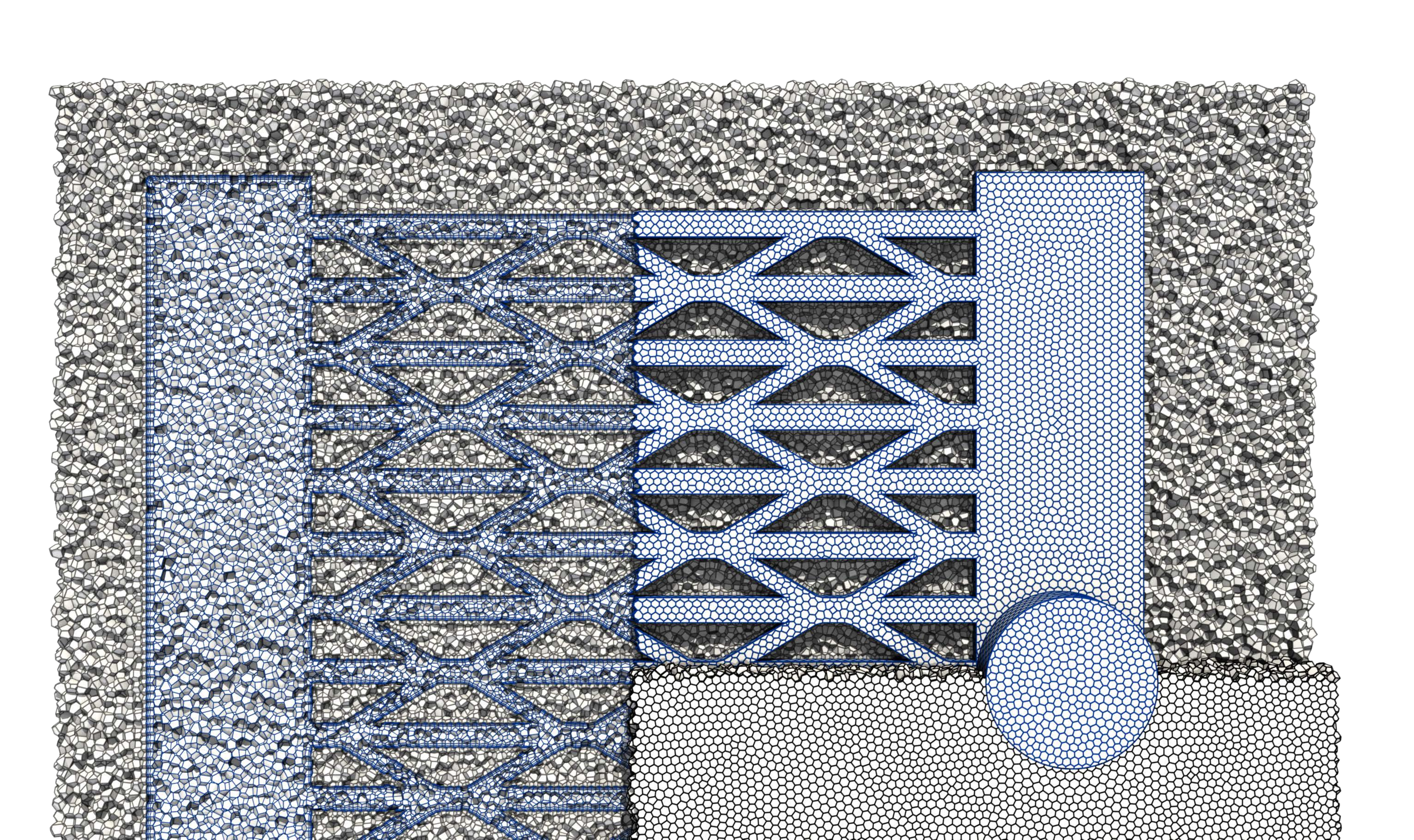}
    \caption{}
    \label{fig:mesh:a}
\end{subfigure}
\begin{subfigure}[b]{0.495\textwidth}
    \centering
    \includegraphics[trim={4cm 4cm 6cm 6cm}, clip, width=\textwidth]{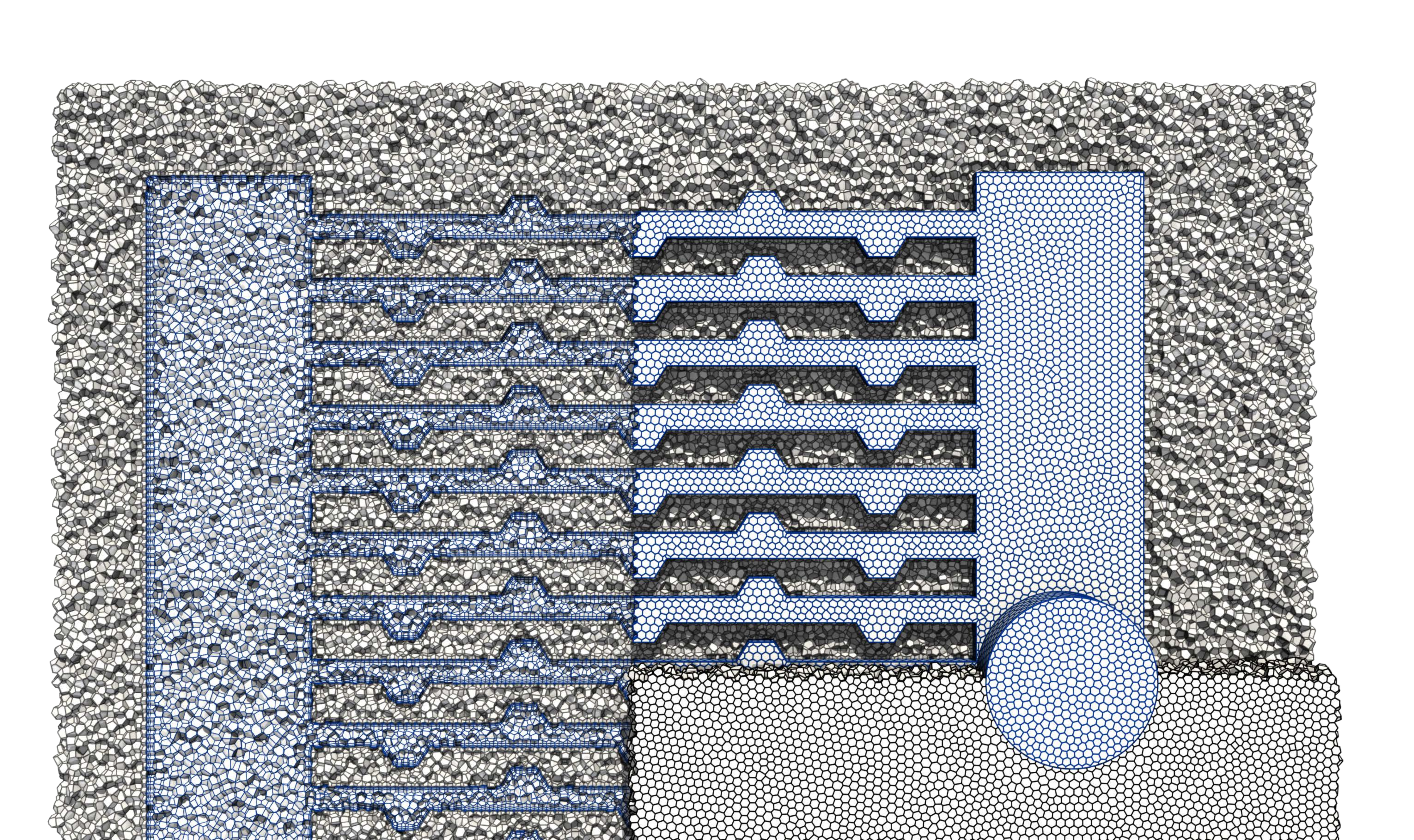}
    \caption{}
    \label{fig:mesh:b}
\end{subfigure}
\caption{Automated grid generation using TetGen and Fluent; (a) mesh for an SC-MC case, (b) mesh for an R-MC case.}
\label{fig:mesh}
\end{figure}

An automated mesh generation procedure was employed in this paper. Since the accuracy of numerical results varies greatly depending on the numerical grid, finalized grids were the result of several iterative processes. Candidate geometry was created using the FreeCAD \cite{freecad} Python API for each MCHS case considered in multi-objective and machine learning workflows. Initial meshing was done using TetGEN's \cite{tetgen} Delaunay tetrahedral generator. This deliberate step was implemented to ensure adequate feature capture. An external control algorithm ensured that each grid contained fewer than $10^7$ cells and that the mesh generation procedure was completed in less than 300 s. Automated coarsening and re-meshing functions were used for complex geometries that did not satisfy noted requirements. The finalized polyhedral mesh was created using the Ansys Fluent mesher. The default cell size was set to $l_{g,c} = r_f \cdot 5 \cdot 10^{-5}$ m, where $r_f$ is a mesh scaling factor. Three layers of prismatic cells with an arbitrary height $l_{g,w} = 5 \cdot 10^{-6}$ m were created at the fluid-solid interface to ensure a better mesh transition. Examples of numerical grids for SC-MC and R-MC are shown in Figure \ref{fig:mesh}.

\subsection{Multi-objective optimization}
\label{subsec:mo_optimization}

Thermal resistance and pumping power, two conflicting objective functions, were used to evaluate the performance of the microchannels studied in this paper. Thermal resistance, $R_{t}$, defines the MCHS's heat transfer capabilities and can be expressed as:
\begin{linenomath}
\begin{gather}
    R_{t} = \frac{T_{max} - T_{f,in}}{q_w A_{sub}}
\end{gather}
\end{linenomath}
where $T_{max}$ is the measured maximal temperature and $A_{sub}$ surface area of the substrate exposed to the heat flux $q_w$. Pumping power, $P_{p}$, denotes the force required to push the fluid through the microchannels:
\begin{linenomath}
\begin{gather}
    P_{p} = Q \cdot \Delta p
\end{gather}
\end{linenomath}
where $\Delta p$ is the pressure drop across the microchannel heat sink. It is evident that, by increasing the flow to dissipate the heat quickly, an increase in pumping power is inevitable. Complex and elaborate channel designs that facilitate heat transfer, on the other hand, produce a greater pressure drop, which has the same effect. The goal of the considered multi-objective optimization procedure is to minimize the objective functions, $R_{t}$ and $P_{p}$, at a constant flow rate. In general, a constrained multi-objective optimization problem can be defined as:
\begin{linenomath}
\begin{equation}
    \begin{aligned}
    & \underset{\bm{x}}{\text{minimize}}
    & & F(\bm{x}) = \bigl(R_t (\bm{x}), P_p (\bm{x})\bigl)^T \\
    & \text{subject to}
    & & d \leq d_m
    \end{aligned}
\end{equation}
\end{linenomath}
where $\bm{x} = (x_1,...,x_n)^T$ is an $n$-dimensional vector of design variables in decision space $\mathbb{R}^n$ while $F(\bm{x})$ is an $m$-dimensional objective vector such that $F(\bm{x}) \in \mathbb{R}^m$. Conducted multi-objective optimizations were constrained. For both designs, the constraint $d \leq d_m$ was set, where $d$ and $d_m$ are geometric characteristics shown in Figure \ref{fig:channel_geometry}. All solutions that satisfy set constraint are considered feasible. If they are non-dominated, they are considered Pareto optimal solutions. Solutions presented in this paper are chosen Pareto optimal solutions that satisfy one of these conditions: 
\begin{itemize}
    \item $R_{t}$ and $P_{p}$ must be improved; relative improvement for $R_{t}$ is larger (variant A),
    \item $R_{t}$ and $P_{p}$ must be improved; relative improvement for $P_{p}$ is larger (variant B),
    \item lowest possible $R_{t}$ where $P_{p}$ is, at worst, equal to the original design (variant C).
\end{itemize}

For the SC-MC case, seven design variables have been considered in the optimization procedure. These variables are: the number of microchannels $n$, the number of secondary channels per fin $m$, microchannel width $d$, secondary channel width $d_m$, secondary channel angle $\alpha$, and two rotational parameters, $r_s$ and $r_f$. Noted variables are shown in Figure \ref{fig:channel_geometry:a} in Section \ref{sec:problem_definition_methodology}. Unlike other variables, $r_{s}$ and $r_{f}$ are binary. $r_s$ controls the orientation of neighboring secondary channels. Secondary channels can thus be parallel or converging, resulting in fin segments that are parallelogram-like or trapezoid-like. The variable $r_{f}$ provides the option to flip the entire secondary channel segment so that two neighboring sequences are symmetrical about the bounded microchannel's centerline. All design variables and bounds are included in Table \ref{tab:design_variables}.

Design variables for the R-MC can be seen in Figure \ref{fig:channel_geometry:b} in Section \ref{sec:problem_definition_methodology}. Six variables were examined: the number of microchannels $n$, the number of ribs $m$, microchannel width $d$, base width of the rib $d_m$, top width of the rib $d_w$ and rib height $d_h$. Ribs are generated alternately on the opposite sides of the microchannel. An overview of all design variables with included upper and lower limits is given in Table \ref{tab:design_variables}.

\begin{table}[H]
\centering
\footnotesize
\caption{Design variables and respective bounds.}
\label{tab:design_variables}
\begin{tabular}{@{}cccccccccccccc@{}}
\toprule
& \multicolumn{7}{c}{SC-MC} & \multicolumn{6}{c}{R-MC} \\
$\bm{x}$ & $n$ & $m$ & \makecell[c]{~$d$ \\[-2pt] \added{(mm)}} & \makecell[c]{~$d_m$ \\[-2pt] \added{(mm)}} & \makecell[c]{~$\alpha$ \\[-2pt] \added{($^{\circ}$)}} & $r_s$ & $r_f$ & $n$ & $m$ & \makecell[c]{~$d$ \\[-2pt] \added{(mm)}} & \makecell[c]{~$d_m$ \\[-2pt] \added{(mm)}} & \makecell[c]{~$d_w$ \\[-2pt] \added{(mm)}} & \makecell[c]{~$d_h$ \\[-2pt] \added{(mm)}} \\ \cmidrule(r){1-1} \cmidrule(lr){2-8} \cmidrule(l){9-14}
LB & 5 & 0 & 0.1 & 0.1 & 10 & 0 & 0 & 5 & 0 & 0.1 & 0.1 & 0.02 & 0.02 \\
UB & 30 & 30 & 0.3 & 0.5 & 90 & 1 & 1 & 30 & 30 & 0.3 & 0.5 & 0.3 & 0.5 \\ \bottomrule
\end{tabular}
\end{table}

The upper bounds of noted design variables were arbitrarily chosen; the number of microchannels (and accordingly ribs and secondary channels) was to be less than in the original design. Other variables had sensibly large yet feasible values for upper bounds. Lower bounds were either set low to ensure constant hydraulic diameter (widths) or were set low enough so that they should never affect the optimization process.

\added{In all considered cases, the geometric parameters defined in Table \ref{tab:geometric_characteristics} have been kept constant. This includes MCHS thickness, $h$, as well as microchannel height, $l_z$. This was done deliberately to reduce complexity and computational effort as well as ensure comparability with Xia et al. \cite{xia2015a}. Implemented design choices should not affect the overall workflow/methodology.}

\begin{figure}[H]
\centering
\includegraphics[trim={0cm 0cm 0cm 0cm}, clip, width=0.5\textwidth]{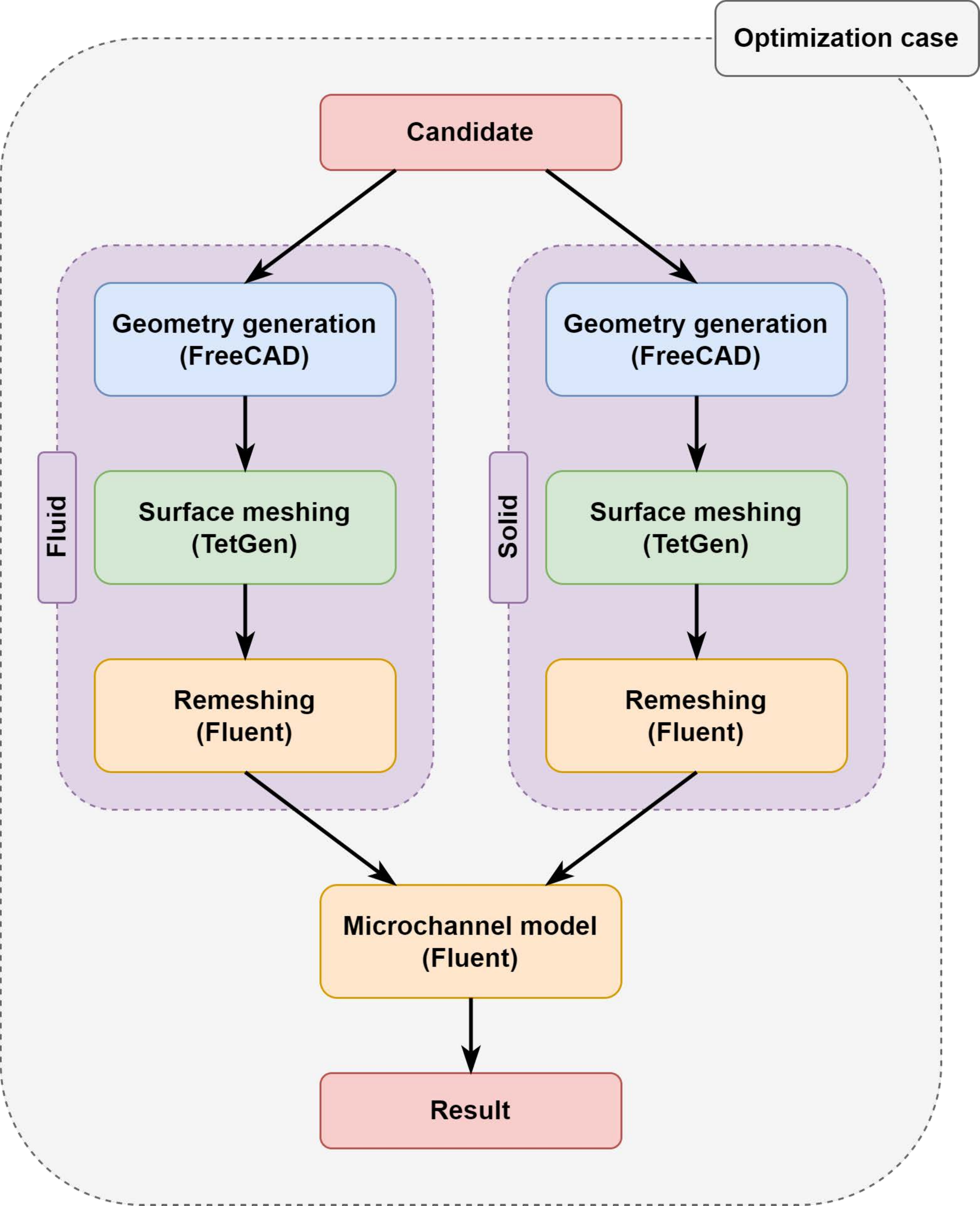}
\caption{Multi-objective optimization workflow based on CFD and NSGA-II algorithm. The geometry generation and meshing routines are given a candidate comprised of design variables as an input.}
\label{fig:workflow_MO}
\end{figure}

Python was used to carry out the optimization procedure. Implementation of the NSGA-II algorithm from the jMetalPy v1.5.5 \cite{jmetalpy} framework was employed to determine the Pareto optimal solutions i.e. Pareto front. The population size was set to 100. The optimization procedure ran for 200 generations. Simulated binary crossover probability, $p_c$, and polynomial mutation probability, $p_m$, were set to $0.9$ and $0.15$, respectively. All other parameters were kept at their default values. Optimization workflow is depicted in Figure \ref{fig:workflow_MO}. For each optimization instance, a candidate i.e. vector of design variables $\bm{x}$ defined in Table \ref{tab:design_variables}, is passed to the geometry generation routine. To improve the performance, parallel geometry generation and meshing instances for the fluid and solid parts were defined for each candidate. The resulting numerical grids were subsequently integrated into a unified model which was evaluated in an HPC environment according to the numerical methodology defined in Section \ref{subsec:numerical_modelling}.

\subsection{Machine learning surrogates}
\label{subsec:ml_surrogates}

Utilized machine learning algorithms were trained on data obtained through numerical simulations. Latin hypercube sampling technique was employed to generate near-random samples from the decision space $\mathbb{R}^n$. Sampling using LHS ensures that the respective ranges for each variable in an $n$-dimensional problem are split into $m$ intervals of equal probability. Each interval is then sampled and the result is combined using random grouping with other components to form an $n$-dimensional sample (i.e. candidate). In total, 1500 candidates per design were generated. Candidates were generated randomly from a decision space bounded by the upper and lower bounds defined in Table \ref{tab:design_variables}. This means that each candidate adheres to the same limits set for the optimization process. Thus defined candidates were evaluated using CFD. The outliers were filtered using the interquartile range filter. Filtered datasets were subsequently used to train multi-output regression models.

Four machine learning algorithms were evaluated in this study: random forest (RF) implementation in scikit-learn v1.1.1 library \cite{scikit}, gradient boosting algorithms CatBoost (CB) v1.0.6 \cite{catboost} and LightGBM (LGBM) v3.3.2 \cite{lightgbm}, and TensorFlow v2.8.1 based ANNs \cite{tensorflow}. Ensemble algorithms (RF, CB, LGBM) were chosen as they allow seamless and quick development of respectable ML models. Noted algorithms rely on a similar concept; baseline models (i.e. decision trees) are trained and subsequently, a collection of models is combined to provide a prediction. The fundamental difference between these algorithms is in the manner they are created and how they tackle the problem of overfitting; RF utilizes fully grown decision trees (which can overfit) and reduces variance, whereas boosting algorithms use weak learners (thus minimizing overfitting), are built additively and aim to reduce the bias. ANNs represent a traditional approach to machine learning whereby the predictive accuracy and overfitting are dependent on the network design and its parameters. All models were trained on an 80/20 split dataset. Input features were independently standardized by subtracting the mean and scaling to unit variance using the StandardScaler function. EarlyStopping was employed to prevent overfitting. For RF, CB and LGBM, rough estimates of the hyperparameters were obtained using a randomized search that ran for 20 iterations. scikit-optimize's \cite{scikitOpt} Bayesian search was thereafter used to fine-tune the hyperparameters. The search ran 500 iterations. Robustness during tuning was assessed using a 5-fold cross-validation on the training dataset.

ANN models were tuned using the fireworks algorithm (FWA) in the indago v0.2.4 module \cite{indago}. Use of metaheuristic algorithms such as particle swarm optimization (PSO) for hyperparameter tuning is becoming more prominent \cite{ye2017pso, lorenzo2017pso, wang2019pso}. Since FWA shows better overall performance than PSO \cite{tan2010}, it was chosen for this task. Five separate optimization runs per ANN were carried out, each lasting 200 iterations. Optimization variables were the number of hidden layers, number of neurons, batch size and learning rate. Additionally, different activation functions and initializers were considered in the optimization procedure. Variables are briefly summarized in Table \ref{tab:design_variables_ANN}. A stochastic gradient-descent-based optimizer, Adam, was used to minimize the loss.

\begin{table}[H]
\centering
\footnotesize
\caption{Optimization variables for the FWA-based ANN tuning.}
\label{tab:design_variables_ANN}
\begin{tabular}{@{}lc@{}}
\toprule
No. hidden layers & $L = \{x \in \mathbb{N}~|~1 \leq x \leq 4 \}$ \\
No. neurons per layer & $2^{N},~N = \{x \in \mathbb{N}~|~0 \leq x \leq 8 \}$ \\
Batch size & $2^{B},~B = \{x \in \mathbb{N}~|~5 \leq x \leq 9 \}$ \\
Learning rate & $\{0.0005, 0.00075, 0.001, 0.0025, 0.005\}$ \\
Activation functions & ELU, ReLU, LeakyReLU \\
Initializers & GlorotNormal(Uniform), HeNormal(Uniform), TruncatedNormal \\ \bottomrule
\end{tabular}
\end{table}

The robustness of fully tuned models was investigated by employing a 10-fold cross-validation. The prediction accuracy was assessed using several metrics, namely, mean absolute percentage error (MAPE), root mean squared error (RMSE) and the coefficient of determination (R$^2$). MAPE is a common statistical method for determining prediction accuracy. It is calculated according to:
\begin{linenomath}
\begin{gather}
    \text{MAPE} = \frac{1}{n} \sum\limits_{i=1}^n  \left|\frac{y_i - \hat{y}_i}{y_i}\right| \cdot 100
\end{gather}
\end{linenomath}
where $n$ is the sample size, $\hat{y}_i$ the predicted value and ${y_i}$ observed value. Similarly, RMSE is often used to measure the predictive error of regression models:
\begin{linenomath}
\begin{gather}
    \text{RMSE} = \sqrt{ \sum\limits_{i=1}^n \frac{\bigl( y_i - \hat{y}_i \bigl) ^2 }{n} }.
\end{gather}
\end{linenomath}
R$^2$ indicates the percentage of variance in the dependent variable that can be explained by the model and is calculated as:
\begin{linenomath}
\begin{gather}
    \text{R}^2 = 1 - \frac{ \sum\limits_{i=1}^n \bigl( y_i - \hat{y}_i \bigl) ^2 }{ \sum\limits_{i=1}^n \bigl( y_i - \bar{y}_i \bigl) ^2 }
\end{gather}
\end{linenomath}
where $\bar{y}_i$ is the mean of all values.

\section{Results and discussion}
\label{sec:results_discussion}

\subsection{Grid convergence and validation}

Three grids were created for the grid convergence study: coarse, medium and fine, with $1.8 \cdot 10^{5}$, $3.9 \cdot 10^{5}$ and $7.9 \cdot 10^{5}$ cells, respectively. Cell sizing was selected according to $l_{g,c} = r_f \cdot 5 \cdot 10^{-5}$ m. The scaling factor, $r_f$, was equal to 1 for fine, 1.4 for medium and 1.96 for the coarse grid. Assessment is conducted for a conventional MCHS design i.e. design without secondary channels and ribs. Grid convergence index (GCI) \cite{roache1998} was calculated for two variables: velocity $u$ measured at the centerline of the MCHS and temperature $T_{max}$ at the MCHS. Results are presented in Table \ref{tab:gci}. As is evident, solutions for both variables and noted grids are in the asymptotic range of convergence. Consequently, the coarse grid was employed for future calculations.

\begin{table}[H]
\centering
\footnotesize
\caption{Results of the grid convergence study.}
\label{tab:gci}
\begin{tabular}{@{}lcccccc@{}}
\toprule
& $p_{GCI}$ & $\varepsilon_{c,m}$ (\%) & $\varepsilon_{m,f}$ (\%) & $GCI_{c,m}$ (\%) & $GCI_{m,f}$ (\%) & $\frac{GCI_{m,f}}{r^p \cdot GCI_{c,m}}$ \\ \midrule
$u$ & 2.209 & 0.986 & 0.471 & 1.117 & 0.534 & 0.995 \\
$T_{max}$ & 1.548 & 0.050 & 0.030 & 0.091 & 0.054 & 1.000 \\ \bottomrule
\end{tabular}
\end{table}

\begin{figure}[H]
\centering
\begin{subfigure}[b]{0.495\textwidth}
    \centering
    \includegraphics[trim={0cm 0.8cm 1cm 1cm}, clip, width=\textwidth]{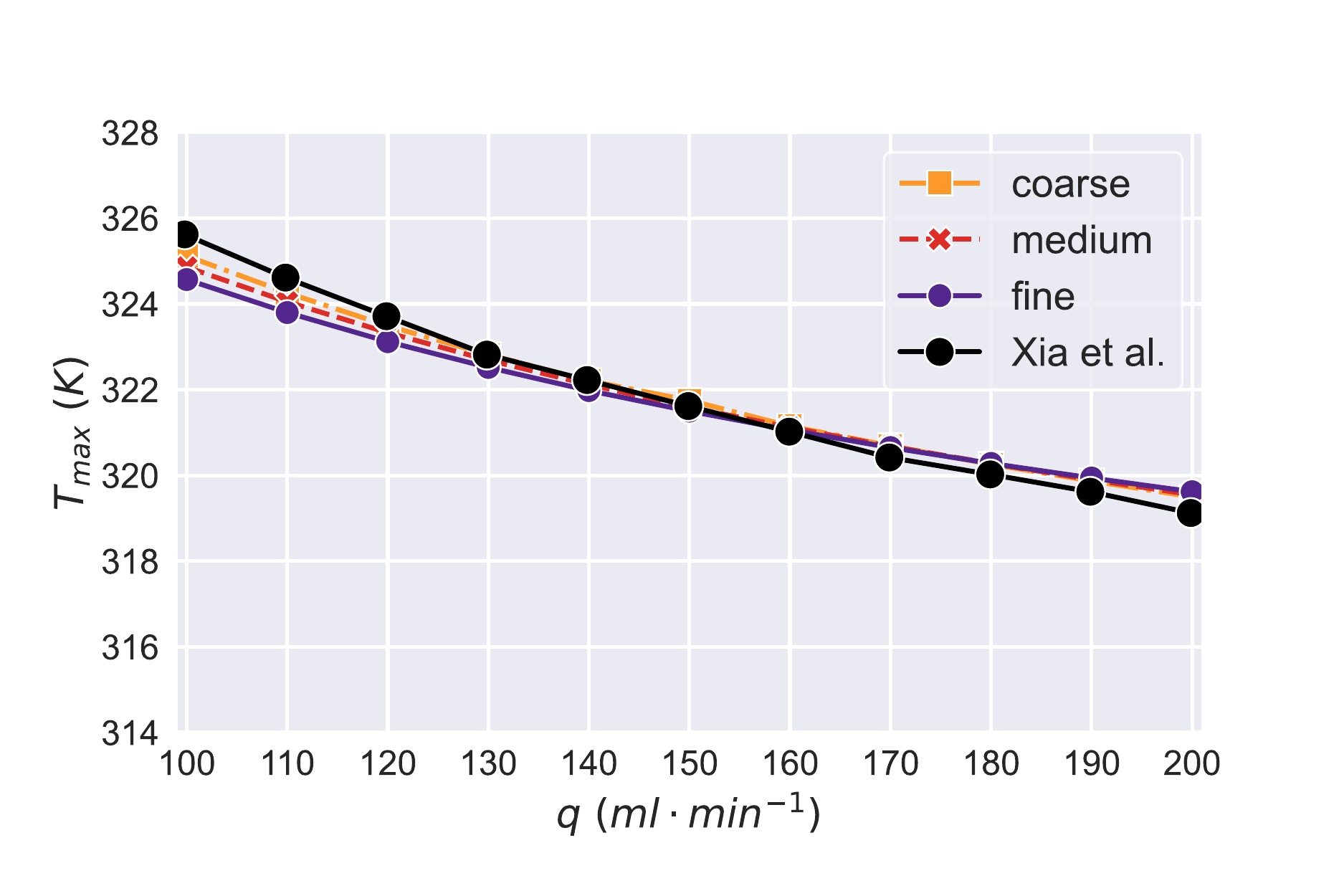}
    \caption{}
    \label{fig:grid_convergence:a}
\end{subfigure}
\begin{subfigure}[b]{0.495\textwidth}
    \centering
    \includegraphics[trim={0cm 0.8cm 1cm 1cm}, clip, width=\textwidth]{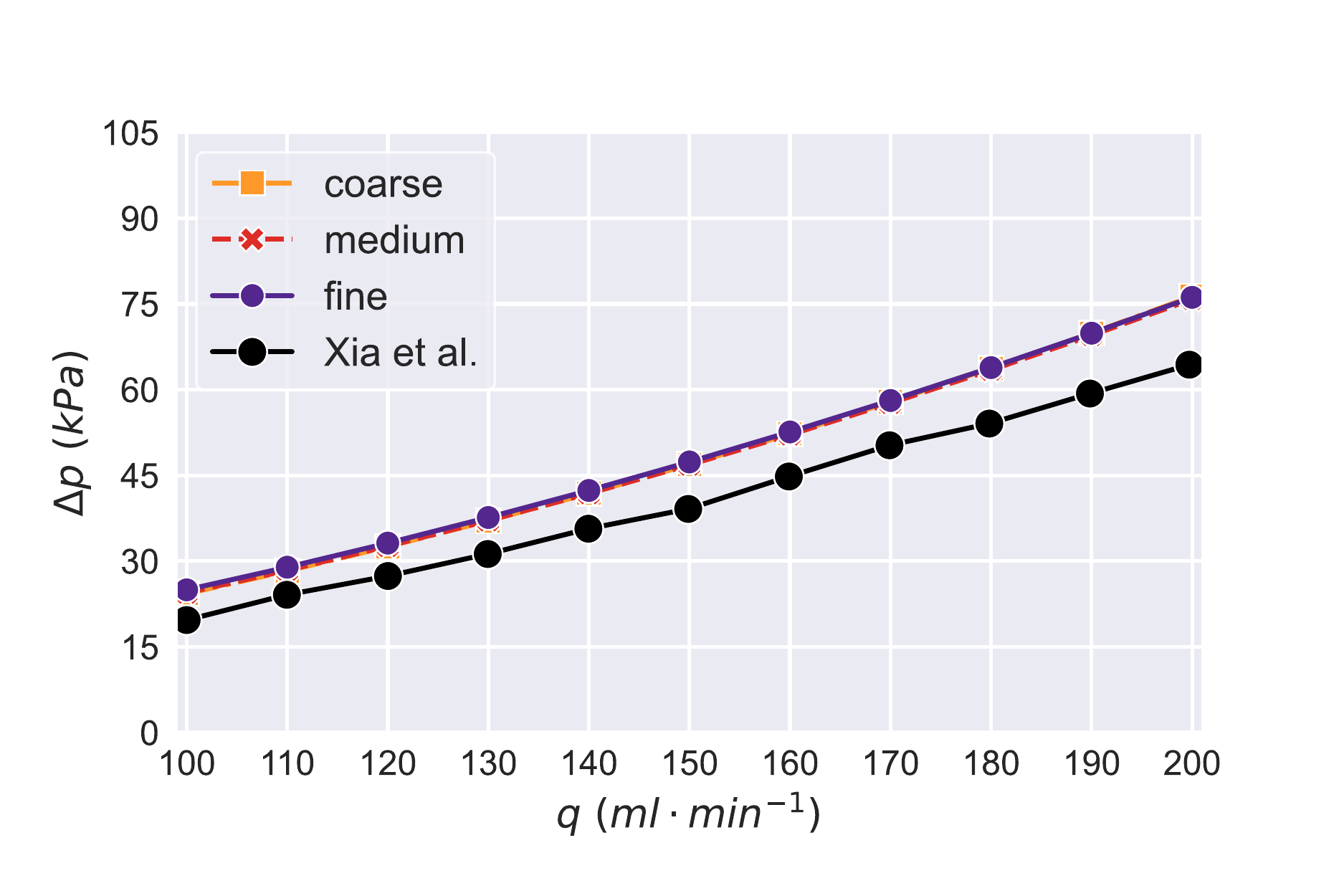}
    \caption{}
    \label{fig:grid_convergence:b}
\end{subfigure}
\caption{Results obtained for different grids compared to the results by Xia et al. \cite{xia2015a}; (a) $T_{max}$, (b) $\Delta P$. \added{The considered model is a conventional rectangular MCHS with the dimensions 10 mm $\times$ 10 mm $\times$ 0.9 mm. MCHS design has 30 microchannels with a hydraulic diameter of 0.15 mm. Remaining parameters are given in Table \ref{tab:geometric_characteristics}. Additional information can be found in the original paper \cite{xia2015a}.}}
\label{fig:grid_convergence}
\end{figure}

To ensure the validity of the numerical procedure, the results were compared to those of a conventional MCHS design presented in a study by Xia et al. \cite{xia2015a}. Despite the introduced simplification (fluid properties are not temperature dependent), the overall agreement is acceptable. The relative error between average and maximum temperatures for all flow rates is below 2\%. The results for pressure show a rather large discrepancy of 26\% at the flow rate range's limits. However, as demonstrated in a later experimental study by Xia et al. \cite{xia2015b}, experimental results produce pressure drops larger than what is typically reported in numerical studies. Furthermore, since all subsequent simulations were performed at $Q = 150$ ml$\cdot$min$^{-1}$, where the temperature disparity is below 0.5\% and the pressure disparity is less than 10\%, and since the pressure trendlines match well, it is reasonable to assume that these comparative results are adequate for a MO workflow. The observed difference can be partially attributed to the numerical grids and setup as well as to the simplification of the physics of the problem. Comparison between CFD results for different grids and results from the literature are given in Figure \ref{fig:grid_convergence}.

\subsection{CFD-based multi-objective optimization}
\label{subsec:cfd_based_section}

\begin{figure}[H]
\centering
\begin{subfigure}[b]{0.495\textwidth}
    \centering
    \includegraphics[trim={0cm 0.8cm 1cm 1cm}, clip, width=\textwidth]{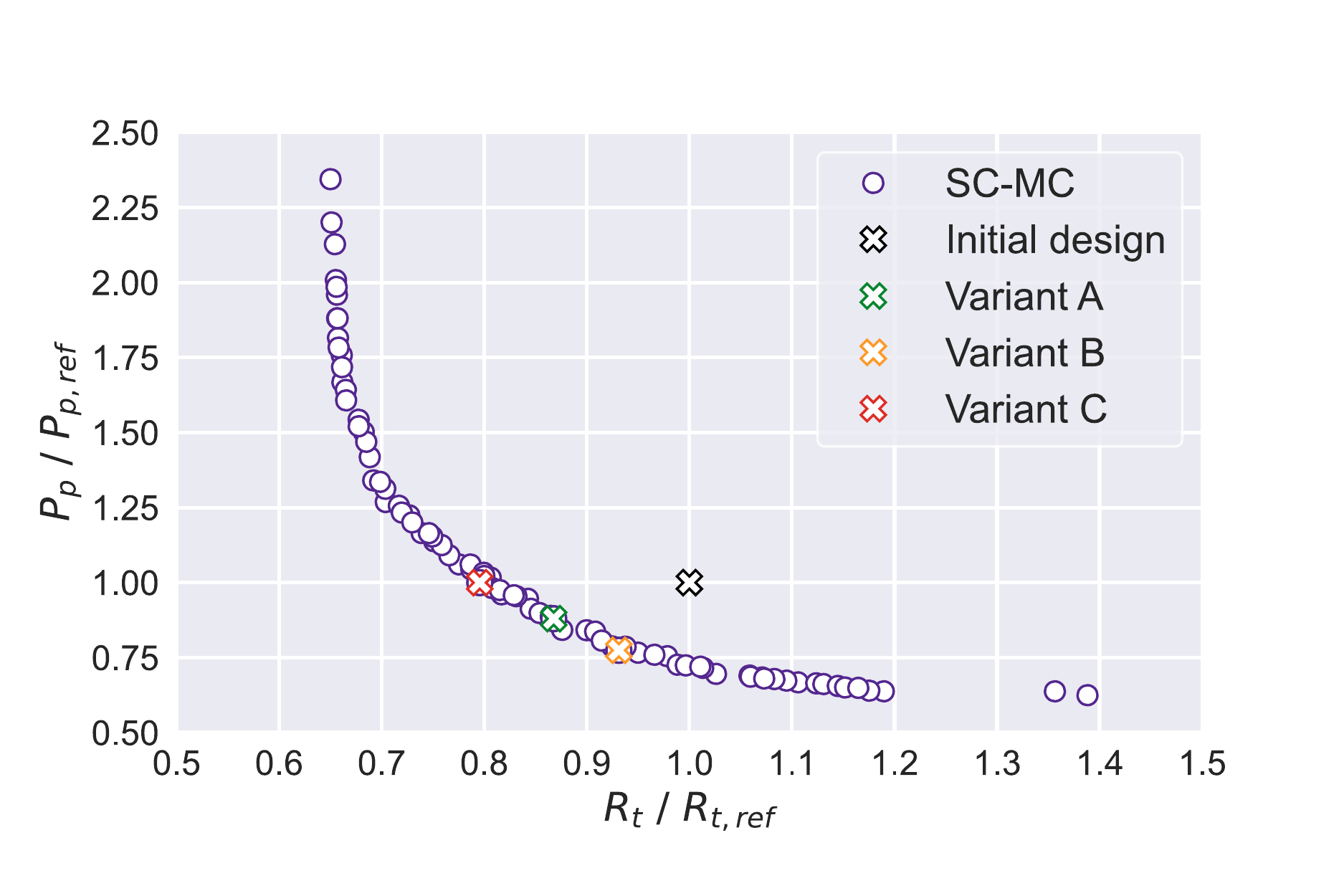}
    \caption{}
    \label{fig:pareto_cfd:a}
\end{subfigure}
\begin{subfigure}[b]{0.495\textwidth}
    \centering
    \includegraphics[trim={0cm 0.8cm 1cm 1cm}, clip, width=\textwidth]{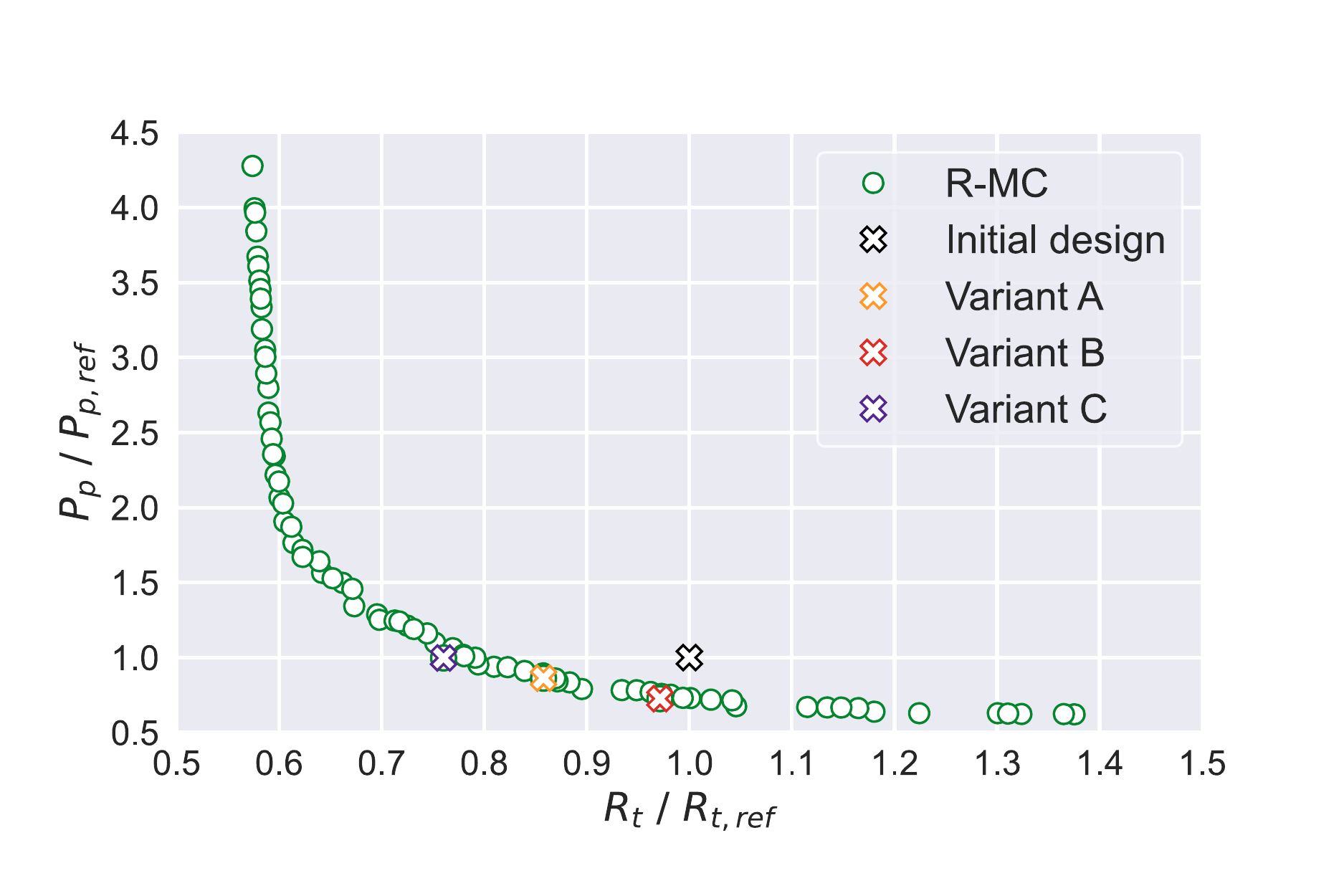}
    \caption{}
    \label{fig:pareto_cfd:b}
\end{subfigure}
\caption{Pareto fronts obtained through the CFD-NSGA-II coupling; (a) front for the SC-MC design, (b) front for the R-MC design.}
\label{fig:pareto_cfd}
\end{figure}

Multi-objective optimizations for each design were executed 10 times. In addition to the previously described partitioning of the workflow, concurrent candidate evaluation was employed to accelerate the optimization process; 25 candidates were evaluated at any time. On average, the complete workflow with 20000 case evaluations took 55 h on an Intel Haswell-based cluster. Obtained Pareto fronts for the SC-MC and R-MC cases are given in Figure \ref{fig:pareto_cfd}. As can be seen, as the normalized thermal resistance increases, the pumping power decreases; at the upper limit, an increase in resistance of roughly 50\% is accompanied by a 50\% reduction in pumping power. Additionally, at equal or lower pumping power ($P_{p} / P_{p,ref} \leq 1$) compared to the baseline design, improvements in thermal resistance are feasible ($R_{t} / R_{t,ref} \leq 1$). As the thermal resistance decreases, a sharp increase in pumping power occurs; this is particularly pronounced for the R-MC where pumping power can increase by more than 4 times. This behavior, however, is expected since complex channel designs which are beneficial for dissipation also induce larger pressure drops.

Optimal designs based on criteria presented in Section \ref{subsec:mo_optimization} for SC-MC and R-MC are given in Table \ref{tab:design_cfd_best}. Pareto solution A considers thermal resistance and pumping power improvements, with a focus on thermal resistance. SC-MC shows improvements of up to 4.5\% and 15.6\% for $T_{max}$ and $\Delta p$ at the design point $Q = 150$ ml$\cdot$min$^{-1}$. The lowest reduction in temperature of 3.1\% is noted for $Q = 100$ ml$\cdot$min$^{-1}$, while for pressure, at $Q = 200$ ml$\cdot$min$^{-1}$, observed reduction is roughly 11.7\%. The R-MC model at the design point shows temperature and pressure drop improvements of 6\% and 17.9\%. Across the entire flow rate range, R-MC achieves better thermal and pressure results; temperatures and pressure are comparatively reduced by approximately 20\% and 12\%, respectively.

Reduction in pumping power is emphasized in Pareto B variants. Obtained results for the SC-MC show a negligible improvement of 0.5\% in temperature at the design point. At the same time, the pressure is reduced by more than 25\%. Improvements in $T_{max}$ are marginal and below 1\% at all flow rates. Pressure drop is reduced by 21.4\% at worst and 31.5\% at best for $Q = 200$ ml$\cdot$min$^{-1}$ and $Q = 100$ ml$\cdot$min$^{-1}$, respectively. For the R-MC, temperature improvement at the design point is 2.6\%, with a 23\% reduction in pressure drop. As the flow rate increases, the reduction in $T_{max}$ improves from 0.6\% to 3.7\%. Pressure drops are lower for all flow rates when compared to SC-MC. At $Q = 100$ ml$\cdot$min$^{-1}$ pressure is reduced by 28.6\%; this value drops and reaches 19.1\% for $Q = 200$ ml$\cdot$min$^{-1}$.

Pareto optimal solutions denoted as C variants provide the greatest reductions in thermal resistance while maintaining comparable pumping power to the baseline design. At the design point, $T_{max}$ is reduced by 9.2\% for SC-MC and by 8.1\% for R-MC. Similarly, $\Delta p$ is reduced by 2.5\% and 7.1\% for SC-MC and R-MC, respectively. Pressure drop is well below the baseline value for all cases (up to 10\% when $Q = 100$ ml$\cdot$min$^{-1}$). As the flow rate approaches $Q = 200$ ml$\cdot$min$^{-1}$, the pressure drop is around 1\% and 4.2\% for SC-MC and R-MC.

\begin{table}[H]
\centering
\footnotesize
\caption{Model parameters for chosen optimal designs.}
\label{tab:design_cfd_best}
\begin{tabular}{@{}cccccccccccccc@{}}
\toprule
& \multicolumn{7}{c}{SC-MC} & \multicolumn{6}{c}{R-MC} \\
$\bm{x}$ & $n$ & $m$ & \makecell[c]{~$d$ \\[-2pt] \added{(mm)}} & \makecell[c]{~$d_m$ \\[-2pt] \added{(mm)}} & \makecell[c]{~$\alpha$ \\[-2pt] \added{($^{\circ}$)}} & $r_s$ & $r_f$ & $n$ & $m$ & \makecell[c]{~$d$ \\[-2pt] \added{(mm)}} & \makecell[c]{~$d_m$ \\[-2pt] \added{(mm)}} & \makecell[c]{~$d_w$ \\[-2pt] \added{(mm)}} & \makecell[c]{~$d_h$ \\[-2pt] \added{(mm)}} \\ \cmidrule(r){1-1} \cmidrule(lr){2-8} \cmidrule(l){9-14}
A & 24 & 4 & 0.14 & 0.33 & 18 & 1 & 0 & 22 & 20 & 0.17 & 0.17 & 0.07 & 0.05 \\
B & 21 & 4 & 0.18 & 0.24 & 23 & 1 & 0 & 22 & 11 & 0.17 & 0.18 & 0.06 & 0.07 \\
C & 26 & 6 & 0.13 & 0.13 & 39 & 1 & 0 & 26 & 15 & 0.13 & 0.13 & 0.03 & 0.13 \\ \bottomrule
\end{tabular}
\end{table}

\begin{figure}[H]
\centering
\begin{subfigure}[b]{0.325\textwidth}
    \centering
    \includegraphics[trim={30cm 10cm 30cm 0cm}, clip, width=0.8\textwidth]{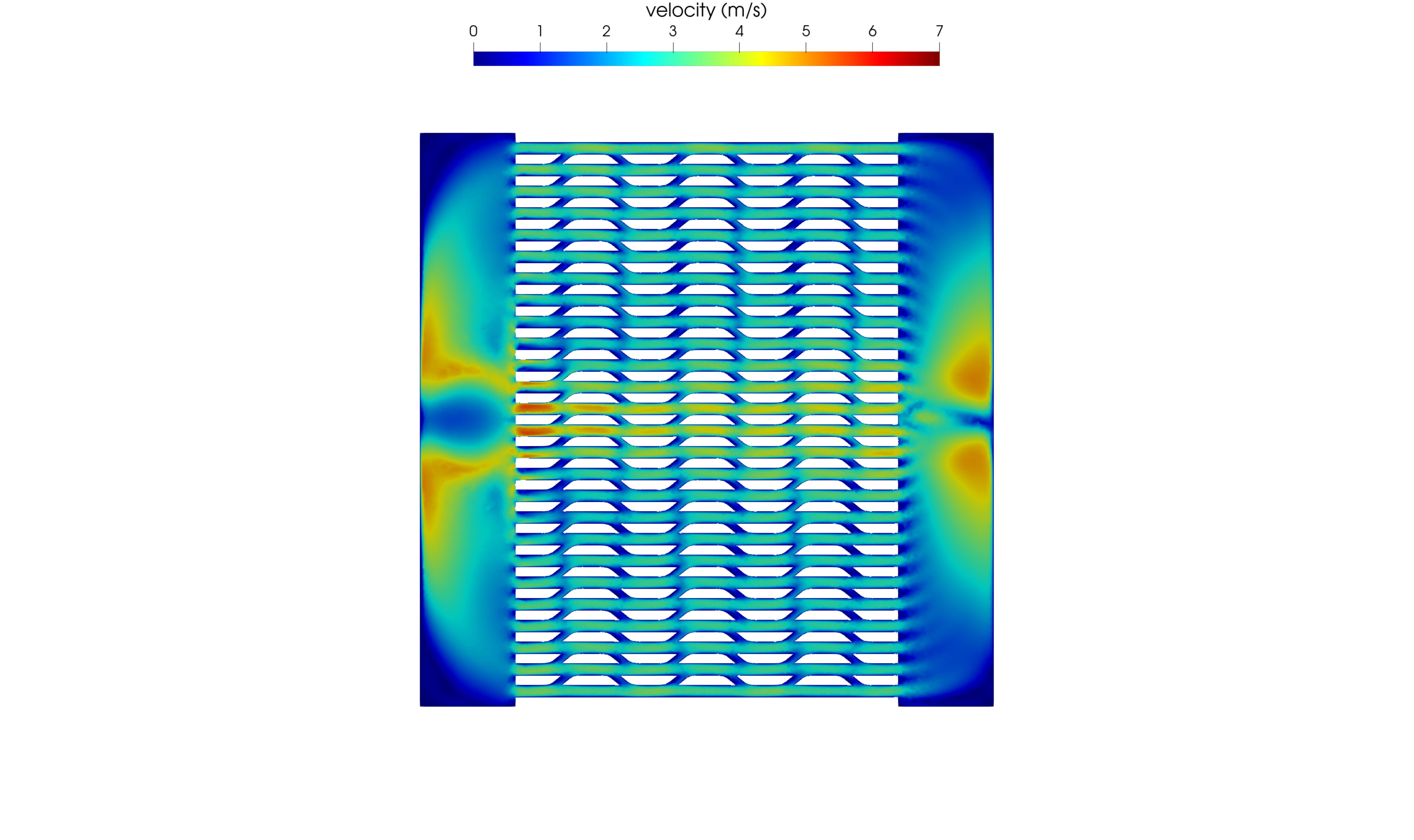}
    \caption{}
    \label{fig:CFD_MO_fields:a}
\end{subfigure}
\begin{subfigure}[b]{0.325\textwidth}
    \centering
    \includegraphics[trim={30cm 10cm 30cm 0cm}, clip, width=0.8\textwidth]{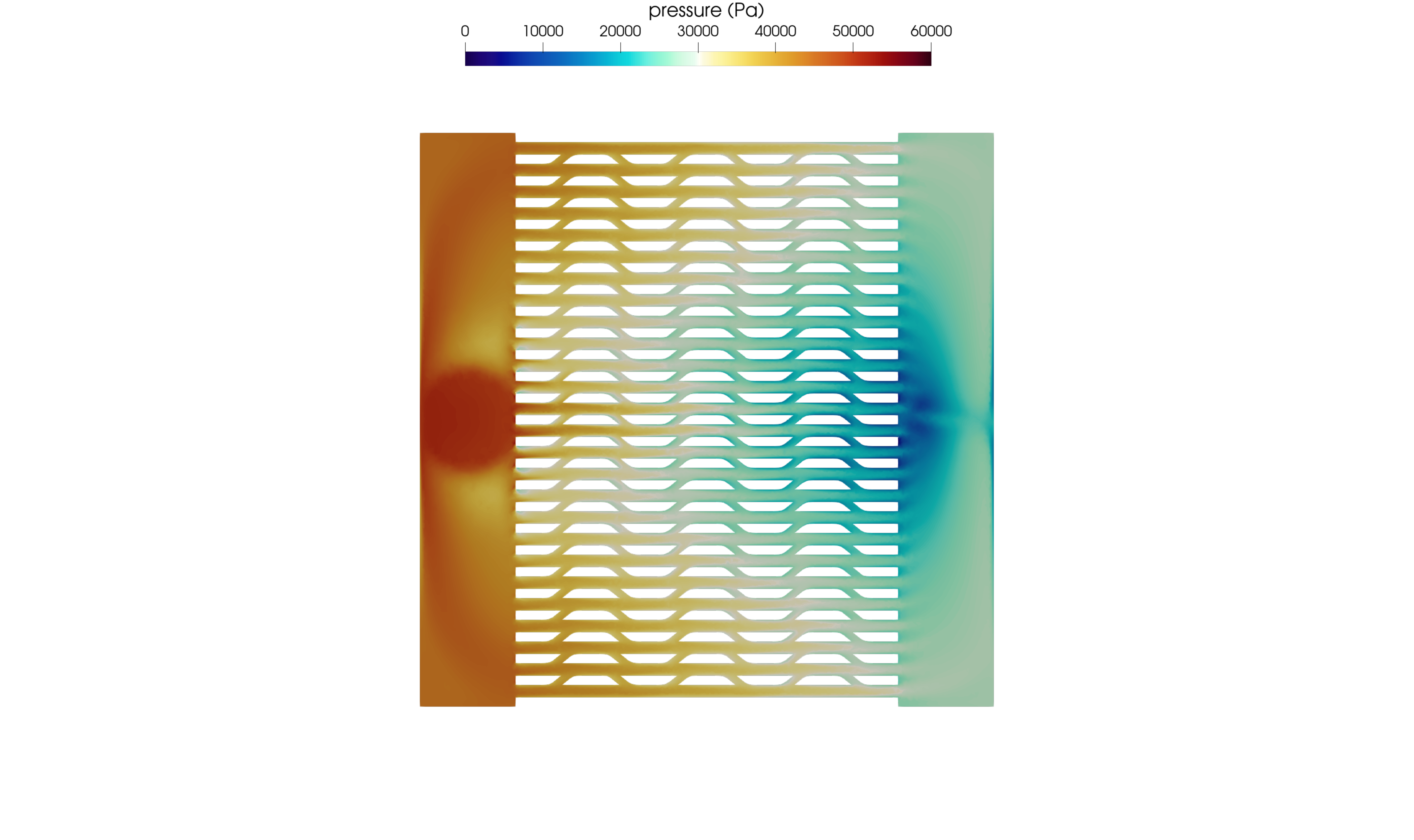}
    \caption{}
    \label{fig:CFD_MO_fields:b}
\end{subfigure} 
\begin{subfigure}[b]{0.325\textwidth}
    \centering
    \includegraphics[trim={30cm 10cm 30cm 0cm}, clip, width=0.8\textwidth]{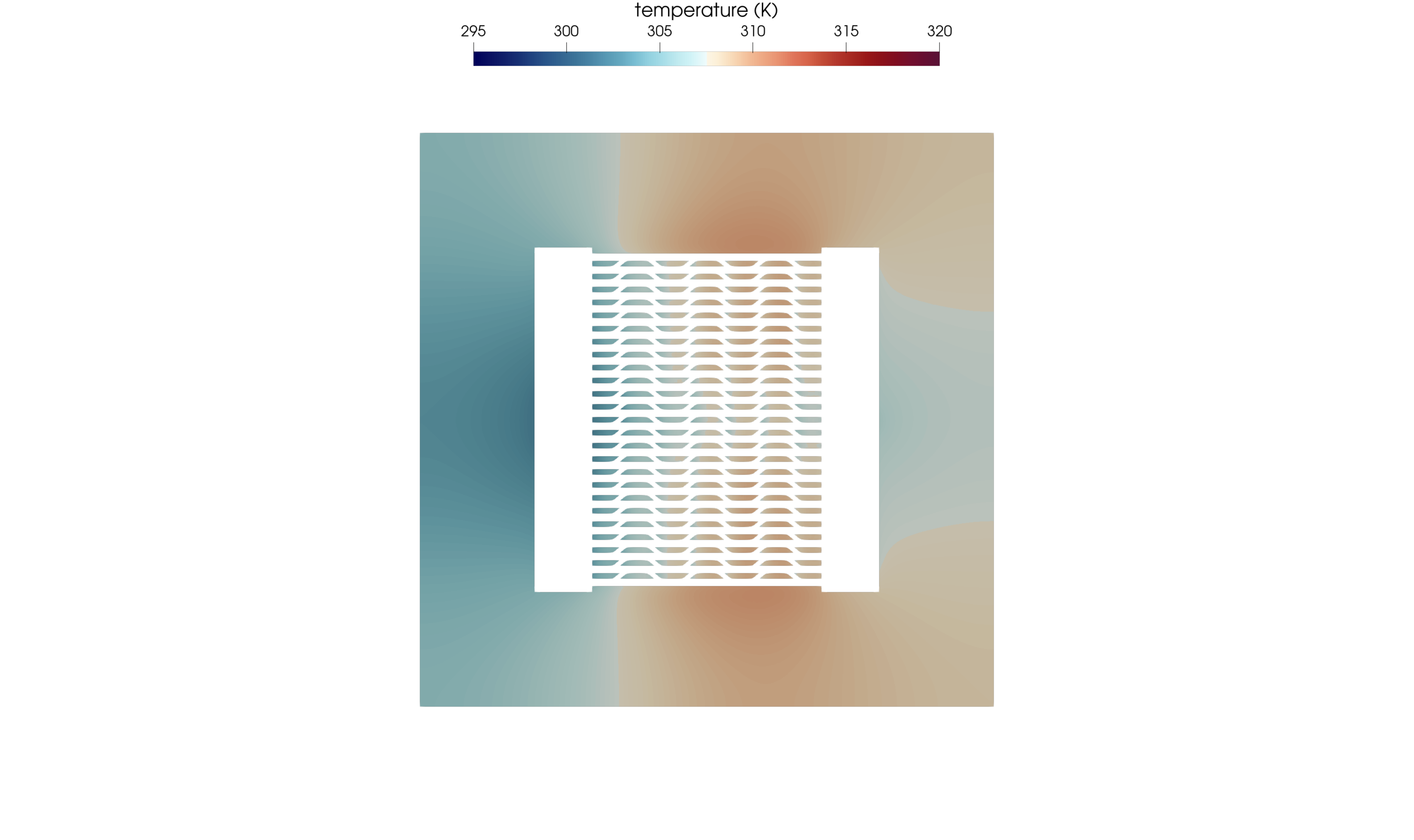}
    \caption{}
    \label{fig:CFD_MO_fields:c}
\end{subfigure}\\
\begin{subfigure}[b]{0.325\textwidth}
    \centering
    \includegraphics[trim={30cm 10cm 30cm 0cm}, clip, width=0.8\textwidth]{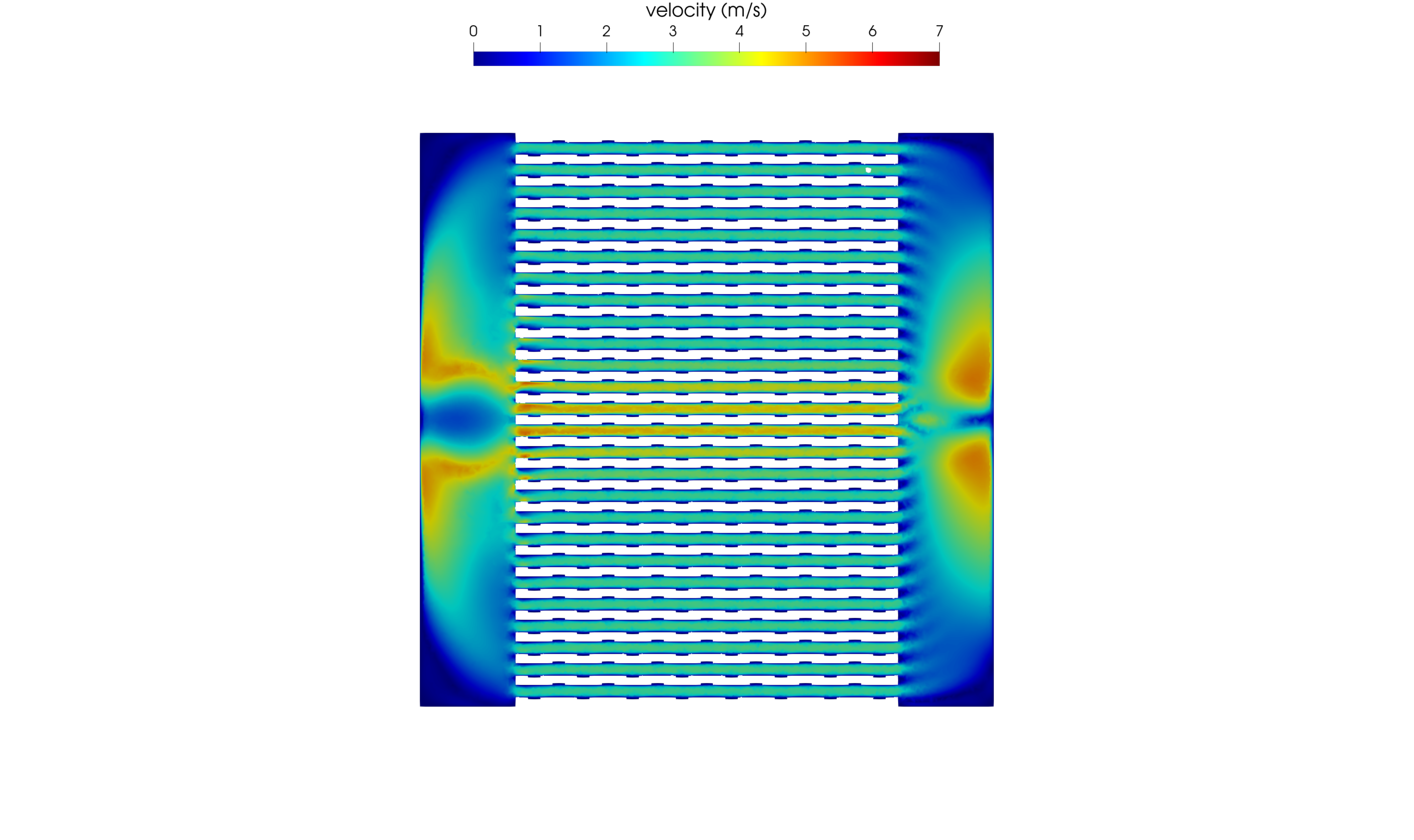}
    \caption{}
    \label{fig:CFD_MO_fields:d}
\end{subfigure}
\begin{subfigure}[b]{0.325\textwidth}
    \centering
    \includegraphics[trim={30cm 10cm 30cm 0cm}, clip, width=0.8\textwidth]{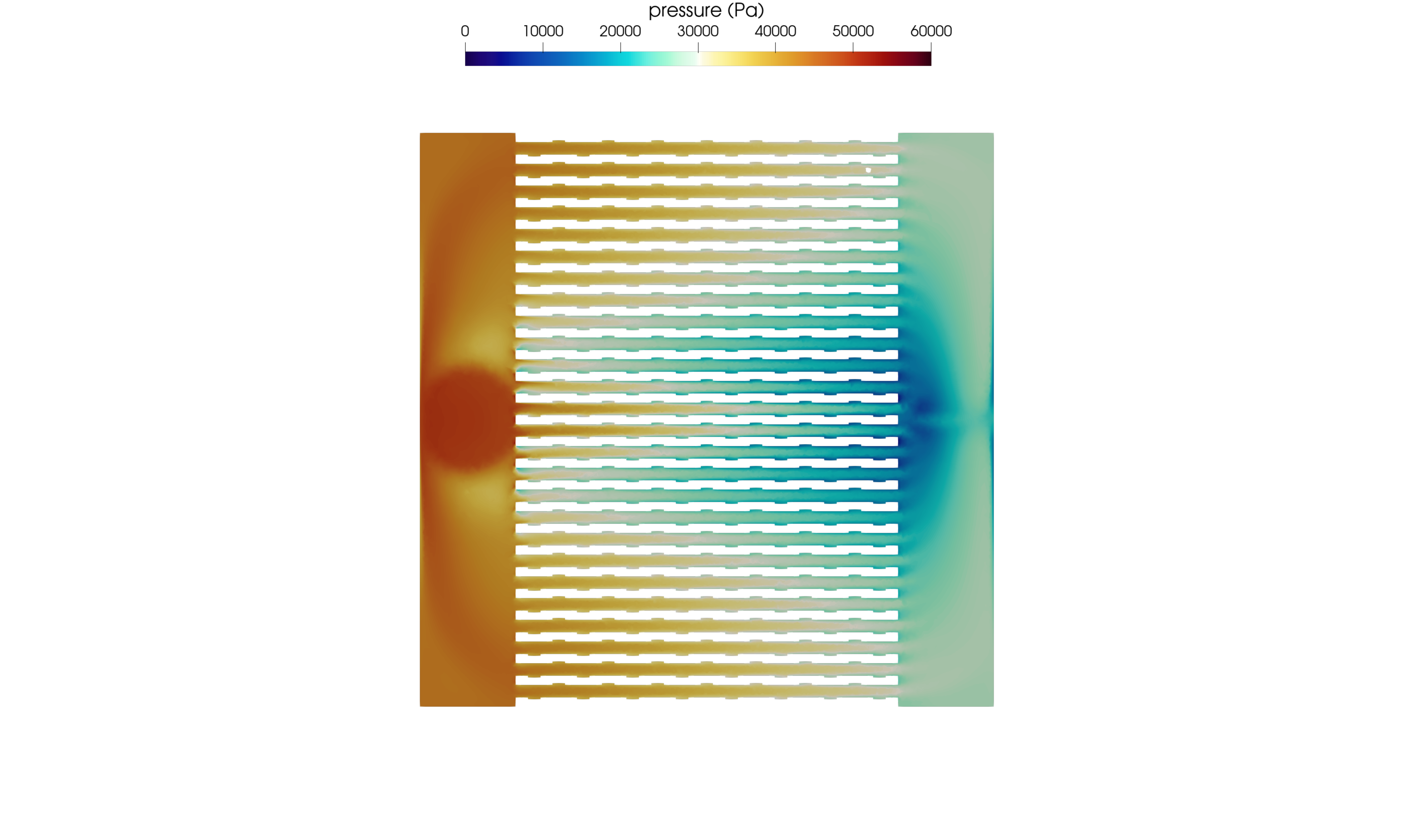}
    \caption{}
    \label{fig:CFD_MO_fields:e}
\end{subfigure}
\begin{subfigure}[b]{0.325\textwidth}
    \centering
    \includegraphics[trim={30cm 10cm 30cm 0cm}, clip, width=0.8\textwidth]{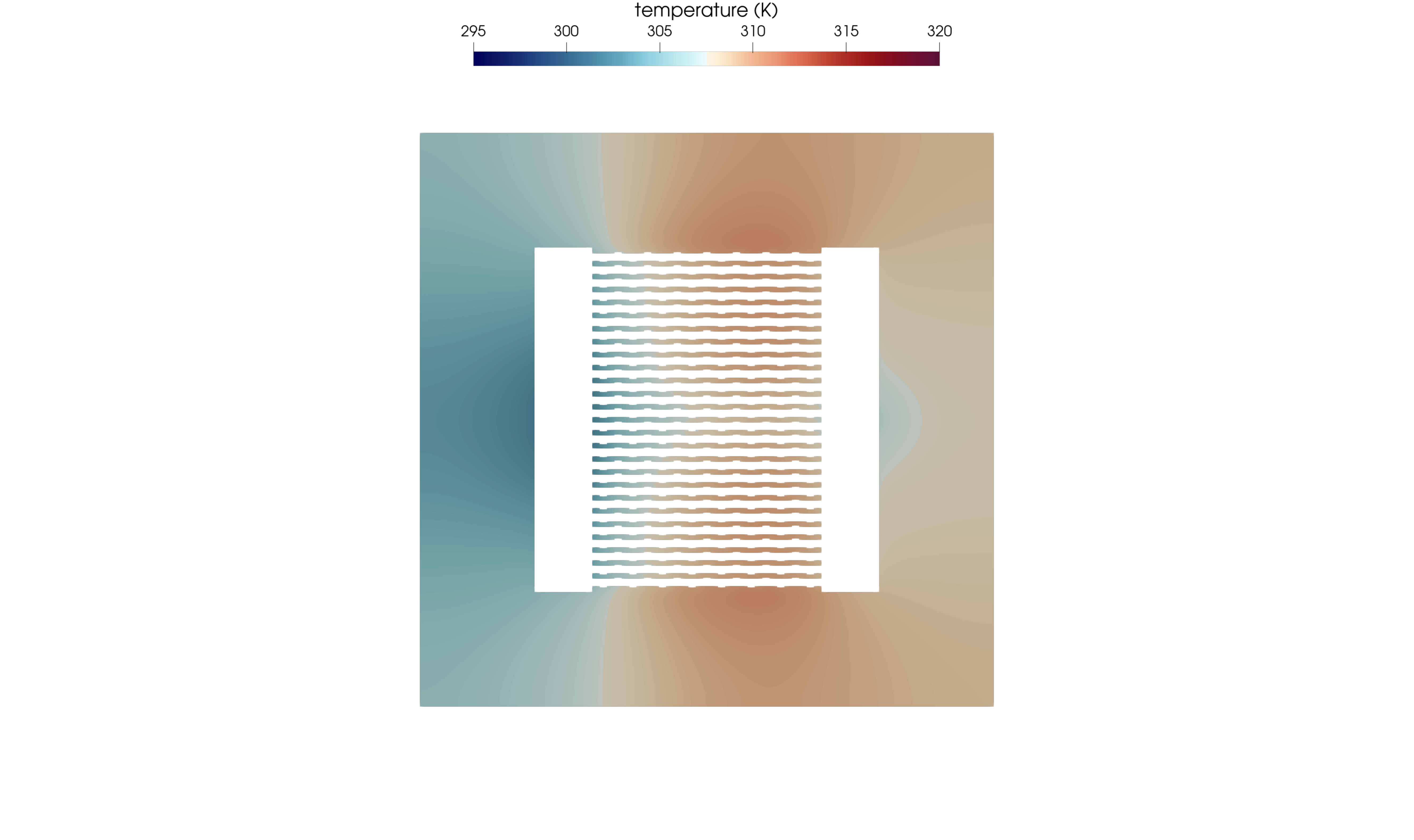}
    \caption{}
    \label{fig:CFD_MO_fields:f}
\end{subfigure}
\caption{Results for the SC-MC and R-MC C Pareto optimal variants at the midplane ($z=0$); (a) SC-MC velocity field, (b) SC-MC pressure field, (c) temperature distribution for the SC-MC, (d) R-MC velocity field, (e) R-MC pressure field, (f) temperature distribution for the R-MC.}
\label{fig:CFD_MO_fields}
\end{figure}

\ref{sec:appendix:a} includes three-dimensional models of all optimal solutions presented in Table \ref{tab:design_cfd_best}. Additionally, velocity, pressure and temperature fields at the midplane ($z=0$) for SC-MC and R-MC designs that achieve the largest temperature reduction (C variant) are given in Figure \ref{fig:CFD_MO_fields}. Pressure and velocity fields are mostly symmetrical about the $yz$ plane due to the positioning of the inlet and outlet. The bulk of the flow takes place near the centerline and is accompanied by a substantial pressure drop. The secondary channels of the SC-MC only account for a small portion of the flow, but, like ribs, they introduce enough disturbance and increase the overall surface area to facilitate heat dissipation.

\subsection{Multi-objective optimization using ML surrogate}
\label{subsec:ann_based_section}

Input datasets for surrogate modeling were made up of numerical results for distinct candidates defined through Latin hypercube sampling. Candidates that violated the MO constraints noted in Section \ref{subsec:mo_optimization} were excluded from datasets as they could hinder the training procedure. Additionally, since these were simple and known (user-defined) relations that would be used in the subsequent MO procedure, they did not need to be considered in the ML model. Following this, interquartile filter and feature scaling were employed. The tuning procedure for all considered ML models has been described in Section \ref{subsec:ml_surrogates}. Metrics for ML models trained using the SC-MC LHS data are given in Table \ref{tab:metrics_scmc}.

\begin{table}[H]
\centering
\footnotesize
\caption{Metrics and standard deviations for different ML models trained on the SC-MC dataset. Noted values represent the worst results from both output features ($R_t$ and $P_p$). Values in bold are the most accurate results for a given metric.}
\label{tab:metrics_scmc}
\begin{tabular}{@{}ccccc@{}}
\toprule
 & RF & CB & LGBM & ANN \\ \midrule
MAPE & 7.762 $\pm$ 0.679 & 4.568 $\pm$ 0.330 & 6.850 $\pm$ 0.649 & \textbf{1.409 $\pm$ 0.123} \\
RMSE & 0.023 $\pm$ 0.004 & 0.016 $\pm$ 0.003 & 0.020 $\pm$ 0.003 & \textbf{0.007 $\pm$ 0.001} \\
R$^2$ & 0.911 $\pm$ 0.006	& 0.956 $\pm$ 0.004 & 0.933 $\pm$ 0.005 & \textbf{0.983 $\pm$ 0.001} \\ \bottomrule
\end{tabular}
\end{table}

Based on the presented data, it is evident that the tuned ANN model achieves the overall best results. In terms of MAPE, the error is below 1.5\%. Given that the values of the output features are larger than 0.5, an RMSE of 0.007 is respectable. A high R$^2$ score indicates that the trained ANN model sufficiently explains the dependent variables. Low standard deviations are indicative of the robustness of the model. Metrics for models trained on the R-MC LHS dataset are given in Table \ref{tab:metrics_rmc}.

\begin{table}[H]
\centering
\footnotesize
\caption{Metrics and standard deviations for different ML models trained on the R-MC dataset. Noted values represent the worst results from both output features ($R_t$ and $P_p$). Values in bold are the most accurate results for a given metric.}
\label{tab:metrics_rmc}
\begin{tabular}{@{}ccccc@{}}
\toprule
 & RF & CB & LGBM & ANN \\ \midrule
MAPE & 8.000 $\pm$ 0.633 & 5.503 $\pm$ 0.466 & 7.217 $\pm$ 0.644 & \textbf{1.925 $\pm$ 0.248} \\
RMSE & 0.031 $\pm$ 0.005 & 0.024 $\pm$ 0.006 & 0.027 $\pm$ 0.005 & \textbf{0.012 $\pm$ 0.006} \\
R$^2$ & 0.939 $\pm$ 0.015 & 0.964 $\pm$ 0.010 & 0.952 $\pm$ 0.012 & \textbf{0.980 $\pm$ 0.004} \\ \bottomrule
\end{tabular}
\end{table}

Metrics for the R-MC ML models are comparatively slightly worse than for SC-MC models. Still, ANN achieves the best results with a high R$^2$ score. The presented data is indicative of the suitability of considered ML models as design surrogates. ANN models were hence used in the multi-objective optimization procedure. Relevant hyperparameters for tuned ANNs are given in Table \ref{tab:parameters_ANN}. Graphical depictions of the neural network designs are included in \ref{sec:appendix:a}.

\begin{table}[H]
\centering
\footnotesize
\caption{Tuned parameters for ANN models. All other parameters are kept at default values.}
\label{tab:parameters_ANN}
\begin{tabular}{@{}lcc@{}}
\toprule
 & SC-MC & R-MC \\ \midrule
No. hidden layers & 4 & 3 \\
No. neurons per layer & 256,128,8,4 & 256,128,64 \\
Batch size & 64 & 64 \\
Learning rate & 0.001 & 0.00075 \\
Activation functions & ELU & ReLU \\
Initializers & TruncatedNormal & TruncatedNormal \\ \bottomrule
\end{tabular}
\end{table}

Learning curves for both ANN models have been calculated and included in \ref{sec:appendix:a} (Figure \ref{fig:learning_curve}) to further validate the models and determine the influence of the dataset size on the model accuracy. It is evident that for $N > 1000$, obtained R$^2$ score for 5-fold validation is larger than 95\%. This implies that the sample size of 1500 is sufficient for the next stage of the workflow. It is important to note that for $N < 500$, accuracy for both models drops significantly hence it is sensible to conclude that smaller datasets would be inadequate to properly model the MCHS.

\begin{figure}[H]
\centering
\includegraphics[trim={0cm 0cm 0cm 0cm}, clip, width=0.5\textwidth]{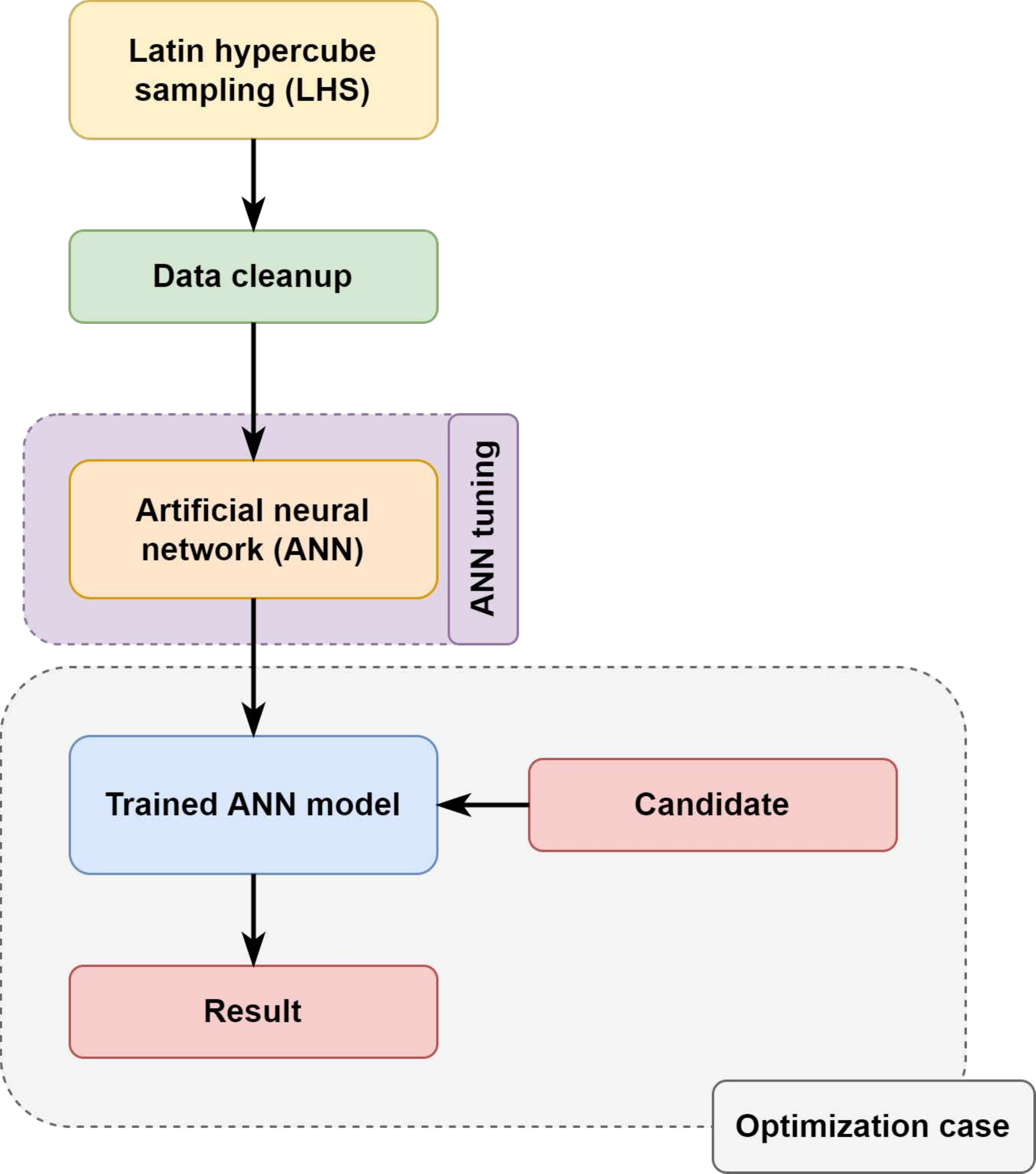}
\caption{Machine learning-based workflow.}
\label{fig:workflow_ML}
\end{figure}

A simplified overview of the proposed machine-learning-based MCHS optimization workflow is given in Figure \ref{fig:workflow_ML}. Numerical results for LHS-defined candidates are filtered, scaled, and used to train an ANN model. The model is fine-tuned using the FWA algorithm and subsequently integrated into an NSGA-II-based optimization procedure. Pareto optimal solutions of the multi-objective optimization are considered when proposing the best designs, per methodology defined in Section \ref{subsec:mo_optimization}.

\begin{figure}[H]
\centering
\begin{subfigure}[b]{0.495\textwidth}
    \centering
    \includegraphics[trim={0cm 0.8cm 1cm 1cm}, clip, width=\textwidth]{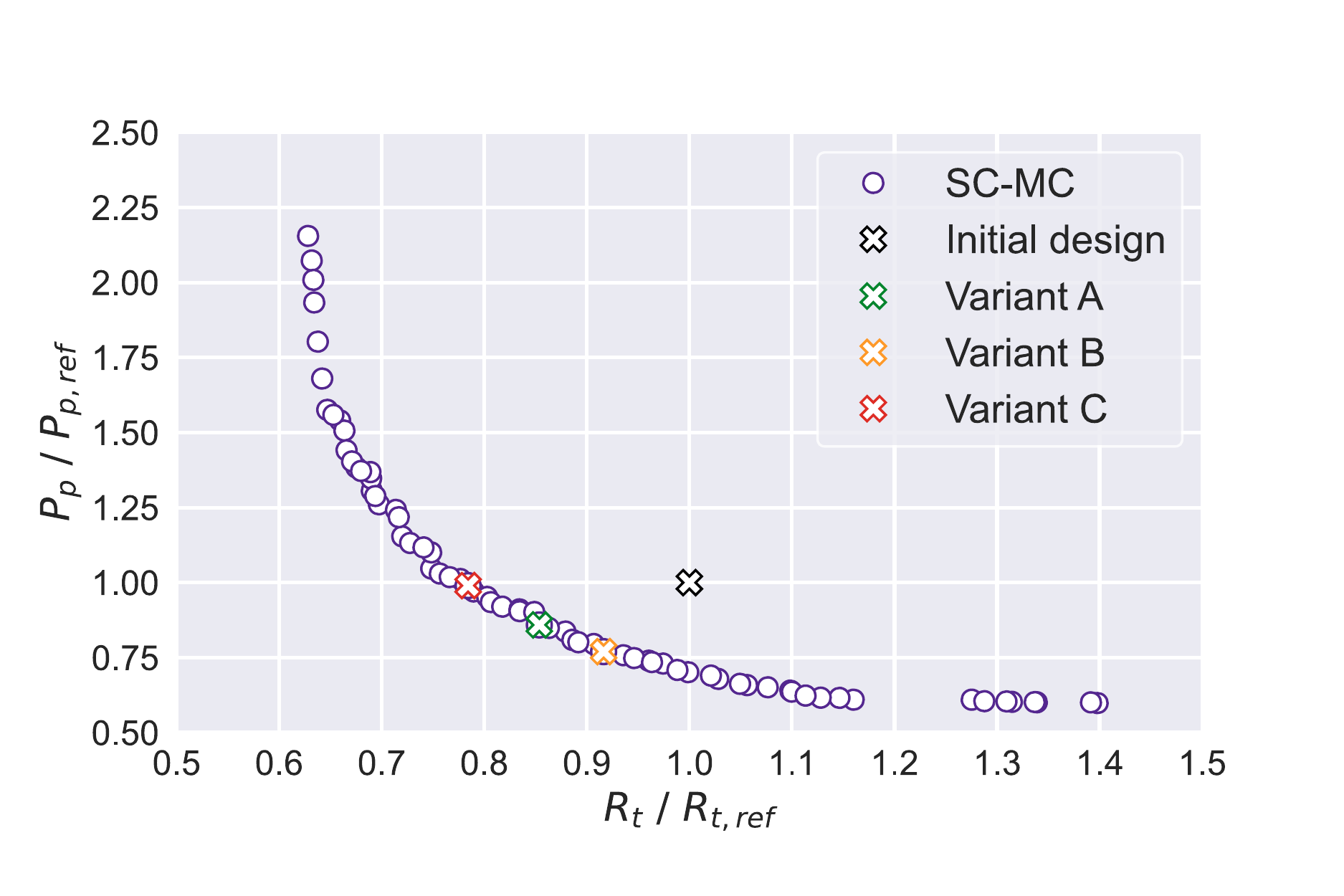}
    \caption{}
    \label{fig:pareto_ann:a}
\end{subfigure}
\begin{subfigure}[b]{0.495\textwidth}
    \centering
    \includegraphics[trim={0cm 0.8cm 1cm 1cm}, clip, width=\textwidth]{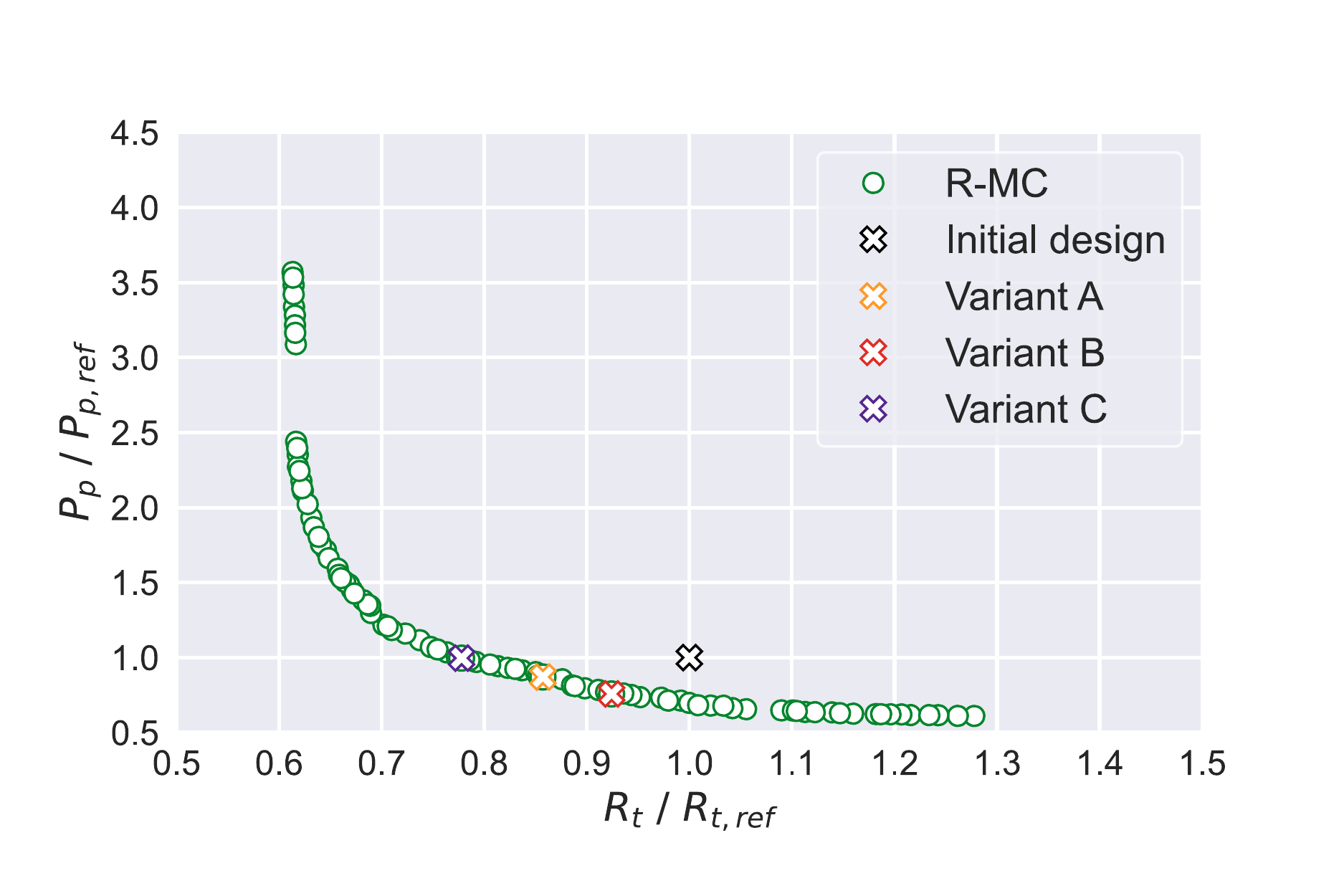}
    \caption{}
    \label{fig:pareto_ann:b}
\end{subfigure}
\caption{Pareto fronts obtained using the ML-based workflow; (a) front for the SC-MC design, (b) front for the R-MC design.}
\label{fig:pareto_ann}
\end{figure}

The Pareto fronts obtained using the aforementioned workflow are shown in Figure \ref{fig:pareto_ann} and exhibit similar trends to those obtained using the conventional optimization approach. The asymptotic behavior of normalized thermal resistance is in accordance with the previous results. Fronts, however, fail to capture the magnitude of the increase in pumping power following the decrease in thermal resistance. As the LHS sampling provides a reduced dataset which may lack information on correlation under these circumstances, the observed behavior is reasonable. Still, this disparity can be partially attributed to the stochastic nature of the MO algorithm.

Proposed designs based on the ANN-NSGA-II approach, that meet the criteria outlined in Section \ref{subsec:mo_optimization}, are given in Table \ref{tab:design_ann_best}. Respective three-dimensional models are included in the \ref{sec:appendix:a}. Optimal designs for both microchannel models show similar performance improvements. At the design point, variants A achieve a $T_{max}$ reduction of 4.6\% for SC-MC and 4.7\% for R-MC. The pressure drop is reduced by 18.6\% and 18\%, respectively. The Pareto B variants provide a greater reduction in pumping power at the expense of thermal resistance. Consequently, the pressure drop is reduced by 23.2\% and 24\%, while the maximum temperature is lowered by 1.8\% and 1.2\% for SC-MC and R-MC designs. Temperature reduction for C variants is around 9.3\% and 8.7\% for considered MCHS models, with pressure drop reduced by approximately 4.7\%.

\begin{table}[H]
\centering
\footnotesize
\caption{Model parameters for chosen optimal designs generated through the ANN-NSGA-II coupling.}
\label{tab:design_ann_best}
\begin{tabular}{@{}cccccccccccccc@{}}
\toprule
& \multicolumn{7}{c}{SC-MC} & \multicolumn{6}{c}{R-MC} \\
$\bm{x}$ & $n$ & $m$ & \makecell[c]{~$d$ \\[-2pt] \added{(mm)}} & \makecell[c]{~$d_m$ \\[-2pt] \added{(mm)}} & \makecell[c]{~$\alpha$ \\[-2pt] \added{($^{\circ}$)}} & $r_s$ & $r_f$ & $n$ & $m$ & \makecell[c]{~$d$ \\[-2pt] \added{(mm)}} & \makecell[c]{~$d_m$ \\[-2pt] \added{(mm)}} & \makecell[c]{~$d_w$ \\[-2pt] \added{(mm)}} & \makecell[c]{~$d_h$ \\[-2pt] \added{(mm)}} \\ \cmidrule(r){1-1} \cmidrule(lr){2-8} \cmidrule(l){9-14}
A & 22 & 8 & 0.17 & 0.17 & 31 & 1 & 0 & 22 & 16 & 0.17 & 0.17 & 0.04 & 0.12 \\
B & 22 & 6 & 0.17 & 0.27 & 20 & 1 & 0 & 21 & 13 & 0.18 & 0.18 & 0.04 & 0.14 \\
C & 26 & 6 & 0.13 & 0.13 & 37 & 1 & 0 & 26 & 14 & 0.13 & 0.13 & 0.03 & 0.09 \\ \bottomrule
\end{tabular}
\end{table}

\subsection{Feature importance}

SHAP is a novel technique that expands on cooperative game theory's Shapely values in order to interpret ML models and determine which input features have the most influence on the output features \cite{lundberg2017}. The fundamental idea behind the technique is to distribute the credit for an output feature among the inputs. Based on this and Shapely values properties, it is possible to calculate the relevance of each input feature.

\begin{figure}[H]
\centering
\begin{subfigure}[b]{0.495\textwidth}
    \centering
    \includegraphics[trim={0cm 0cm 0cm 0cm}, clip, width=\textwidth]{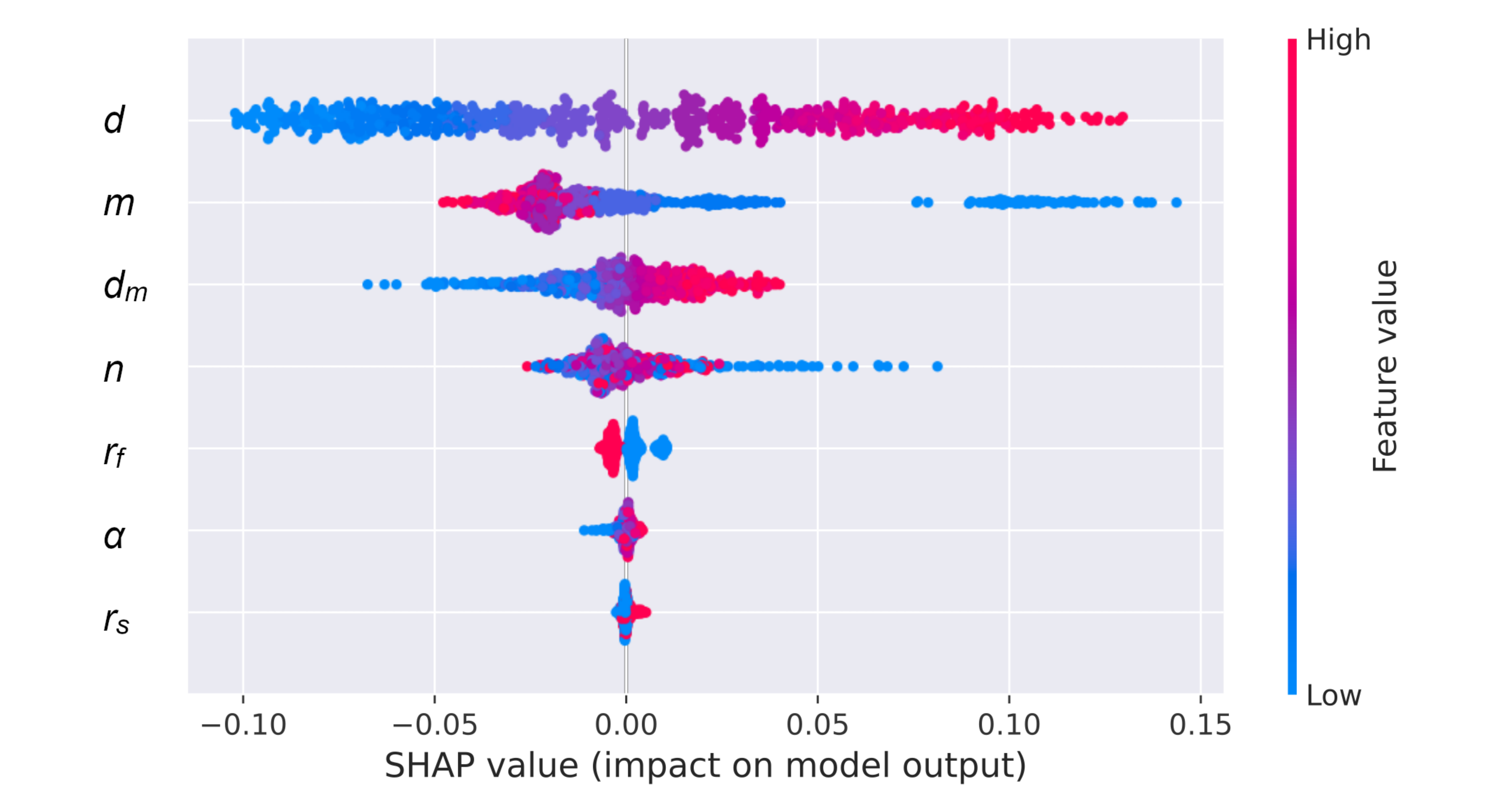}
    \caption{}
    \label{fig:shap:a}
\end{subfigure}
\begin{subfigure}[b]{0.495\textwidth}
    \centering
    \includegraphics[trim={0cm 0cm 0cm 0cm}, clip, width=\textwidth]{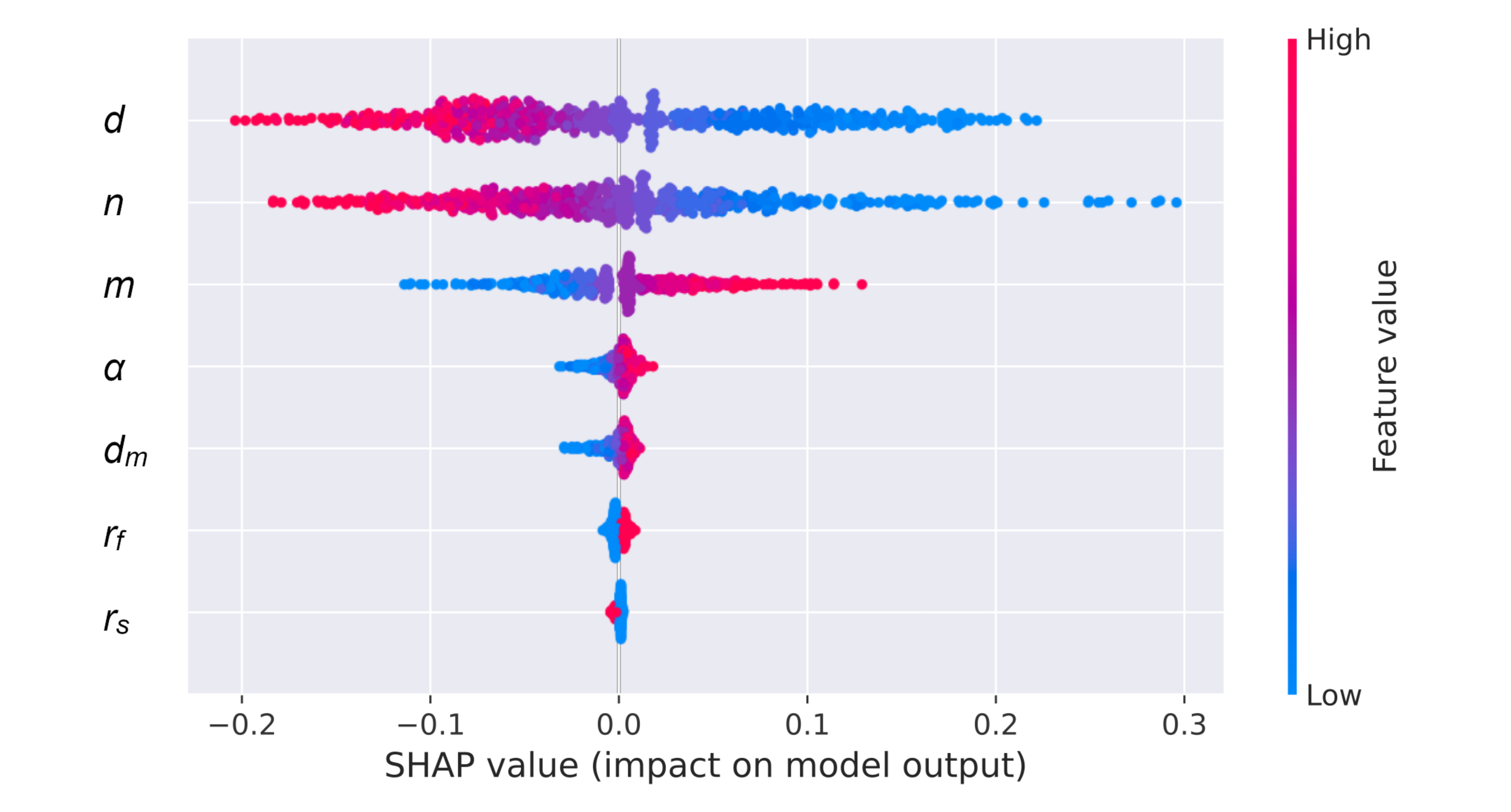}
    \caption{}
    \label{fig:shap:b}
\end{subfigure}\\
\begin{subfigure}[b]{0.495\textwidth}
    \centering
    \includegraphics[trim={0cm 0cm 0cm 0cm}, clip, width=\textwidth]{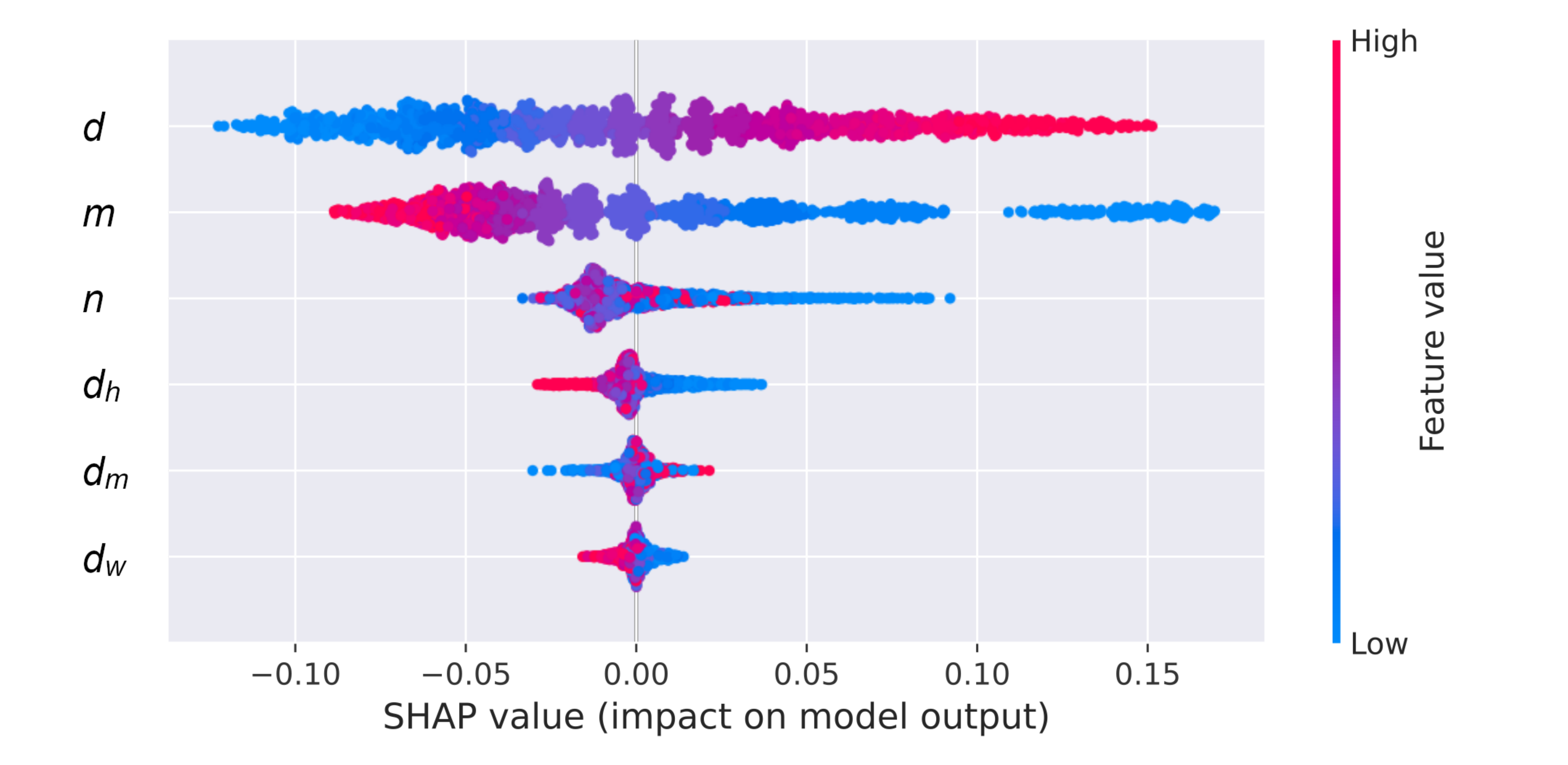}
    \caption{}
    \label{fig:shap:c}
\end{subfigure}
\begin{subfigure}[b]{0.495\textwidth}
    \centering
    \includegraphics[trim={0cm 0cm 0cm 0cm}, clip, width=\textwidth]{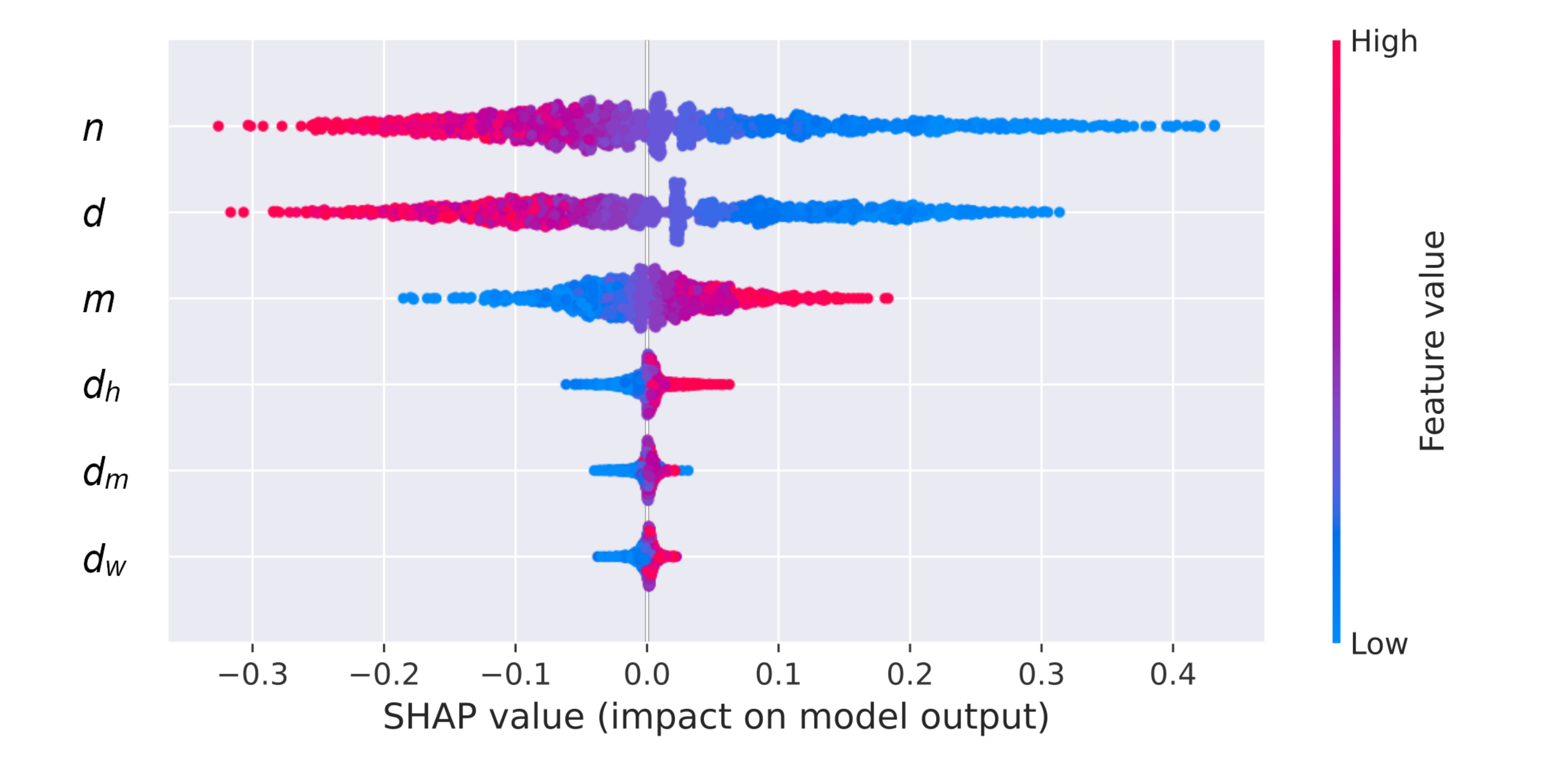}
    \caption{}
    \label{fig:shap:d}
\end{subfigure}
\caption{SHAP feature impact on ML model outputs; (a) feature impact on thermal resistance for the SC-MC, (b) feature impact on pumping power for the SC-MC, (c) feature impact on thermal resistance for the R-MC, (d) feature impact on pumping power for the R-MC.}
\label{fig:shap}
\end{figure}

Since LHS was used to generate the data for training, the inherent applicability is not immediately apparent. However, as it was demonstrated that proposed ANNs accurately model the relationship between the inputs and outputs, SHAP methodology can be employed to determine features i.e. design variables with the most influence on thermal resistance and pumping power. These variables are thus the most important when optimizing MCHS design. Employed KernelExplainer method uses a weighted linear regression to ascertain the feature importance \cite{lundberg2017}. Figure \ref{fig:shap} shows SHAP values for each of the model outputs (for both SC-MC and R-MC). Feature importance is ranked in a descending manner. Graphs can be interpreted as follows: as the feature value for a given feature (variable, $y$-axis) increases (red) or decreases (blue), it can have a positive or a negative influence (SHAP value, $x$-axis) on the output. Microchannel width $d$ and the number of secondary channels/number of ribs $m$ have the biggest influence on thermal resistance for SC-MC and R-MC. Pumping power is predominantly determined by microchannel width $d$ and the number of microchannels $n$ i.e. with their increase $\Delta p$ is reduced. On the other hand, an increase in $m$ leads to an increase in pumping power, which is expected. These conclusions are valid for the conventional optimization approach too. 

The limitations, however, of the SHAP methodology stem from the machine learning approach itself, in which assessed surrogates rely on data that may be limited or not representative, making it impossible to fully understand or interpret the physics of the problem. Recently, however, there have been strides to incorporate well-known physics into machine learning models \cite{karniadakis2021} as well as to use machine learning to discover the underlying physics \cite{karagiorgi2022}.

\subsection{Comparative analysis}

Pareto optimal solutions obtained at the design point and presented in previous chapters will thus be assessed for a range of flow rates and compared to respective counterparts, i.e. best solutions obtained using the CFD-NSGA-II approach will be compared to solutions obtained using the proposed ML-based workflow. Presented graphs include CFD-verified curves for generated solutions.

\begin{figure}[H]
\centering
\begin{subfigure}[b]{0.495\textwidth}
    \centering
    \includegraphics[trim={0cm 0.8cm 1cm 1cm}, clip, width=\textwidth]{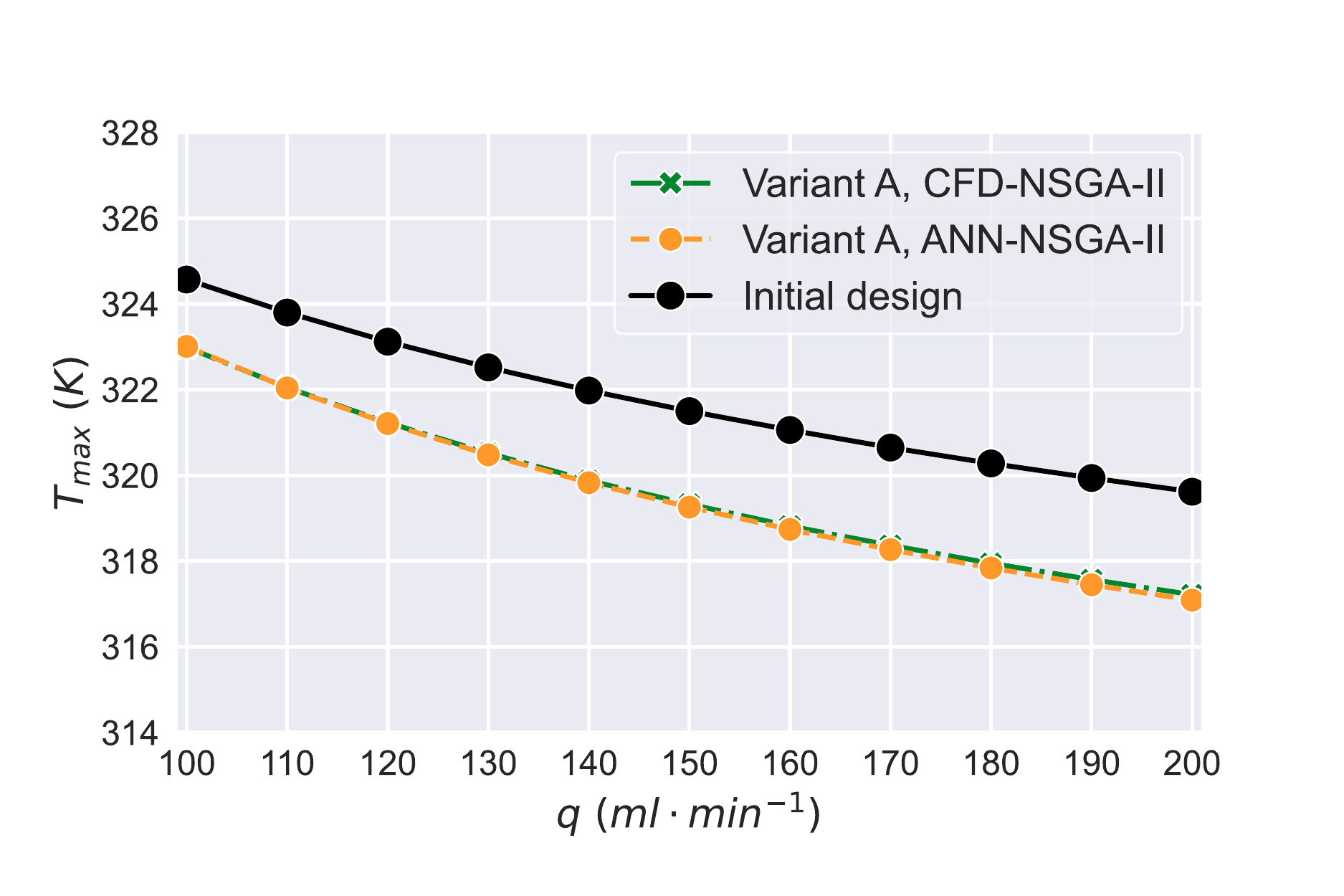}
    \caption{}
    \label{fig:scmc_A_comparison:a}
\end{subfigure}
\begin{subfigure}[b]{0.495\textwidth}
    \centering
    \includegraphics[trim={0cm 0.8cm 1cm 1cm}, clip, width=\textwidth]{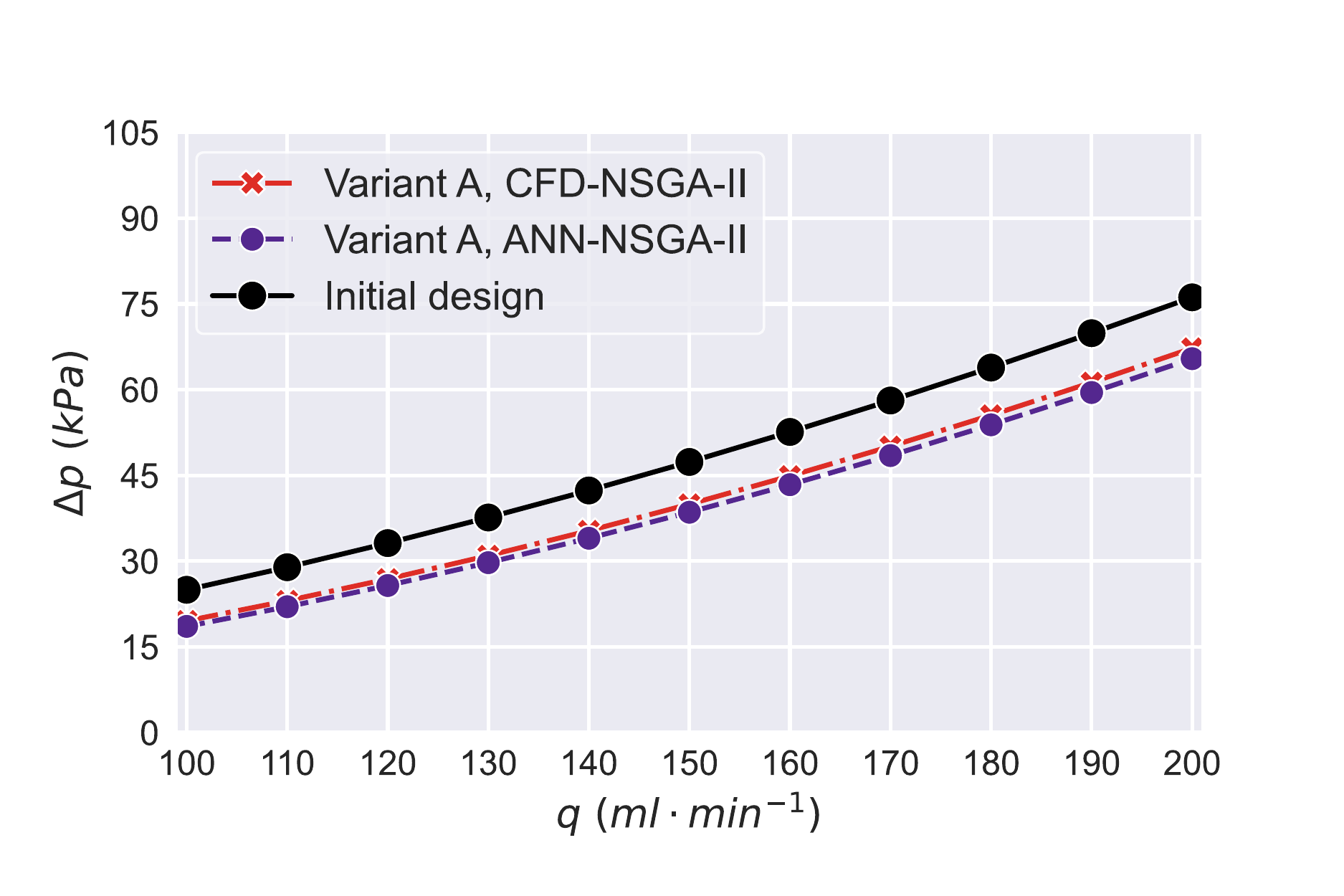}
    \caption{}
    \label{fig:scmc_A_comparison:b}
\end{subfigure}
\caption{Comparison between proposed SC-MC A variants for different flow rates; (a) temperature, (b) pressure drop.}
\label{fig:scmc_A_comparison}
\end{figure}

Proposed A variants for the SC-MC design show negligible variance with respect to the temperature. At best, results obtained using the ML approach are 0.3\% lower. Pressure drop is slightly larger for the CFD-NSGA-II solution, up to 4.9\% at $Q = 100$ ml$\cdot$min$^{-1}$. Noted results are shown in Figure \ref{fig:scmc_A_comparison}. When compared to the initial design, the CFD-NSGA-II variant shows improvements of, on average, 1.6\degree C and 7.3 kPa, for $T_{max}$ and $\Delta p$.

\begin{figure}[H]
\centering
\begin{subfigure}[b]{0.495\textwidth}
    \centering
    \includegraphics[trim={0cm 0.8cm 1cm 1cm}, clip, width=\textwidth]{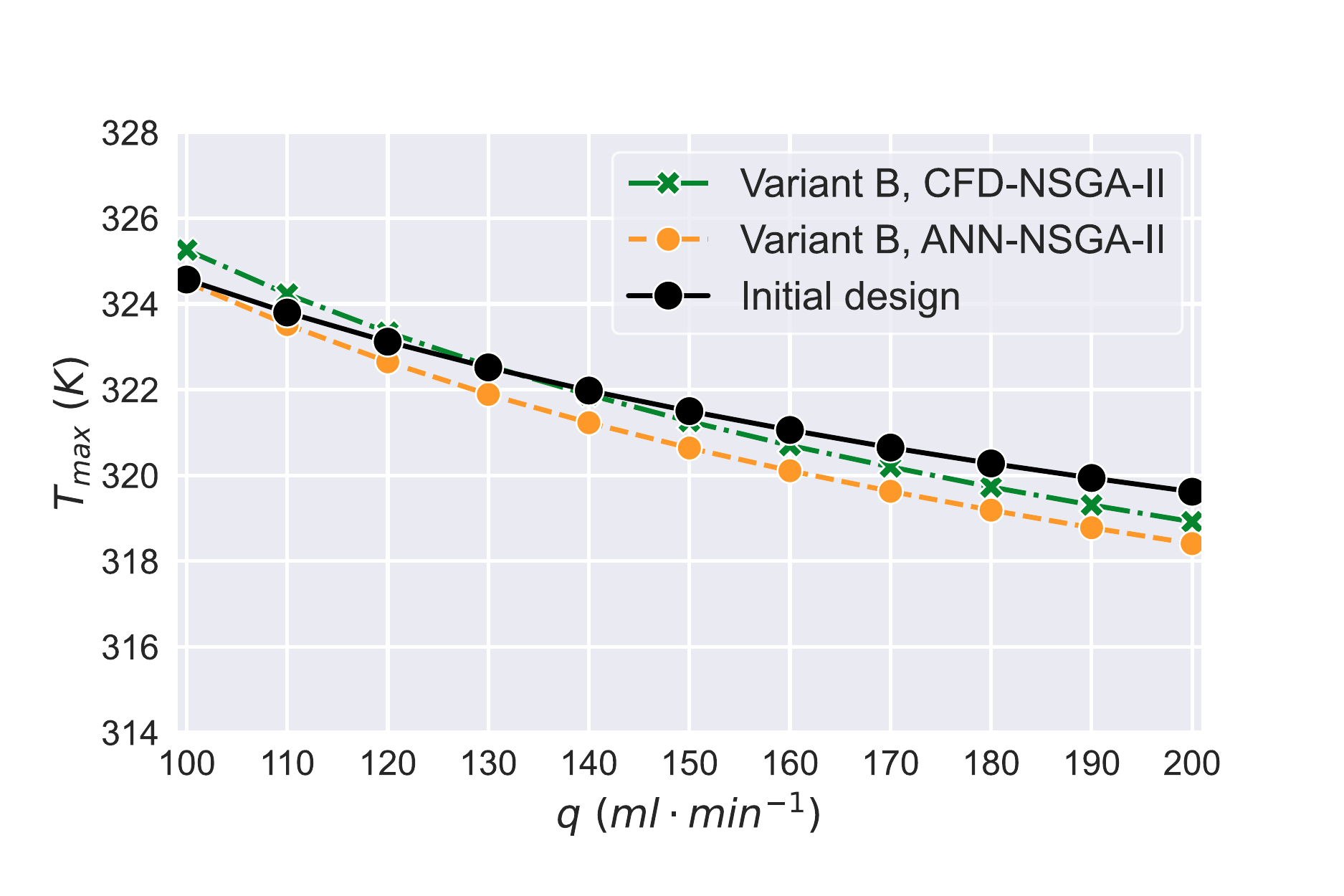}
    \caption{}
    \label{fig:scmc_B_comparison:a}
\end{subfigure}
\begin{subfigure}[b]{0.495\textwidth}
    \centering
    \includegraphics[trim={0cm 0.8cm 1cm 1cm}, clip, width=\textwidth]{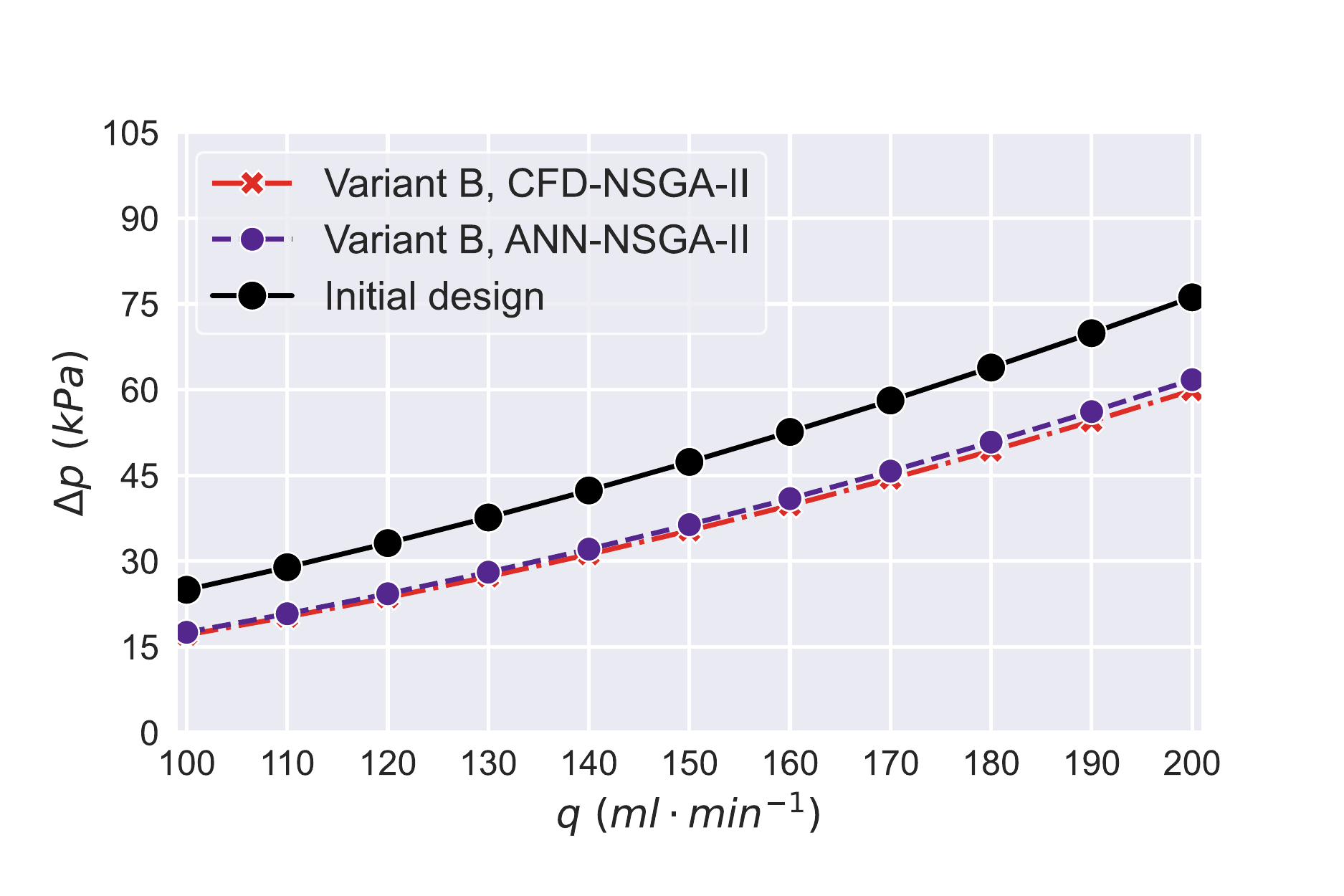}
    \caption{}
    \label{fig:scmc_B_comparison:b}
\end{subfigure}
\caption{Comparison between proposed SC-MC B variants for different flow rates; (a) temperature, (b) pressure drop.}
\label{fig:scmc_B_comparison}
\end{figure}

ANN-NSGA-II solutions are more convincing when comparing B variants for the SC-MC model. Compared to the initial design, $T_{max}$ and $\Delta p$ are lowered by 0.7\degree C and 11 kPa, respectively. This is to be expected, given that the emphasis in B variants is on pressure drop reduction. Temperatures between variants differ by up to 1.4\% and for $Q = 100$ ml$\cdot$min$^{-1}$, the temperature limit of the initial design is respected. In terms of pressure, the CFD-NSGA-II solution shows up to 3.1\% lower drops. Comparison is shown in Figure \ref{fig:scmc_B_comparison}.

\begin{figure}[H]
\centering
\begin{subfigure}[b]{0.495\textwidth}
    \centering
    \includegraphics[trim={0cm 0.8cm 1cm 1cm}, clip, width=\textwidth]{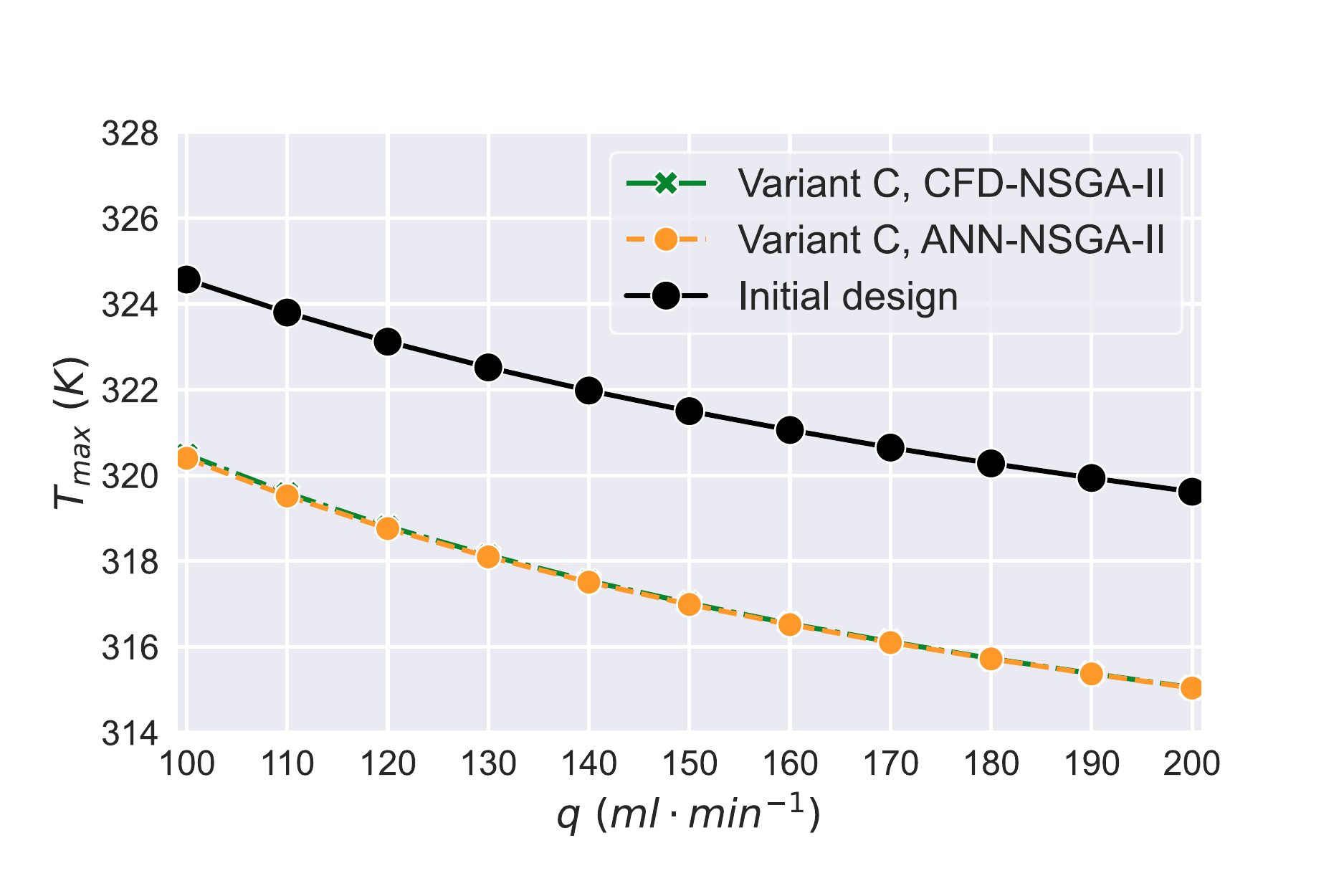}
    \caption{}
    \label{fig:scmc_C_comparison:a}
\end{subfigure}
\begin{subfigure}[b]{0.495\textwidth}
    \centering
    \includegraphics[trim={0cm 0.8cm 1cm 1cm}, clip, width=\textwidth]{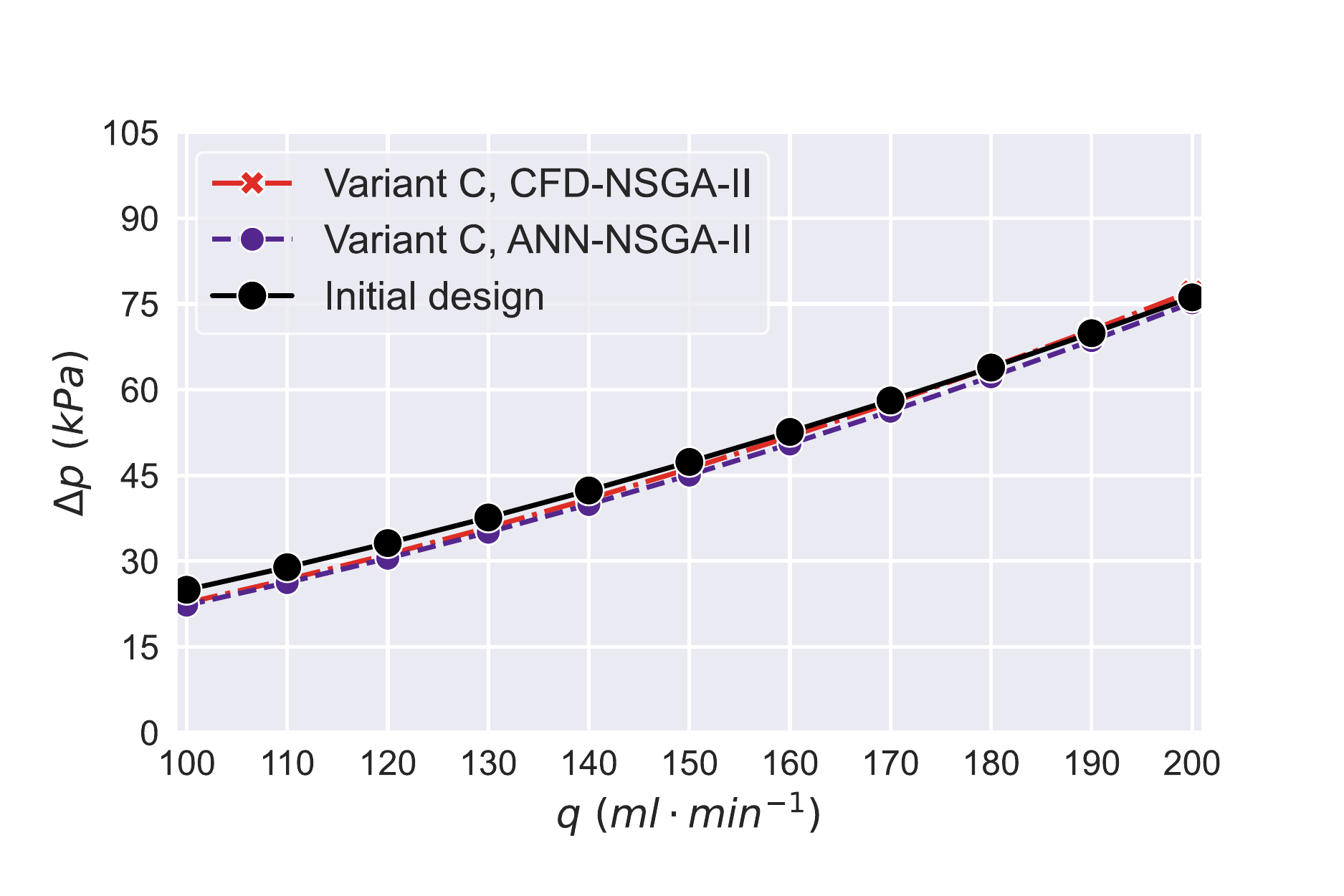}
    \caption{}
    \label{fig:scmc_C_comparison:b}
\end{subfigure}
\caption{Comparison between proposed SC-MC C variants for different flow rates; (a) temperature, (b) pressure drop.}
\label{fig:scmc_C_comparison}
\end{figure}

Both MO optimization approaches produce similar results when determining the lowest attainable temperatures for the SC-MC model. The temperature difference is insignificant. However, CFD-NSGA-II solutions achieve this temperature equivalence due to an increase in pressure drop of up to 2.6\%. Still, the pressure is approximately 0.9 kPa lower than that of the initial design, while the temperatures are improved, on average, by 4.4\degree C. These observations can be seen in Figure \ref{fig:scmc_C_comparison}.

\begin{figure}[H]
\centering
\begin{subfigure}[b]{0.495\textwidth}
    \centering
    \includegraphics[trim={0cm 0.8cm 1cm 1cm}, clip, width=\textwidth]{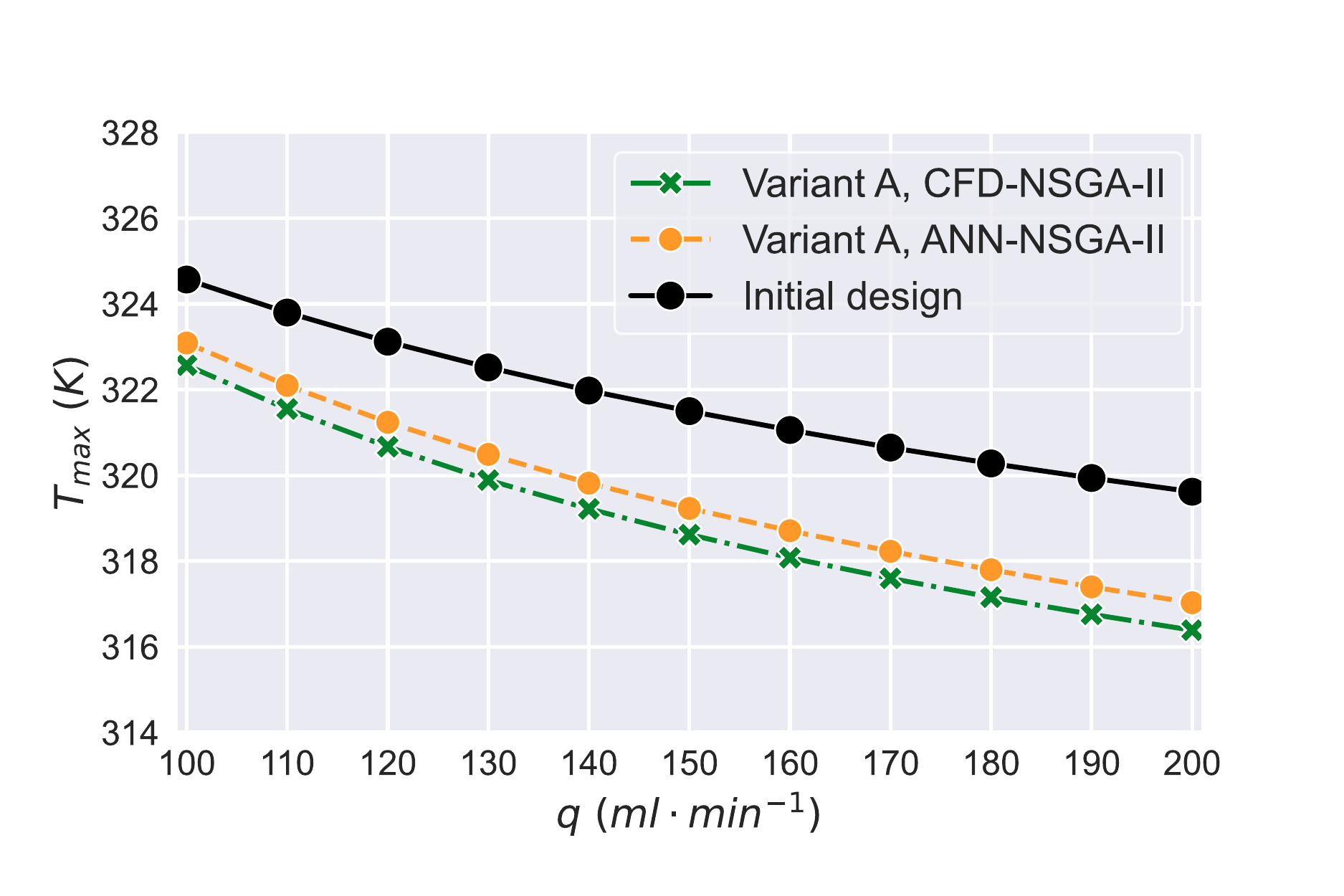}
    \caption{}
    \label{fig:rmc_A_comparison:a}
\end{subfigure}
\begin{subfigure}[b]{0.495\textwidth}
    \centering
    \includegraphics[trim={0cm 0.8cm 1cm 1cm}, clip, width=\textwidth]{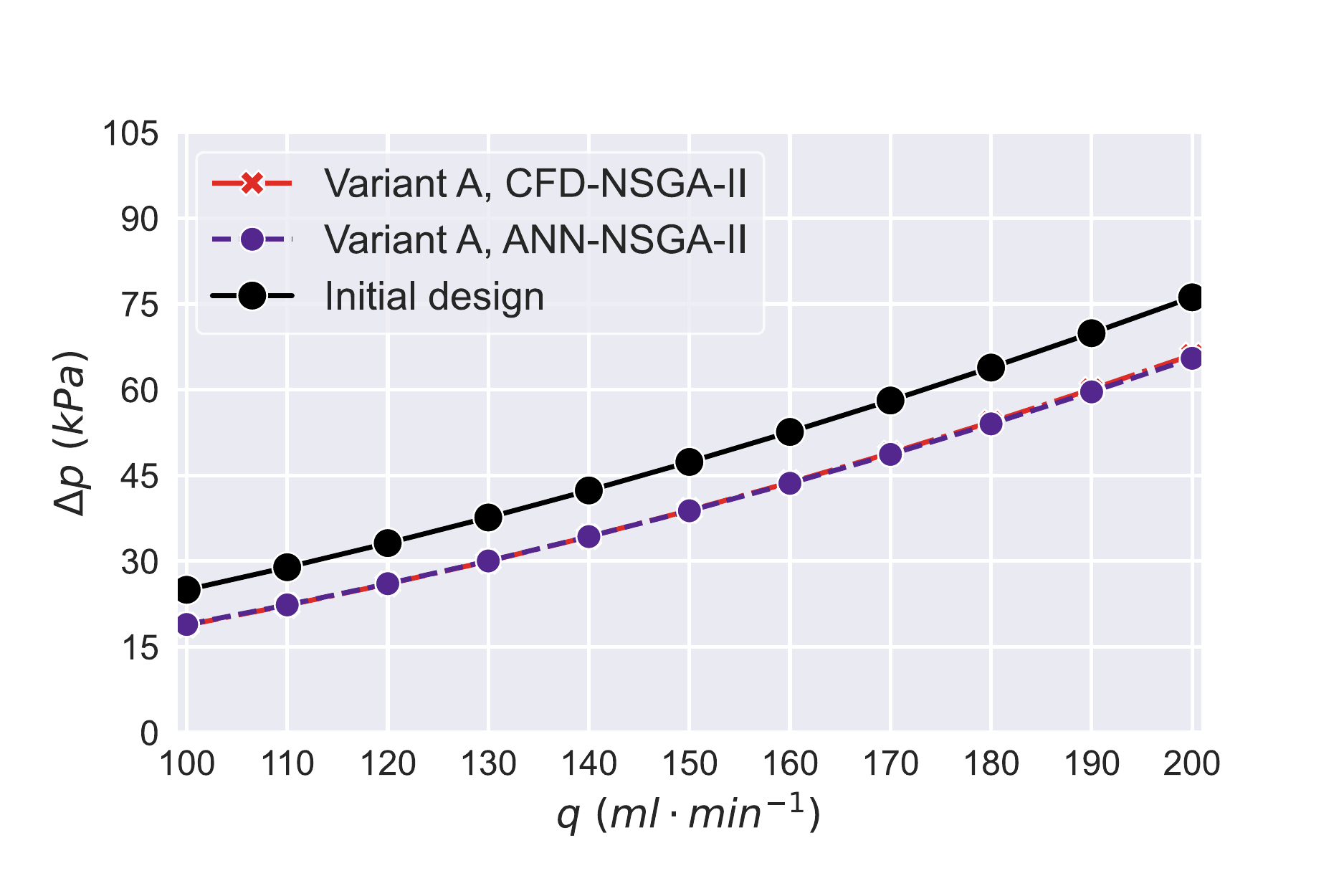}
    \caption{}
    \label{fig:rmc_A_comparison:b}
\end{subfigure}
\caption{Comparison between proposed R-MC A variants for different flow rates; (a) temperature, (b) pressure drop.}
\label{fig:rmc_A_comparison}
\end{figure}

The temperature difference between presented A variants of the R-MC model is less than 1.5\%. The CFD-NSGA-II solution achieves lower temperatures but with a 1\% increase in pressure drop. A slight improvement in thermal performance is thus obtained at the expense of an increase in pressure drop. When compared to the initial design, temperatures and pressure are decreased by, on average, 2.7\degree C and 8.3 kPa. These results are given in Figure \ref{fig:rmc_A_comparison}.

\begin{figure}[H]
\centering
\begin{subfigure}[b]{0.495\textwidth}
    \centering
    \includegraphics[trim={0cm 0.8cm 1cm 1cm}, clip, width=\textwidth]{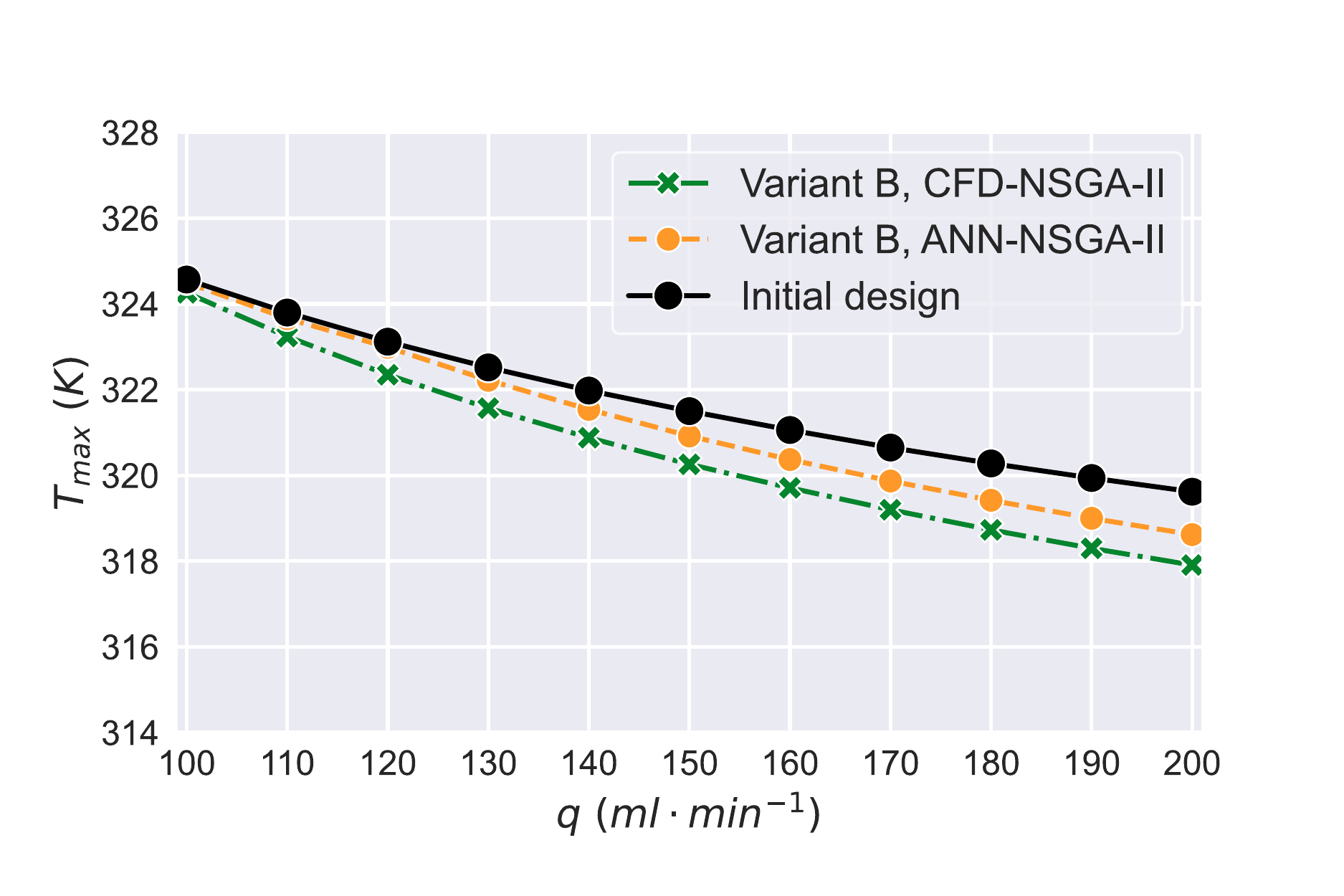}
    \caption{}
    \label{fig:rmc_B_comparison:a}
\end{subfigure}
\begin{subfigure}[b]{0.495\textwidth}
    \centering
    \includegraphics[trim={0cm 0.8cm 1cm 1cm}, clip, width=\textwidth]{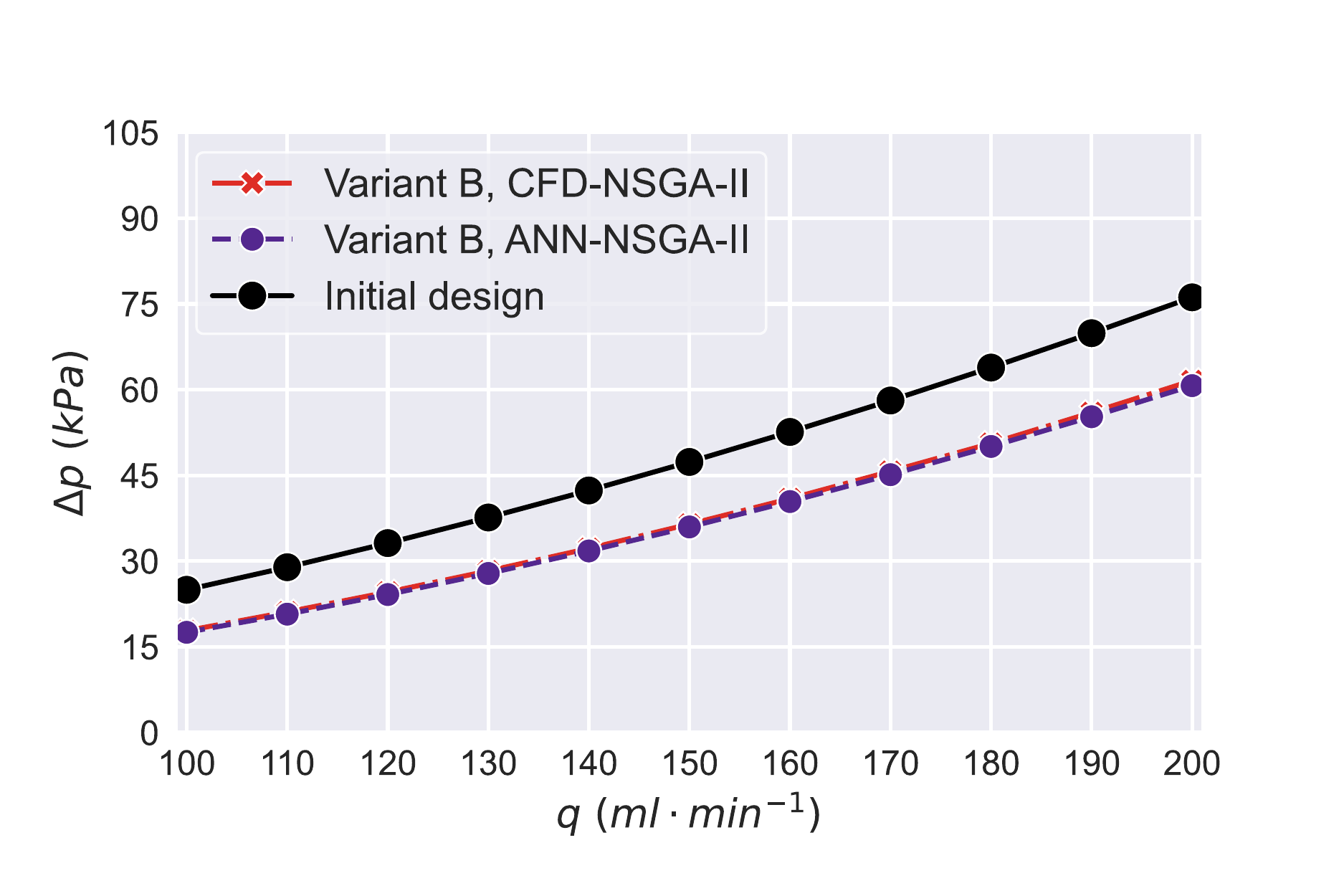}
    \caption{}
    \label{fig:rmc_B_comparison:b}
\end{subfigure}
\caption{Comparison between proposed R-MC B variants for different flow rates; (a) temperature, (b) pressure drop.}
\label{fig:rmc_B_comparison}
\end{figure}

The deviations for the B solutions are equally small. Temperature disparity is below 1.5\%. The difference in pressure drop is slightly increased, now up to 1.6\%, but the overall conclusion remains the same; the CFD-NSGA-II solution provides lower temperatures at the expense of increased pressure. $T_{max}$ and $\Delta p$ for the ANN-NSGA-II solution are on average 0.4\degree C and 11.4 kPa lower than the initial design. Figure \ref{fig:rmc_B_comparison} depicts this observation.

\begin{figure}[H]
\centering
\begin{subfigure}[b]{0.495\textwidth}
    \centering
    \includegraphics[trim={0cm 0.8cm 1cm 1cm}, clip, width=\textwidth]{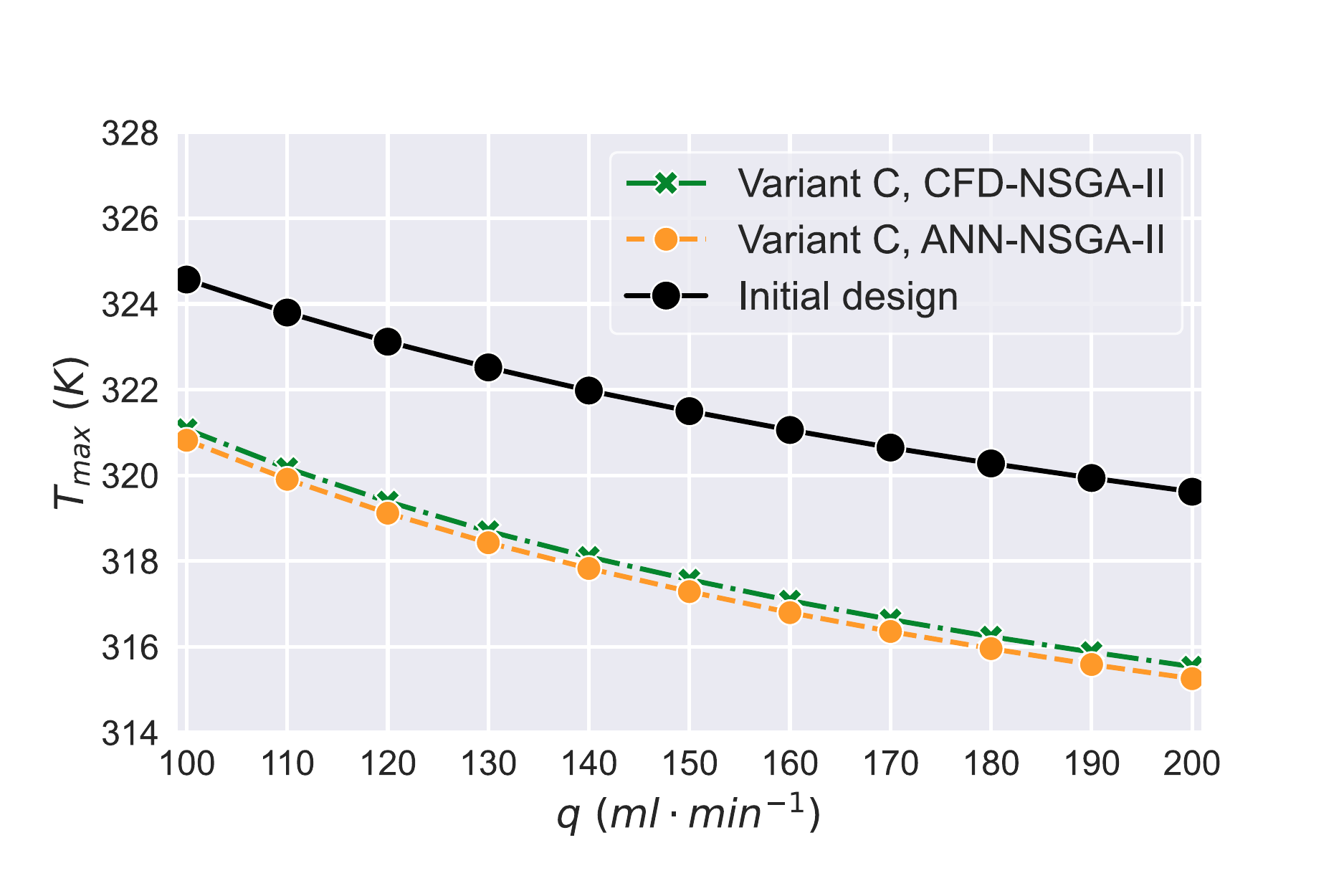}
    \caption{}
    \label{fig:rmc_C_comparison:a}
\end{subfigure}
\begin{subfigure}[b]{0.495\textwidth}
    \centering
    \includegraphics[trim={0cm 0.8cm 1cm 1cm}, clip, width=\textwidth]{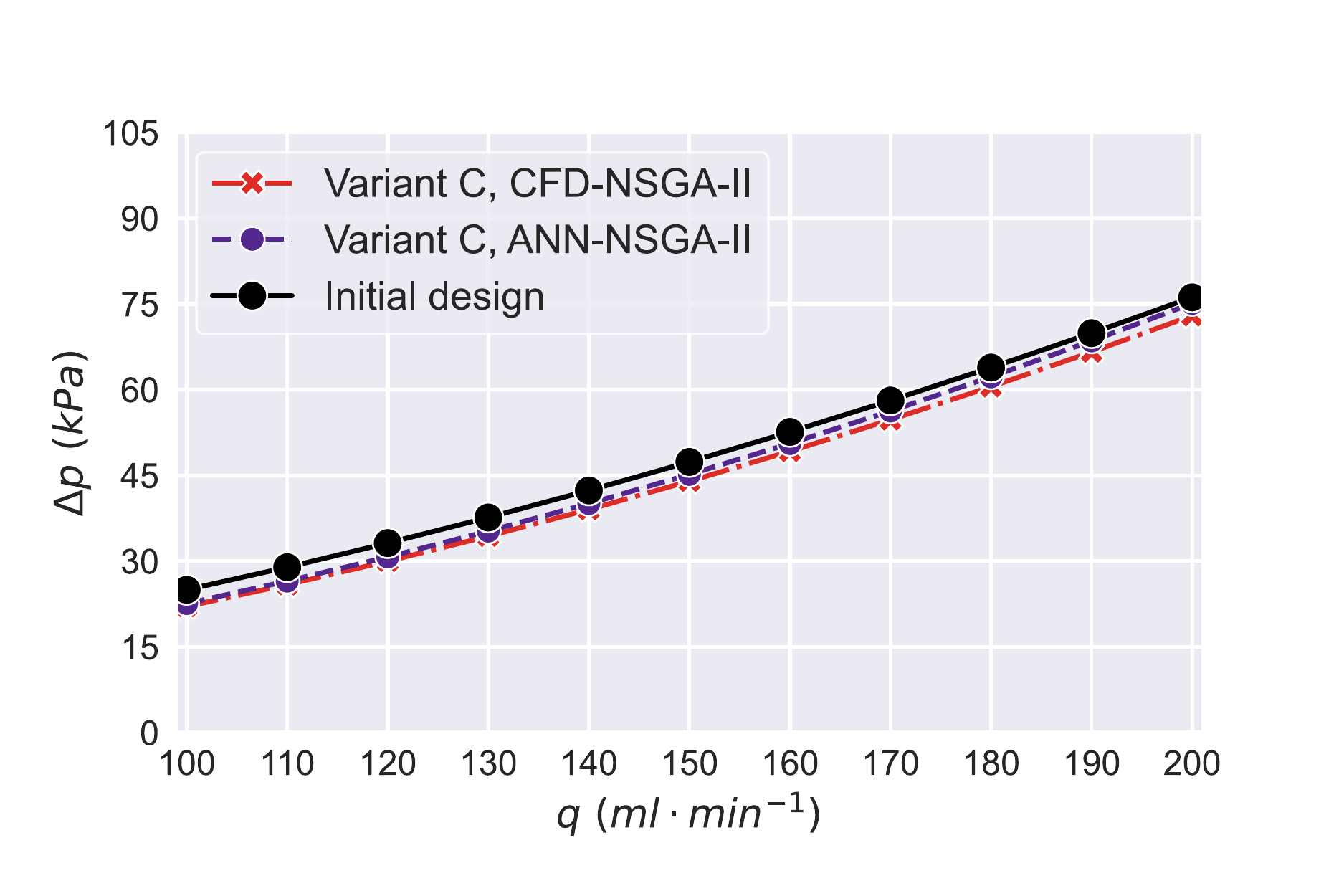}
    \caption{}
    \label{fig:rmc_C_comparison:b}
\end{subfigure}
\caption{Comparison between proposed R-MC C variants for different flow rates; (a) temperature, (b) pressure drop.}
\label{fig:rmc_C_comparison}
\end{figure}

The results shown in Figure \ref{fig:rmc_C_comparison} indicate that the proposed ANN-NSGA-II variant C solution provides the largest temperature decrease. The difference between considered variants is less than 0.7\%, however. The pressure drop is larger for the noted solution, yet it is still well below the initial design guideline. This means that if the pumping power is limited i.e. $P_{p} / P_{p,ref} \approx 1$, the ML-based approach is more flexible. When compared to the initial design, temperature and pressure drop are on average improved by 4.2\degree C and 2 kPa, respectively.

Overall, the presented optimal solutions, while different, provide comparable thermal performance. The difference is, at worst, 1.5\%, which is negligible. Observed pressure differences are slightly larger, up to 4.9\%, but still acceptable. Based on this, it is reasonable to conclude that both approaches have merit and are applicable. Differences in obtained and proposed optimal solutions can be attributed to stochasticity, ML model inadequacies (small dataset) and general CFD-induced inaccuracy (numerical errors).

When compared to the original, conventional MCHS design, the proposed solutions are clear improvements. Depending on the interpretation of the results on the Pareto front, different balanced optimums can be found; however, if we focus on the temperature and limit the pressure, it has been shown that the $T_{max}$ can be reduced by more than 4\degree C.

It should be noted that proposed solutions stem from a dataset and optimization procedure conducted for a flow rate $Q = 150$ ml$\cdot$min$^{-1}$. Since MCHS are sensitive to flow rate, the behavior of the trendlines and overall results might have been different if the optimization process had been conducted at a different flow rate.

\section{Conclusion}
\label{sec:conclusion}

In this paper, multi-objective optimization techniques used to improve the performance of microchannel designs were investigated. The study focused on two different and widely used MCHS design concepts, namely microchannel heat sinks with secondary channels and ribs. A numerical procedure was implemented to evaluate MCHS performance. Validated numerical models were coupled with an NSGA-II-based multi-objective optimization workflow. In such a defined workflow, conflicting goal functions, thermal resistance and pumping power, were used to determine a series of optimal design solutions.

A computationally efficient and effective methodology was hence proposed to generate optimal designs. It uses Latin Hypercube Sampling and CFD to generate a dataset of unique design solutions. Thus obtained data are used to train ANN models which can subsequently be used for a variety of purposes, but primarily in conjunction with an optimization method to provide optimal designs. The use of ML-based surrogates has several advantages over traditional approaches. One important aspect is that the validity of the obtained surrogate model can be easily determined by analyzing the learning curve. Additionally, ML-based surrogates can be seamlessly assessed to determine key aspects/parameters that govern the physics of the problem, even if they do not fully explain the physics of the problem. Key findings are as follows:
\begin{itemize}
    \item Ensemble learning and gradient boosting methods provide respectable results without tuning. However, tuned ANNs were shown to outperform tuned alternatives. The proposed ML-based methodology can generate acceptable solutions while requiring only about 20\% of the computational time. Solutions obtained using CFD-based and ML-based methodologies have different topological designs but similar thermal performance.
    \item Optimal ML-generated solutions provide lower maximum temperatures (by more than 4\degree C) under the same pressure drop limit. A significant reduction in pressure drop (more than 25\%) can be achieved without an increase in temperature. The SC-MC MCHS design outperforms the R-MC design. The differences are, however, below 0.5\degree C.
    \item Conducted SHAP analysis indicated that the channel width $d$ and the number of secondary channels/ribs $m$ are the main variables governing the thermal resistance. Pumping power is mainly driven by the width $d$ and the number of microchannels $n$. The relevance of every design variable in the optimization process is presented.
\end{itemize}

\added{As demonstrated, the proposed methodology provides acceptable solutions and allows for quick turnaround times in the design process. It combines previously established principles and methods with novel machine learning concepts, which should improve the overall understanding of the design variable influence and thus facilitate the optimization and design process, in addition to improving performance. The applicability of various ML algorithms is also evaluated, emphasizing the importance of proper assessment in the context of a given problem.} Applicability should be further investigated by assessing different designs, combining varying models into a unified model, and investigating the influence of the dataset size on predictive accuracy.
\newpage

\appendix
\setcounter{figure}{0}

\section{}
\label{sec:appendix:a}

\begin{figure}[H]
\centering
\begin{subfigure}[b]{0.325\textwidth}
    \centering
    \includegraphics[trim={16cm 1.5cm 17cm 1.5cm}, clip, width=\textwidth]{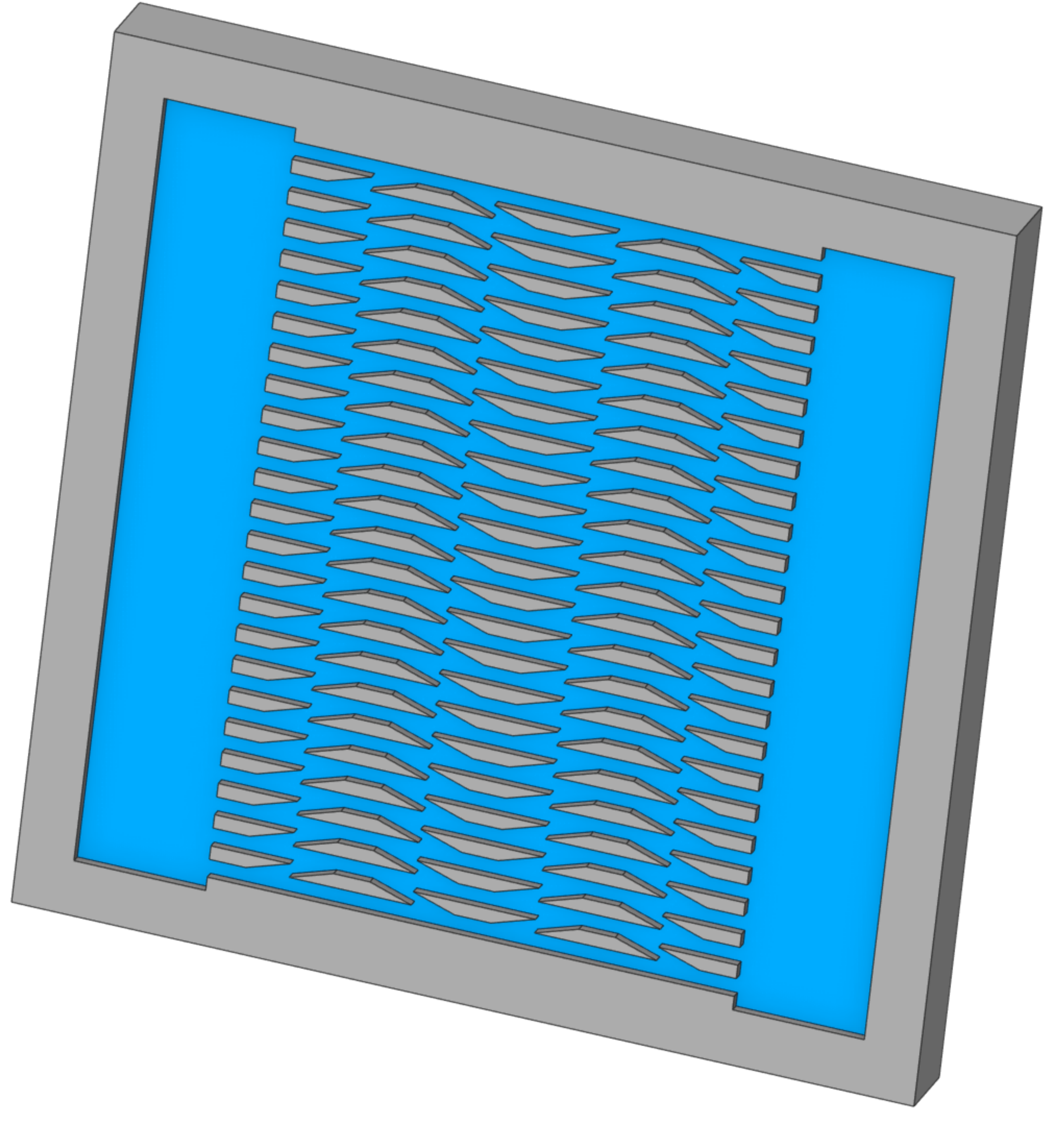}
    \caption{}
    \label{fig:CFD_MO_scmc:a}
\end{subfigure}
\begin{subfigure}[b]{0.325\textwidth}
    \centering
    \includegraphics[trim={16cm 1.5cm 17cm 1.5cm}, clip, width=\textwidth]{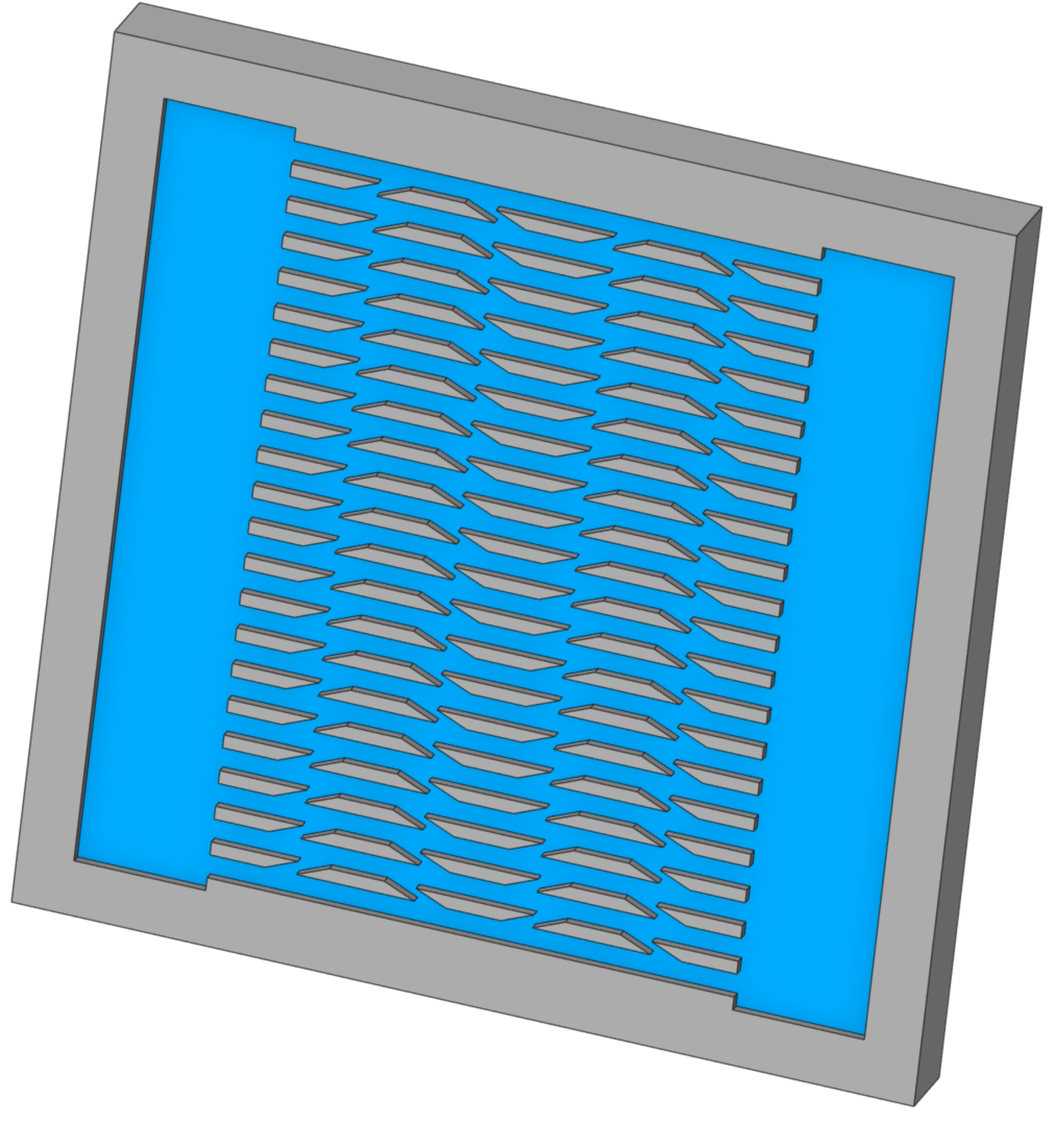}
    \caption{}
    \label{fig:CFD_MO_scmc:b}
\end{subfigure}
\begin{subfigure}[b]{0.325\textwidth}
    \centering
    \includegraphics[trim={16cm 1.5cm 17cm 1.5cm}, clip, width=\textwidth]{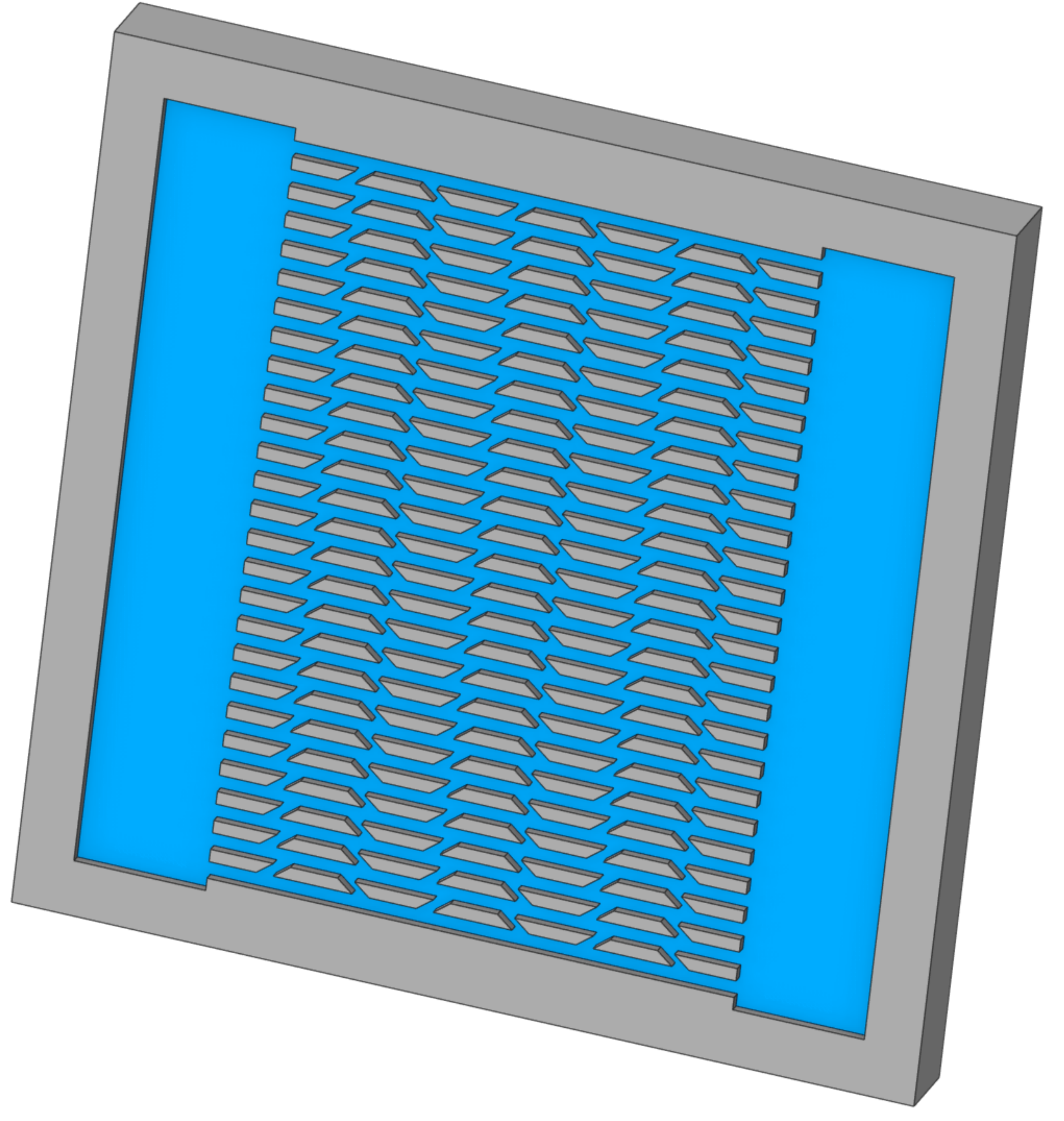}
    \caption{}
    \label{fig:CFD_MO_scmc:c}
\end{subfigure}
\caption{Cross-sectional view of the optimal SC-MC designs obtained using coupled CFD-NSGA-II approach; (a) variant A, (b) variant B, (c) variant C. Relevant parameters for each design are included in Table \ref{tab:design_cfd_best}. Performance specifics of each design are given in Section \ref{subsec:cfd_based_section}. Position on the Pareto front for each variant is given in Figure \ref{fig:pareto_cfd:a}.}
\label{fig:CFD_MO_scmc}
\end{figure}

\begin{figure}[H]
\centering
\begin{subfigure}[b]{0.325\textwidth}
    \centering
    \includegraphics[trim={16cm 1.5cm 17cm 1.5cm}, clip, width=\textwidth]{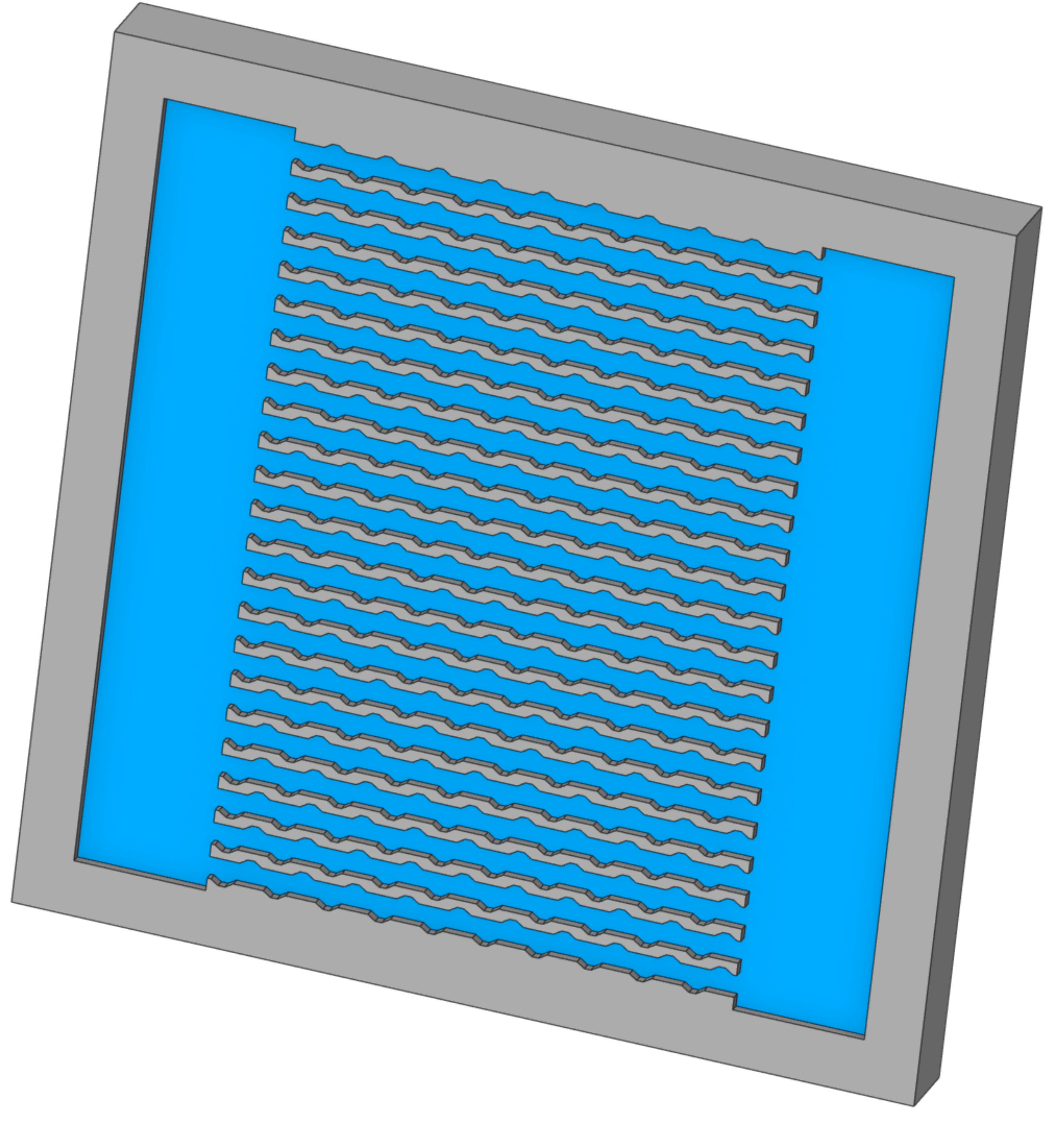}
    \caption{}
    \label{fig:CFD_MO_rmc:a}
\end{subfigure}
\begin{subfigure}[b]{0.325\textwidth}
    \centering
    \includegraphics[trim={16cm 1.5cm 17cm 1.5cm}, clip, width=\textwidth]{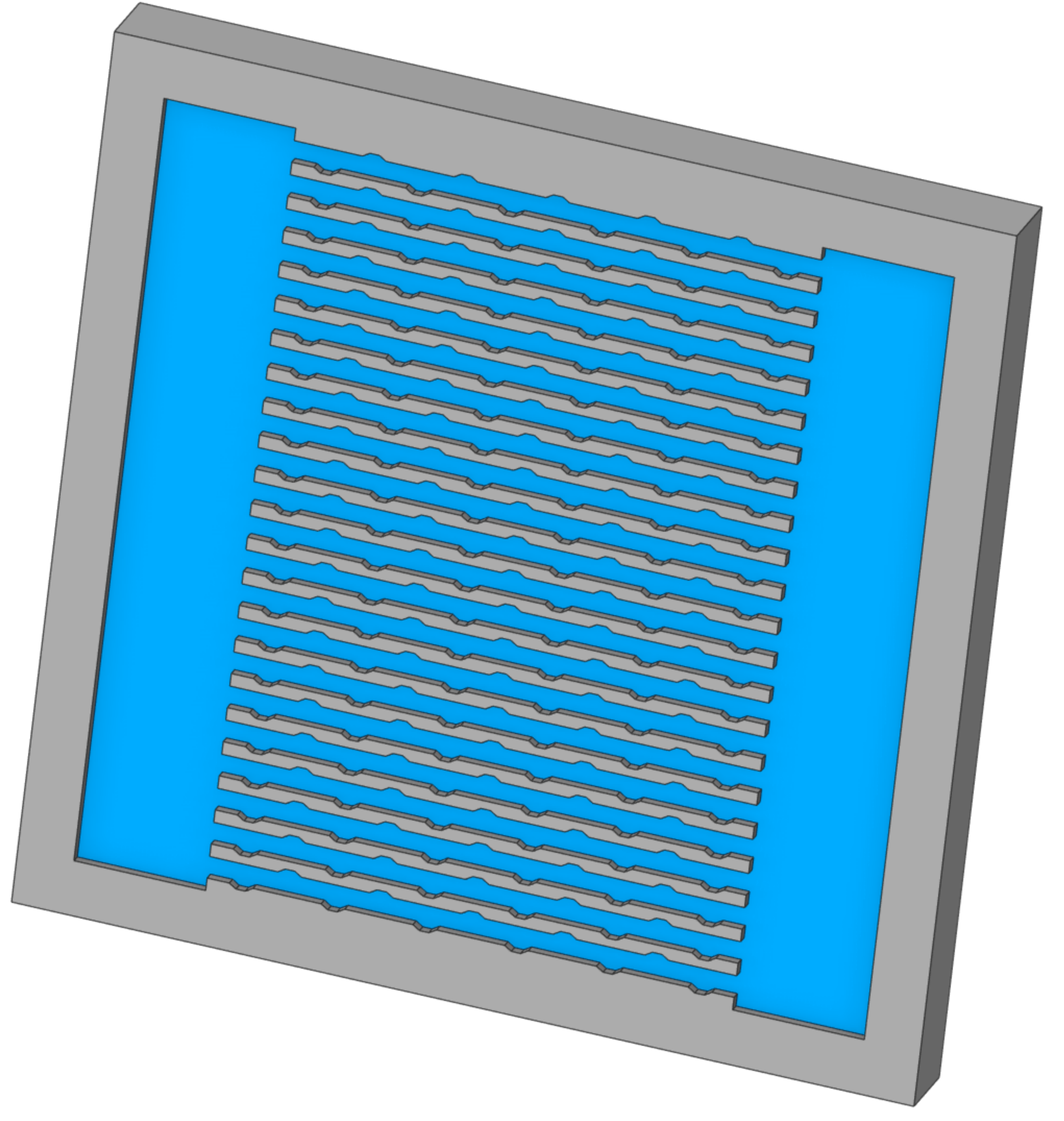}
    \caption{}
    \label{fig:CFD_MO_rmc:b}
\end{subfigure}
\begin{subfigure}[b]{0.325\textwidth}
    \centering
    \includegraphics[trim={16cm 1.5cm 17cm 1.5cm}, clip, width=\textwidth]{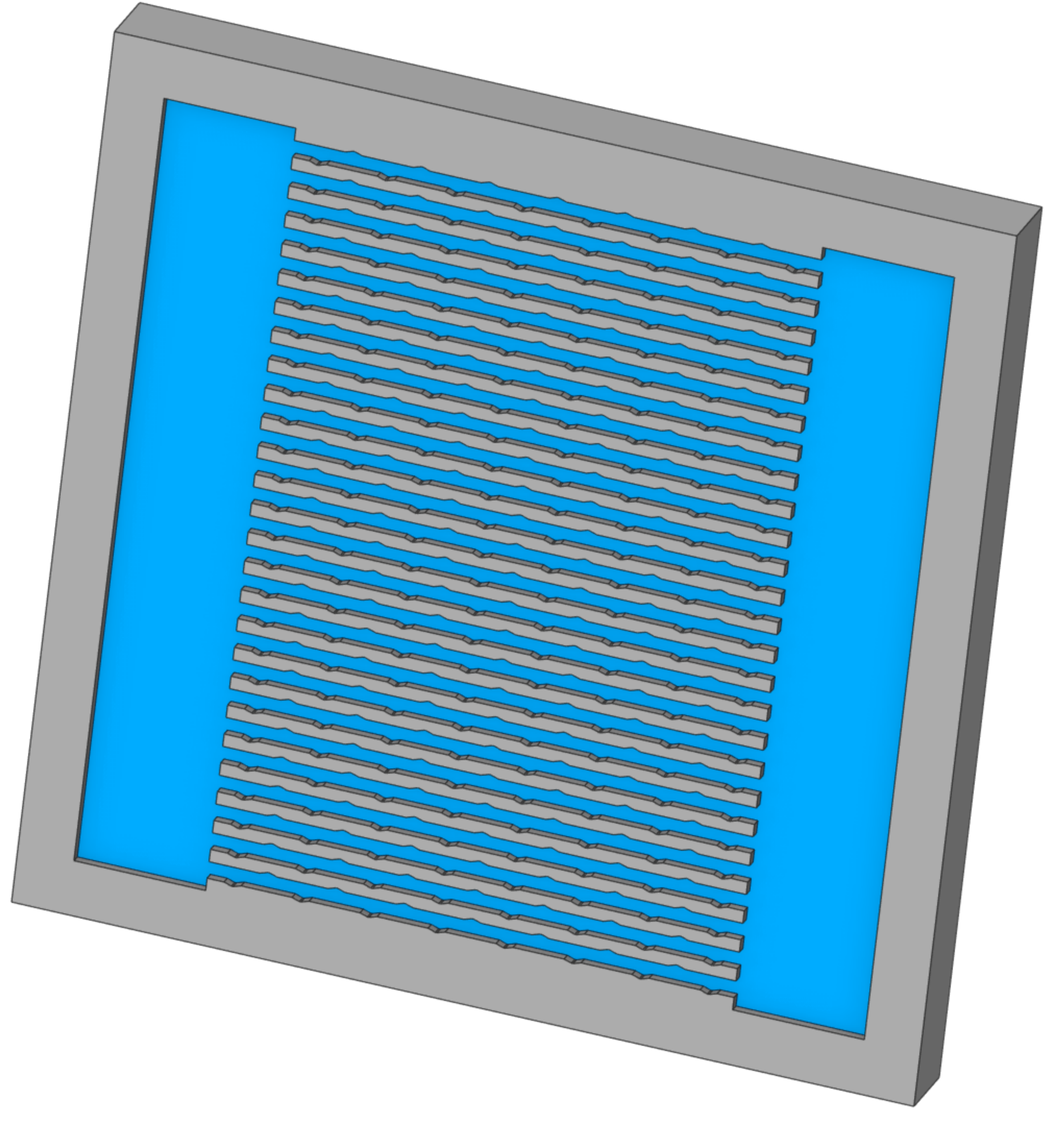}
    \caption{}
    \label{fig:CFD_MO_rmc:c}
\end{subfigure}
\caption{Cross-sectional view of the optimal R-MC designs obtained using coupled CFD-NSGA-II approach; (a) variant A, (b) variant B, (c) variant C. Relevant parameters for each design are included in Table \ref{tab:design_cfd_best}. Performance specifics of each design are given in Section \ref{subsec:cfd_based_section}. Position on the Pareto front for each variant is given in Figure \ref{fig:pareto_cfd:b}.}
\label{fig:CFD_MO_rmc}
\end{figure}

\begin{figure}[H]
\centering
\begin{subfigure}[b]{0.325\textwidth}
    \centering
    \includegraphics[trim={16cm 1.5cm 17cm 1.5cm}, clip, width=\textwidth]{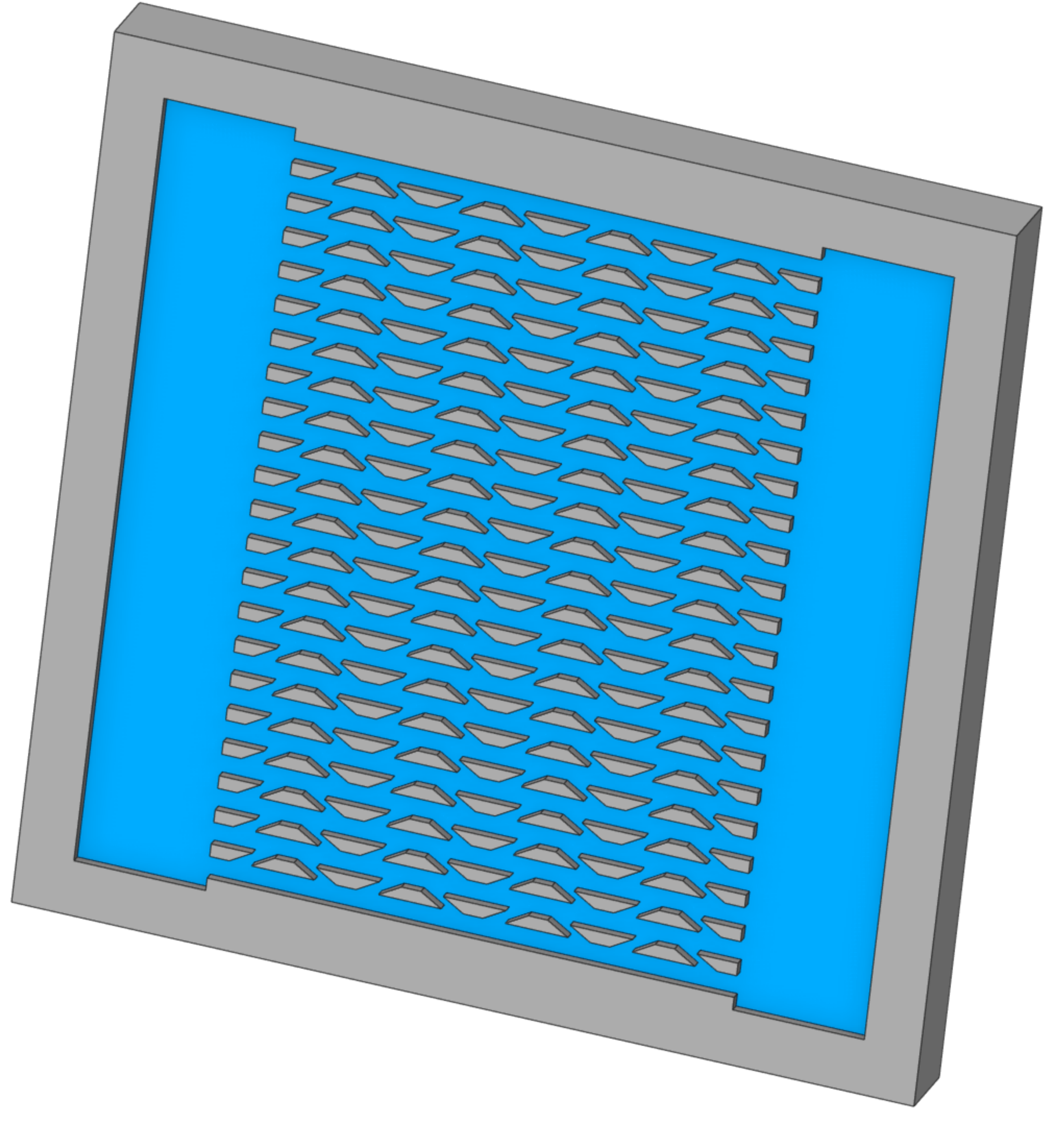}
    \caption{}
    \label{fig:ANN_MO_scmc:a}
\end{subfigure}
\begin{subfigure}[b]{0.325\textwidth}
    \centering
    \includegraphics[trim={16cm 1.5cm 17cm 1.5cm}, clip, width=\textwidth]{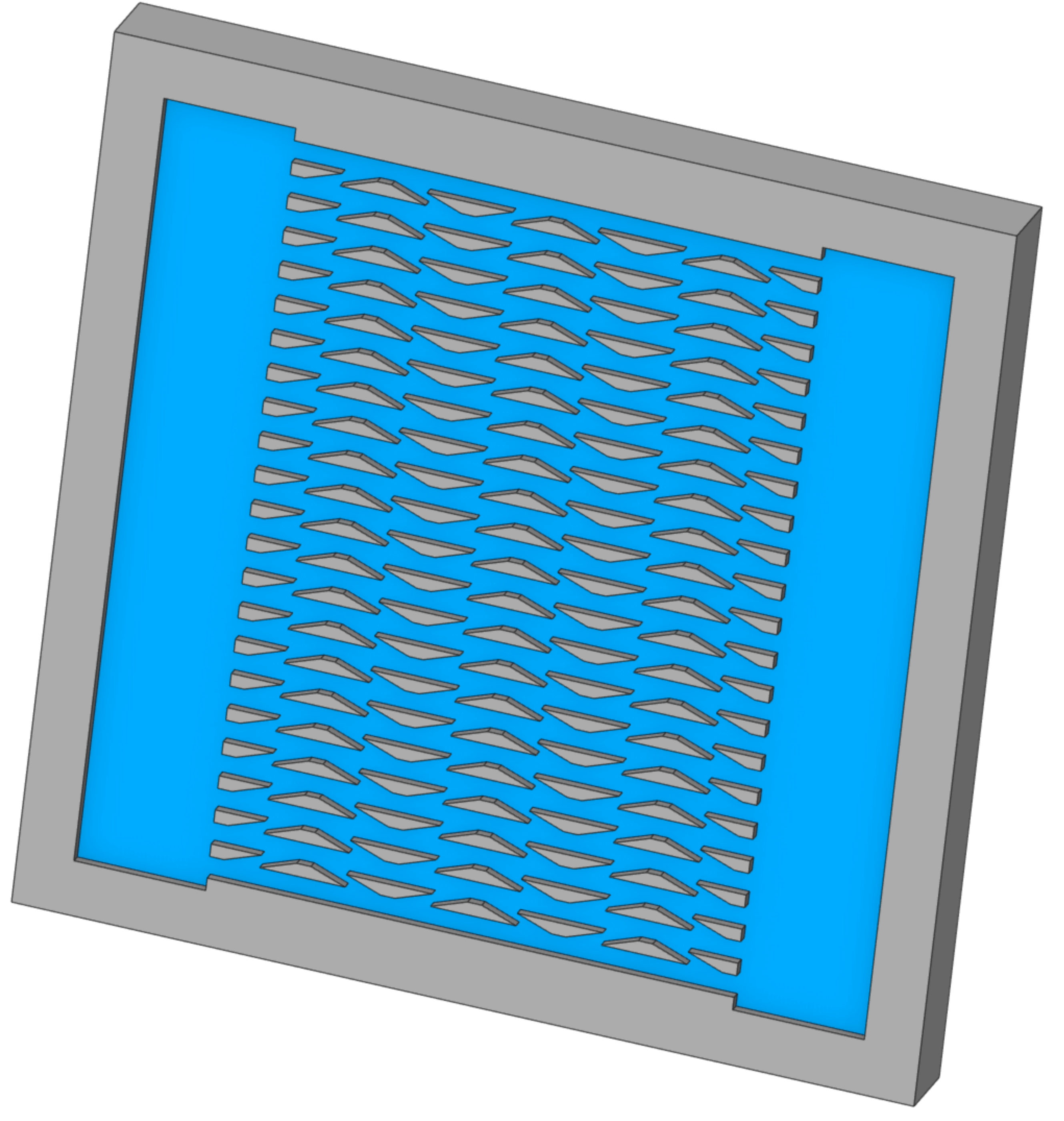}
    \caption{}
    \label{fig:ANN_MO_scmc:b}
\end{subfigure}
\begin{subfigure}[b]{0.325\textwidth}
    \centering
    \includegraphics[trim={16cm 1.5cm 17cm 1.5cm}, clip, width=\textwidth]{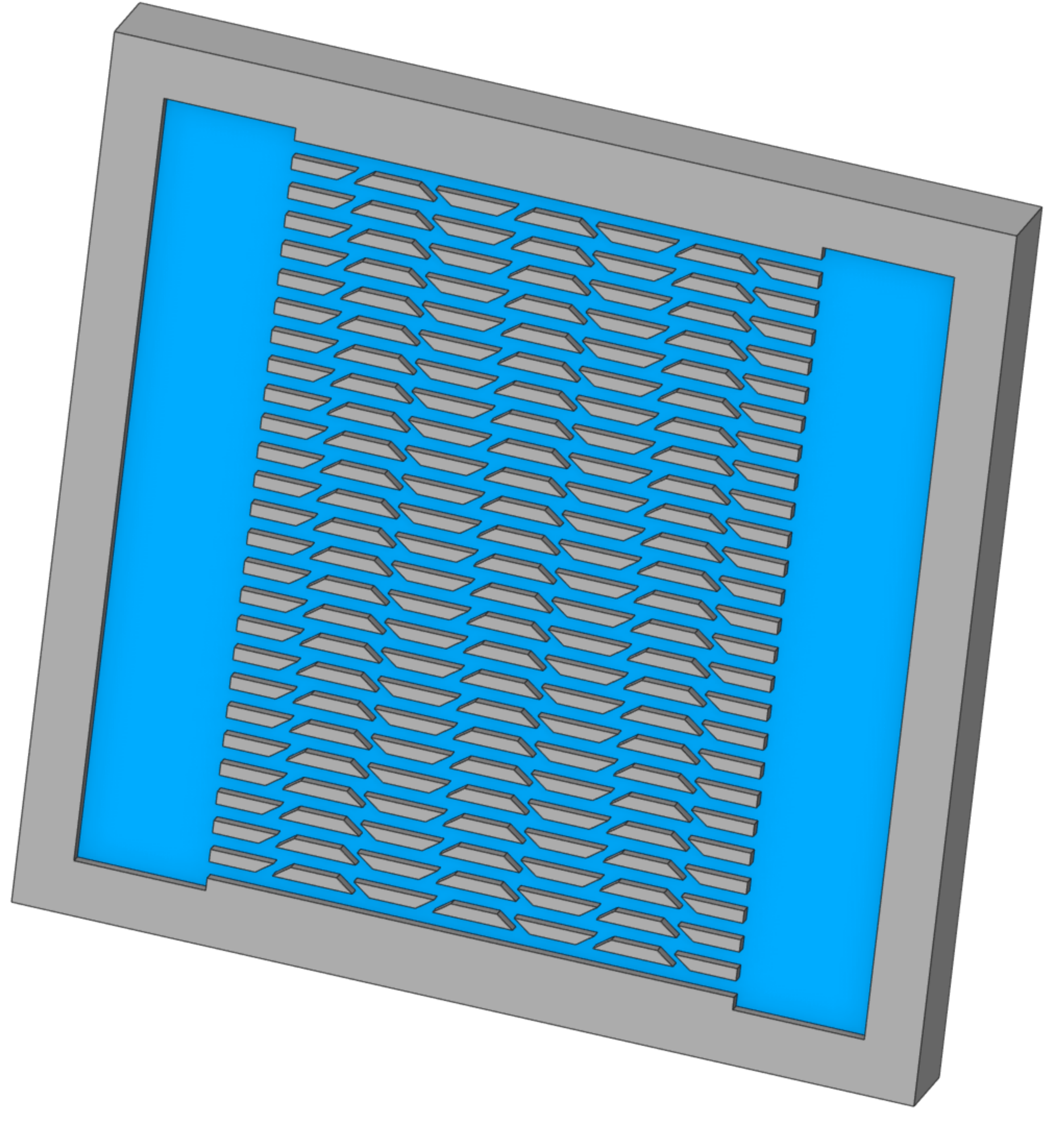}
    \caption{}
    \label{fig:ANN_MO_scmc:c}
\end{subfigure}
\caption{Cross-sectional view of the optimal SC-MC designs obtained using ANN-NSGA-II coupling; (a) variant A, (b) variant B, (c) variant C.  Relevant parameters for each design are included in Table \ref{tab:design_ann_best}. Performance specifics of each design are given in Section \ref{subsec:ann_based_section}. Position on the Pareto front for each variant is given in Figure \ref{fig:pareto_ann:a}.}
\label{fig:ANN_MO_scmc}
\end{figure}

\begin{figure}[H]
\centering
\begin{subfigure}[b]{0.325\textwidth}
    \centering
    \includegraphics[trim={16cm 1.5cm 17cm 1.5cm}, clip, width=\textwidth]{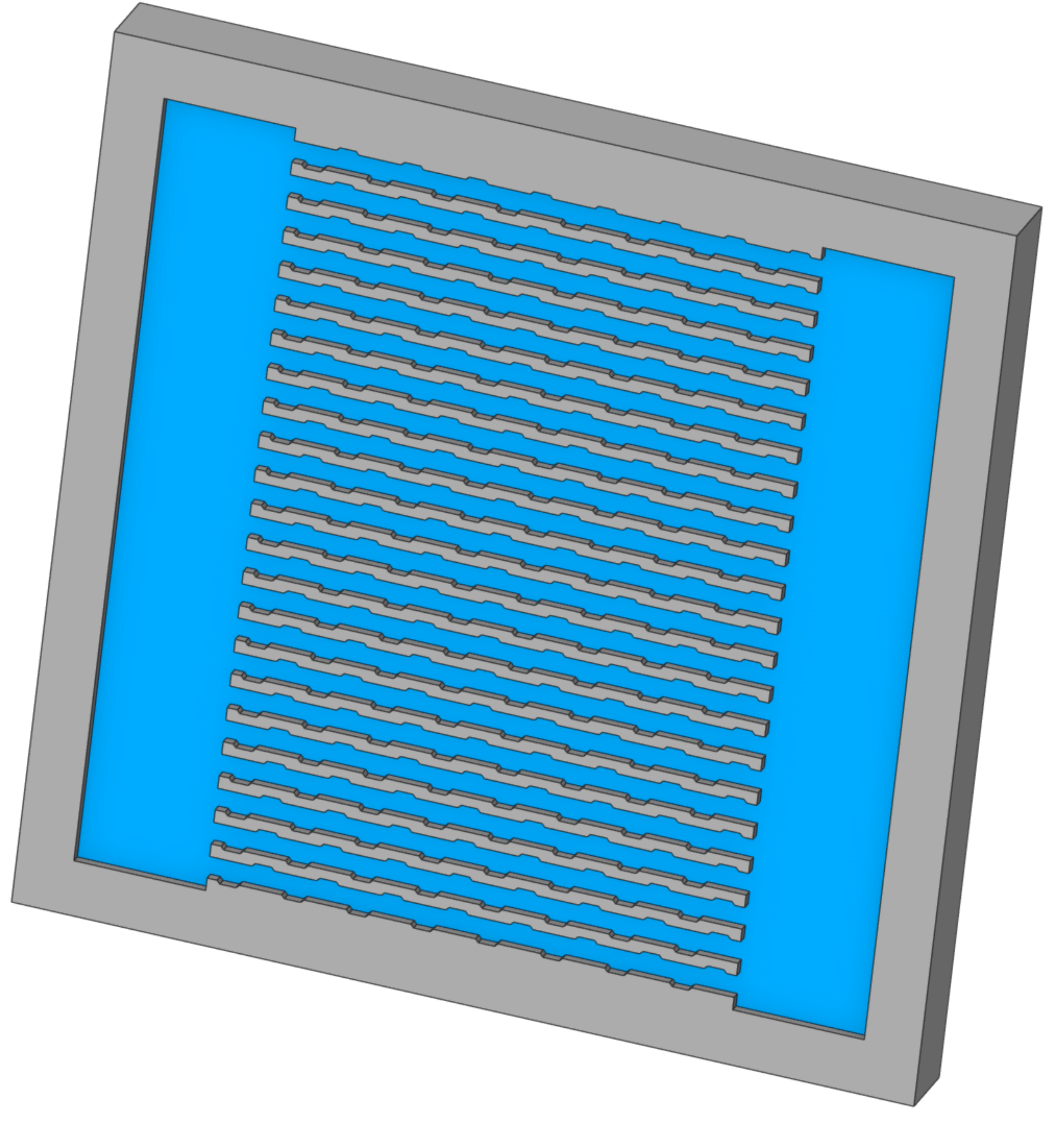}
    \caption{}
    \label{fig:ANN_MO_rmc:a}
\end{subfigure}
\begin{subfigure}[b]{0.325\textwidth}
    \centering
    \includegraphics[trim={16cm 1.5cm 17cm 1.5cm}, clip, width=\textwidth]{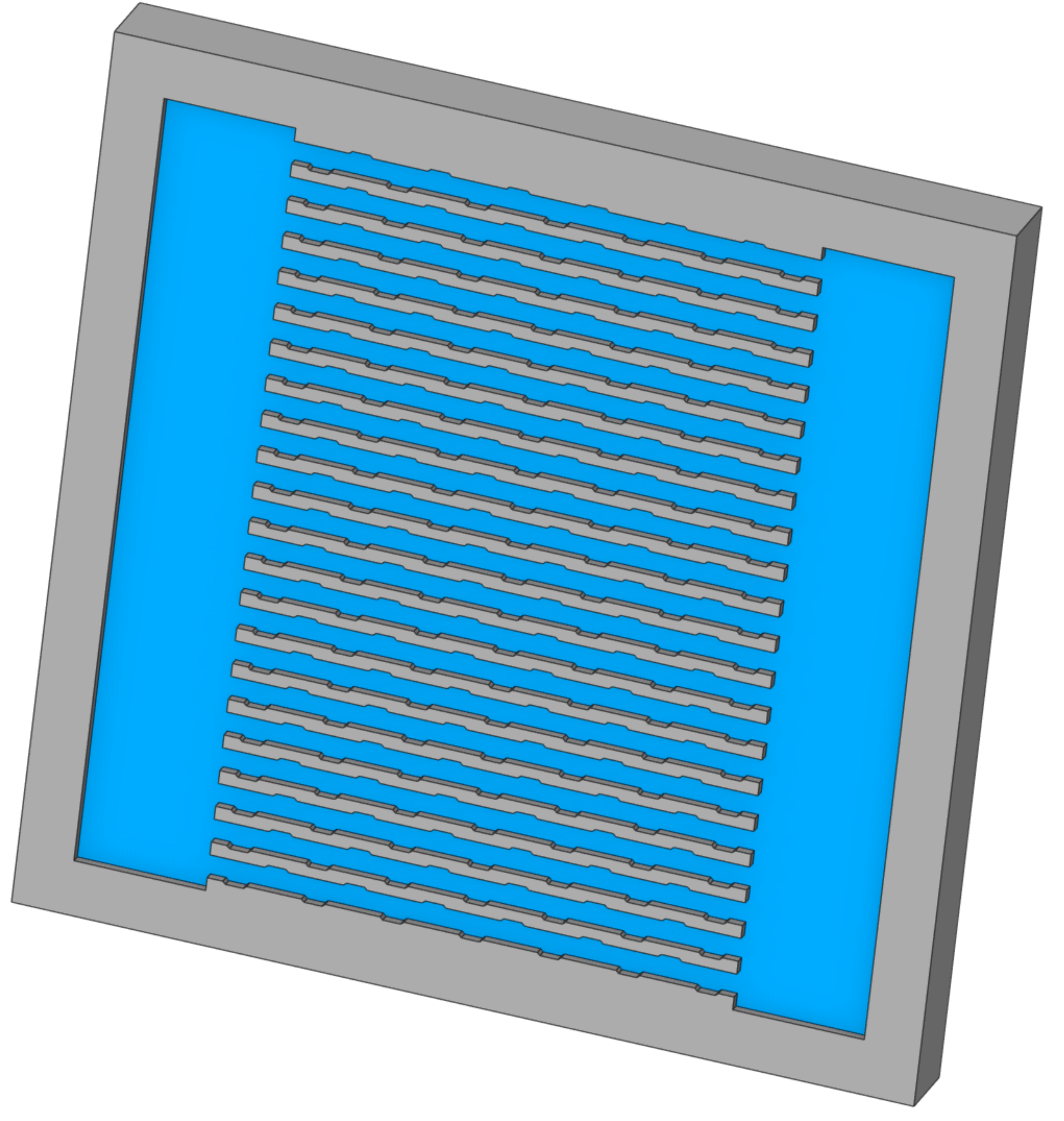}
    \caption{}
    \label{fig:ANN_MO_rmc:b}
\end{subfigure}
\begin{subfigure}[b]{0.325\textwidth}
    \centering
    \includegraphics[trim={16cm 1.5cm 17cm 1.5cm}, clip, width=\textwidth]{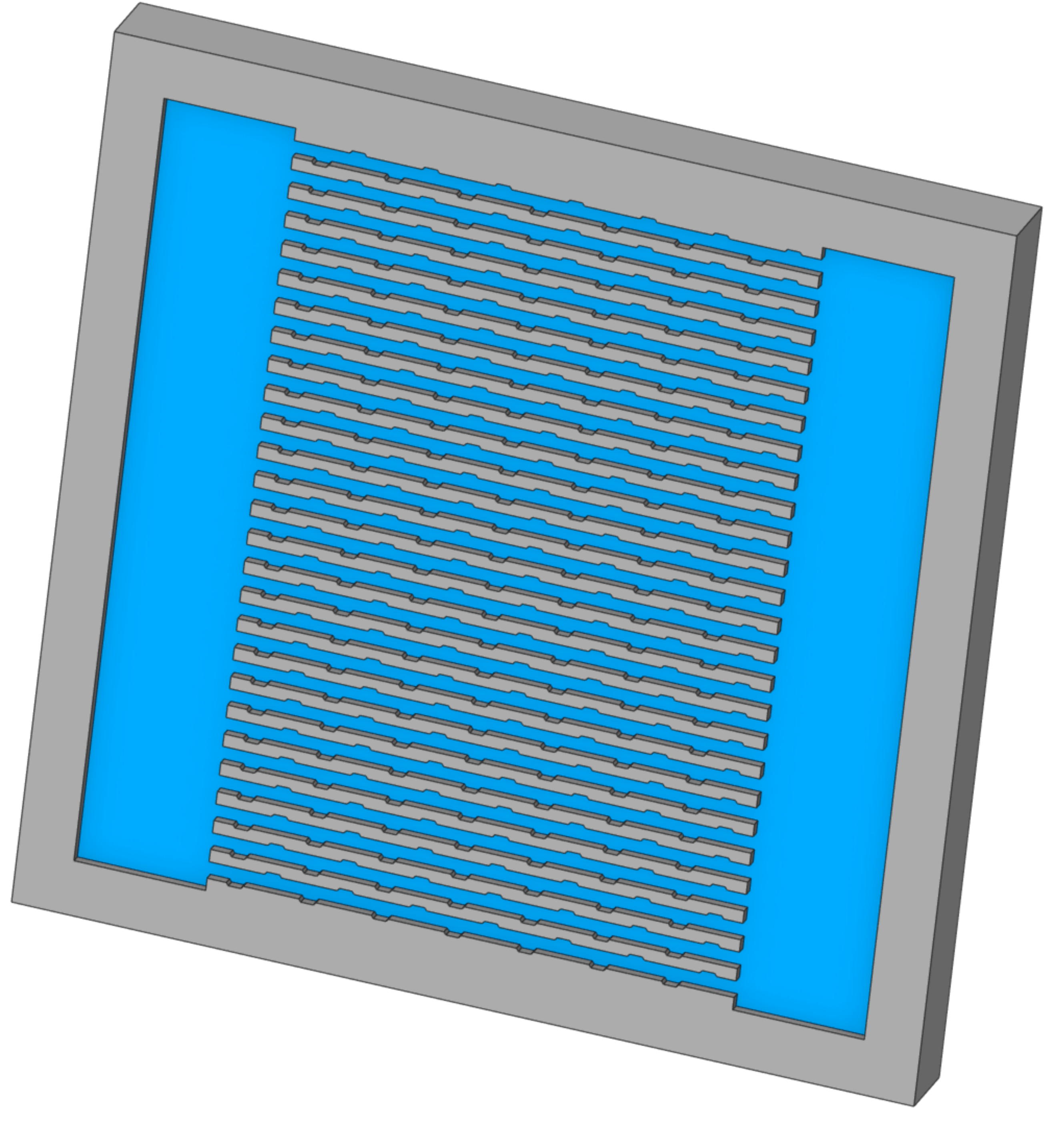}
    \caption{}
    \label{fig:ANN_MO_rmc:c}
\end{subfigure}
\caption{Cross-sectional view of the optimal R-MC designs obtained using ANN-NSGA-II coupling; (a) variant A, (b) variant B, (c) variant C. Relevant parameters for each design are included in Table \ref{tab:design_ann_best}. Performance specifics of each design are given in Section \ref{subsec:ann_based_section}. Position on the Pareto front for each variant is given in Figure \ref{fig:pareto_ann:b}.}
\label{fig:ANN_MO_rmc}
\end{figure}

\begin{figure}[H]
\centering
\includegraphics[trim={0cm 0cm 0cm 0cm}, clip, width=0.75\textwidth]{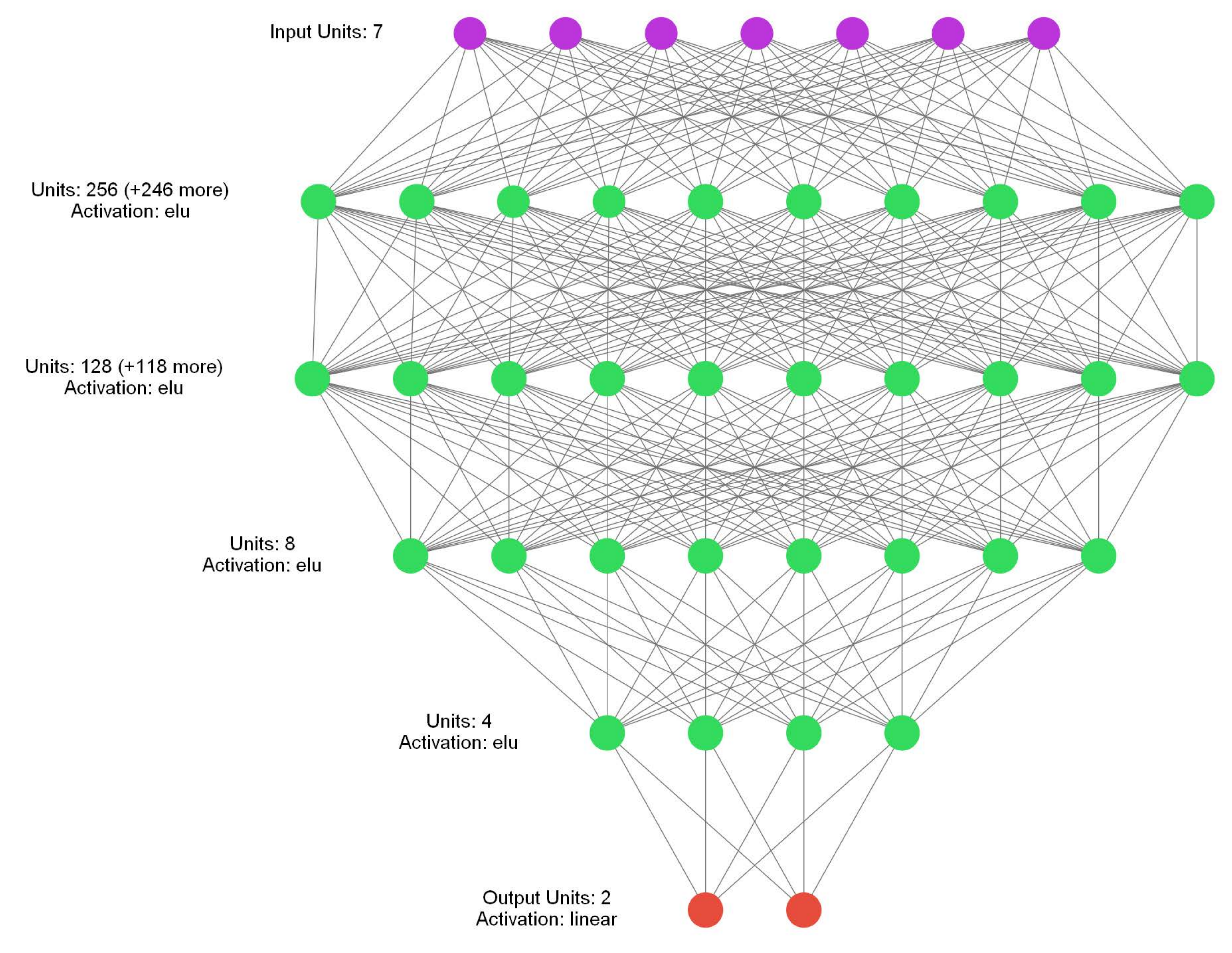}
\caption{Topology of the tuned ANN employed in the SC-MC optimization workflow. An input layer, four hidden layers, and an output layer comprise the network. The number of neurons per layer is shown. In total, 7 features are considered. The model outputs thermal resistance $R_t$ and pumping power $P_p$.}
\label{fig:ANN_network_scmc}
\end{figure}

\begin{figure}[H]
\centering
\includegraphics[trim={0cm 0cm 0cm 0cm}, clip, width=0.75\textwidth]{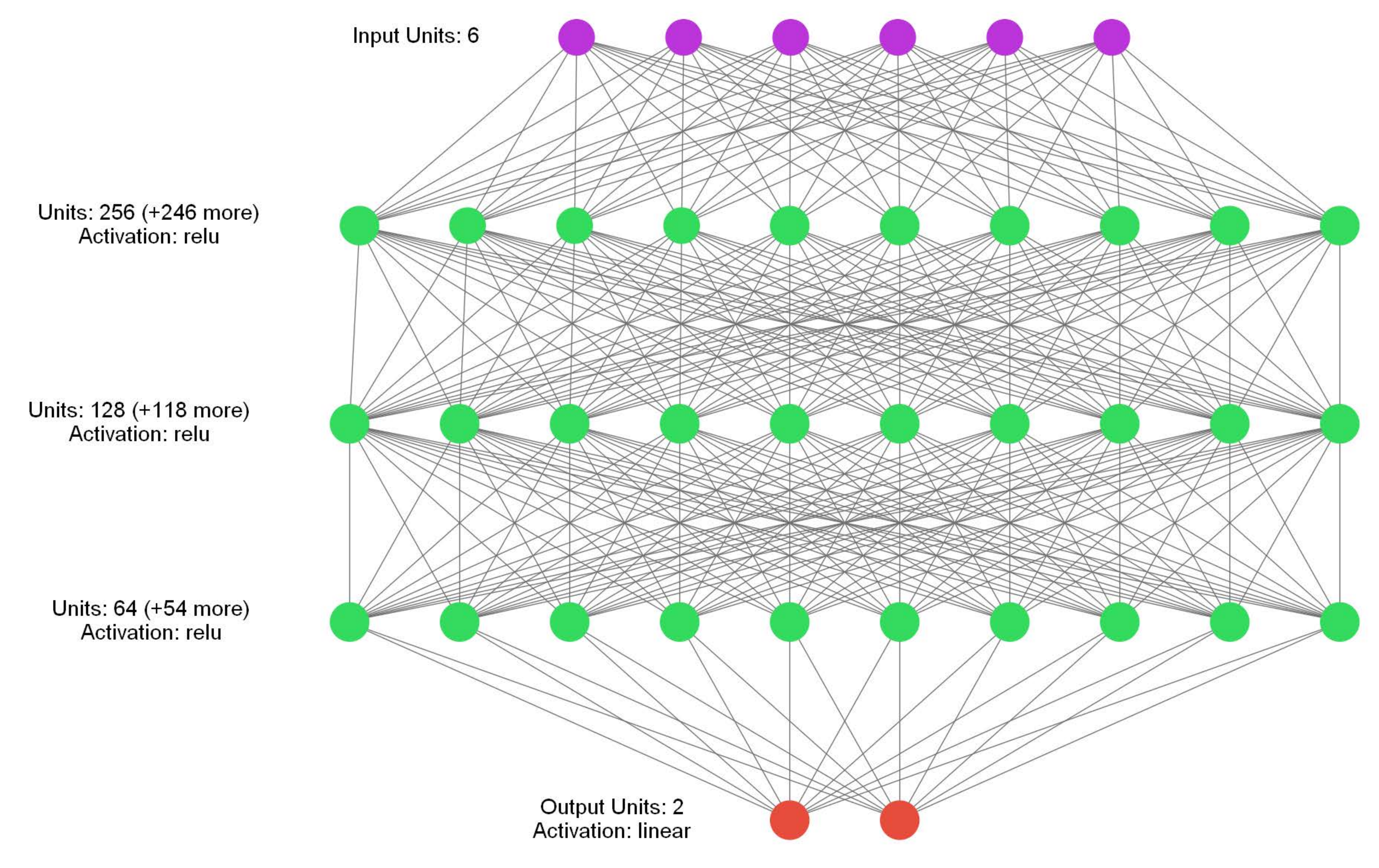}
\caption{Topology of the tuned ANN employed in the R-MC optimization workflow. An input layer, three hidden layers, and an output layer comprise the network. The number of neurons per layer is shown. In total, 6 features are considered. The model outputs thermal resistance $R_t$ and pumping power $P_p$.}
\label{fig:ANN_network_rmc}
\end{figure}

\begin{figure}[H]
\centering
\begin{subfigure}[b]{0.495\textwidth}
    \centering
    \includegraphics[trim={0cm 0.8cm 1cm 1cm}, clip, width=\textwidth]{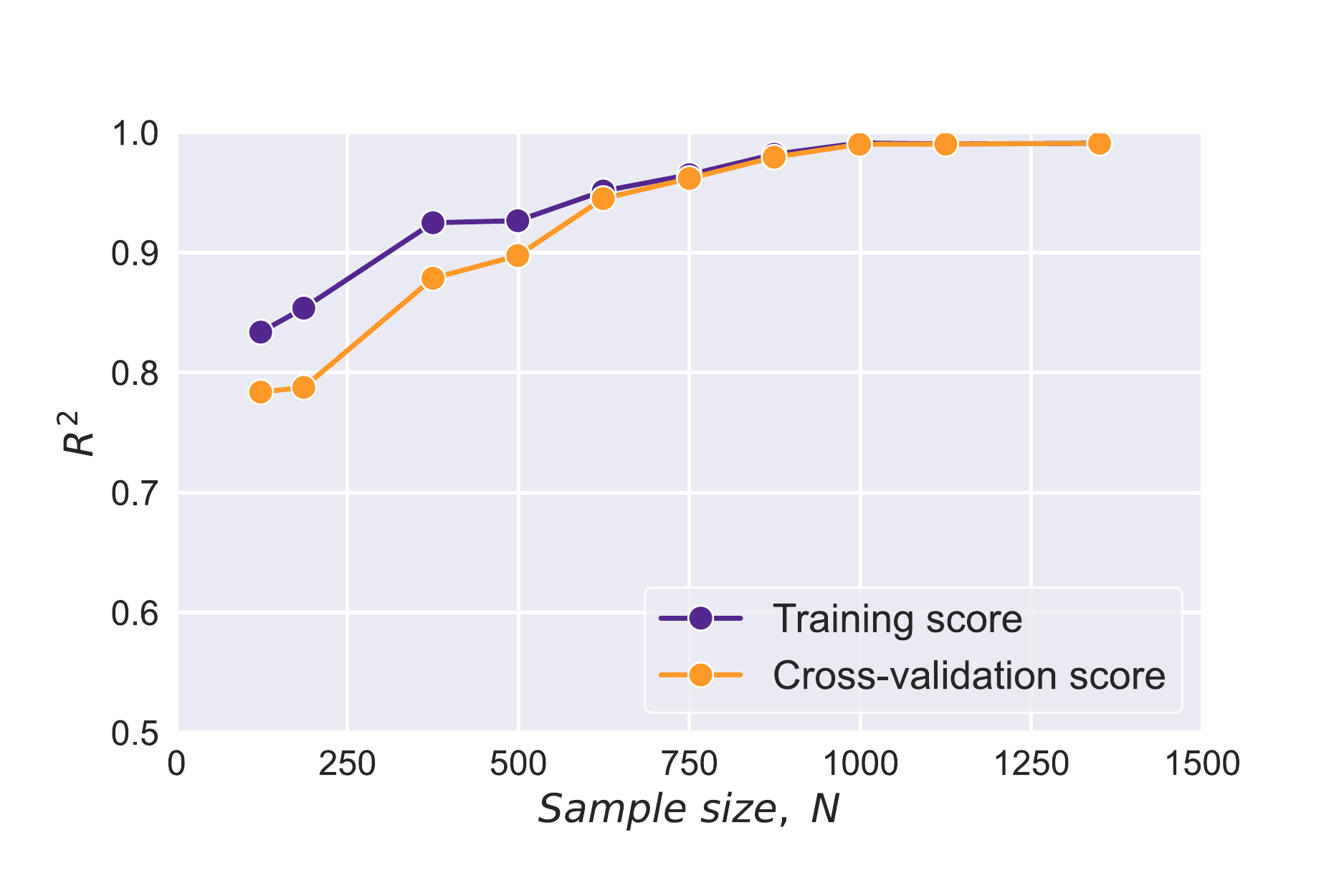}
    \caption{}
    \label{fig:learning_curve:a}
\end{subfigure}
\begin{subfigure}[b]{0.495\textwidth}
    \centering
    \includegraphics[trim={0cm 0.8cm 1cm 1cm}, clip, width=\textwidth]{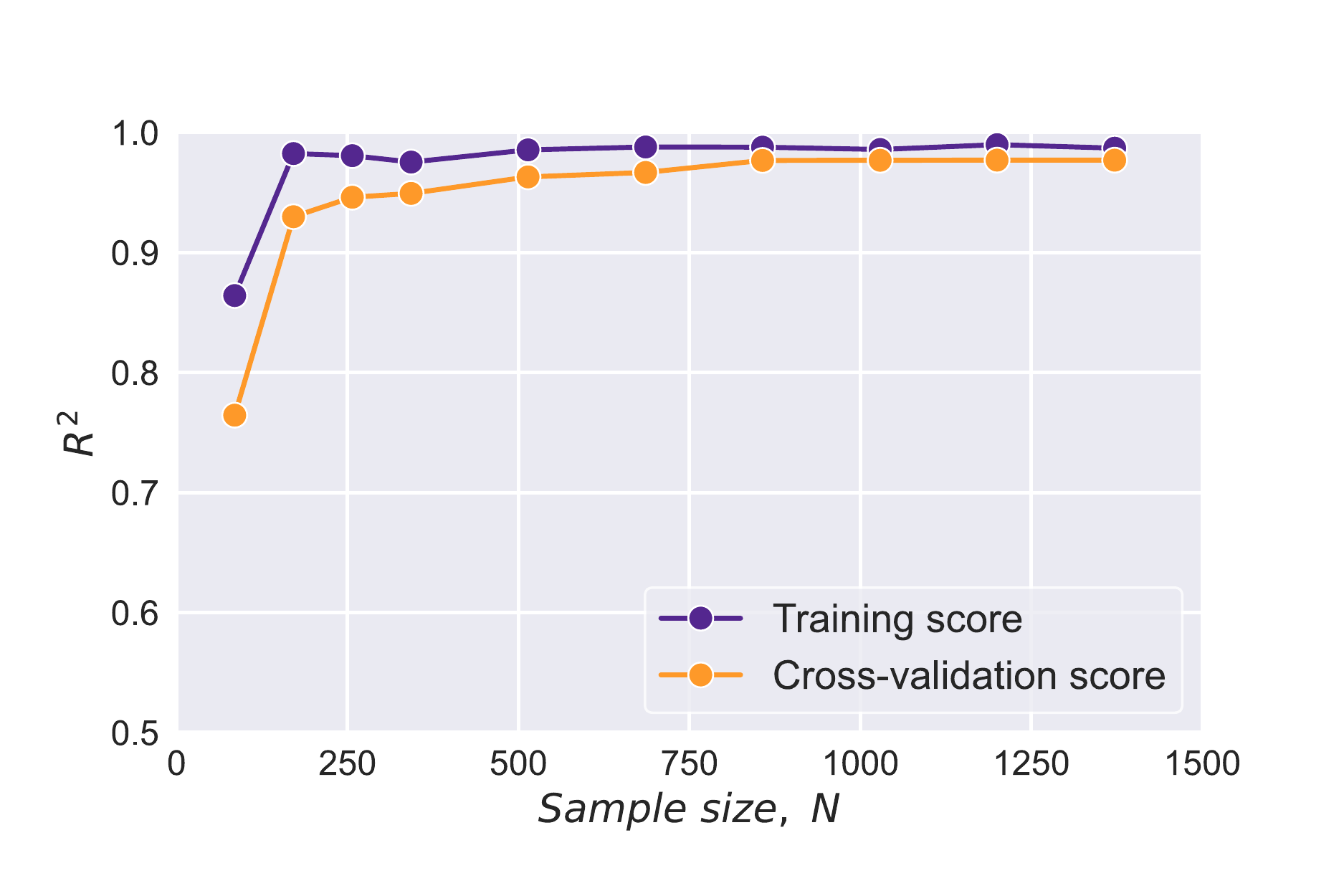}
    \caption{}
    \label{fig:learning_curve:b}
\end{subfigure}
\caption{Learning curves obtained using the tuned ANN models; (a) for the SC-MC design, (b) for the R-MC design. Learning curves depict the change in $R_2$ score as the size of the dataset used to train the model increases. It is evident that for $N > 1000$, both considered models are sufficiently accurate.}
\label{fig:learning_curve}
\end{figure}

\begin{figure}[H]
\centering
\begin{subfigure}[b]{0.495\textwidth}
    \centering
    \includegraphics[trim={0cm 0.8cm 1cm 1cm}, clip, width=\textwidth]{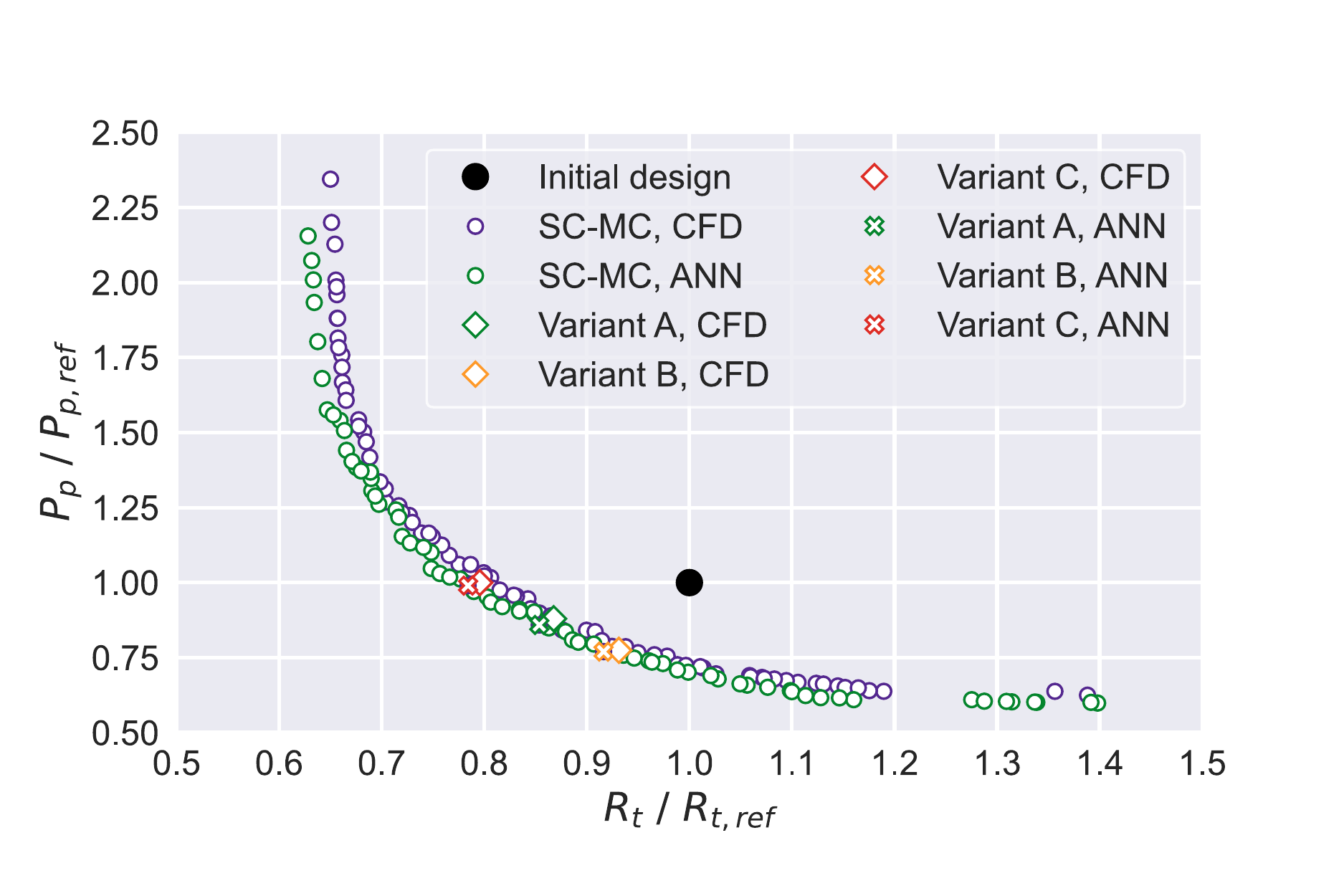}
    \caption{}
    \label{fig:pareto_comp:a}
\end{subfigure}
\begin{subfigure}[b]{0.495\textwidth}
    \centering
    \includegraphics[trim={0cm 0.8cm 1cm 1cm}, clip, width=\textwidth]{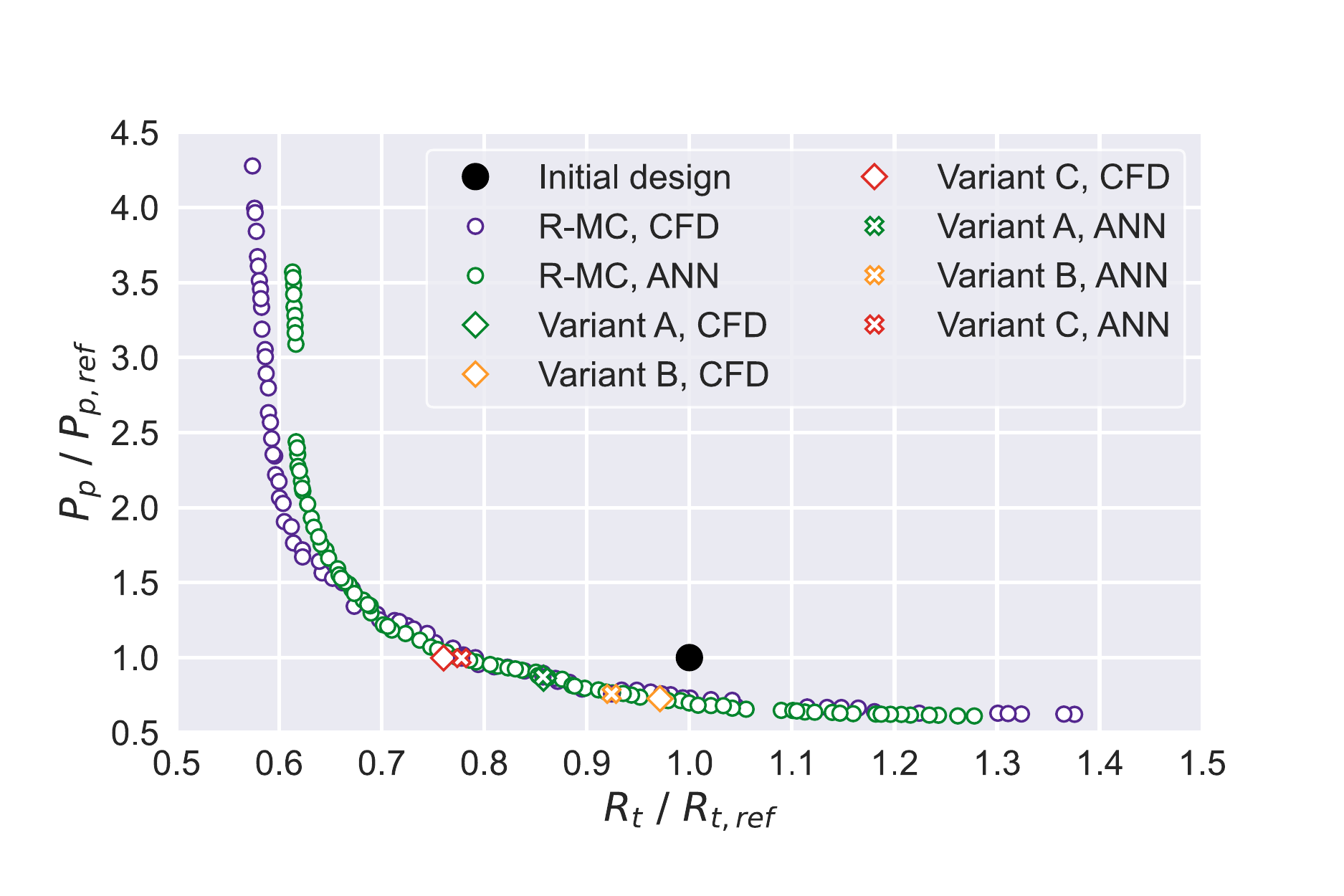}
    \caption{}
    \label{fig:pareto_comp:b}
\end{subfigure}
\caption{Comparison of obtained Pareto fronts; (a) fronts for the SC-MC design, (b) fronts for the R-MC design. Each figure includes fronts obtained using CFD-based and ANN-based approaches as well as optimal variants for each case. The initial design is depicted with a black dot.}
\label{fig:pareto_comp}
\end{figure}

\section{}
\label{sec:appendix:b}
Supplementary data on Pareto fronts discussed in this paper can be found online at \url{https://gitlab.com/sikirica_a/mchs_data}.
\newpage

\bibliographystyle{elsarticle-num-names}
\bibliography{bibliography}

\end{document}